\documentclass[3p]{elsarticle}

\usepackage{lineno}

\usepackage{amsmath}
\usepackage{amssymb}
\usepackage{subfigure}
\usepackage{epsfig}
\usepackage{siunitx}
\DeclareSIUnit\Molar{M}     %

\usepackage{color}

\usepackage{hyperref}   %

\makeatletter %

\def\input@path{{01_sections/}{02_figures/}}

\newcommand{\defeq}{\overset{!}{=}}
\newcommand{\defvariable}{:=}%

\newcommand{\bigO}{\mathcal{O}}%

\newcommand{\secref}[1]{Section~\ref{#1}}%
\newcommand{\figref}[1]{Figure~\ref{#1}}%

\newcommand\DeclareBoldMathCommand[2]{%
  \protected@edef\@tempb{%
    \noexpand\DeclareRobustCommand{\csname #1\endcsname}{\boldsymbol{\ensuremath{#2}}}}
  \@tempb}
\@tfor\AM@letter:=abcdefghijklmnopqrstuvwxyz%
\do{\DeclareBoldMathCommand{v\AM@letter}{\AM@letter}}
\@tfor\AM@letter:=ABCDEFGHIJKLMNOPQRSTUVWXYZ%
\do{\DeclareBoldMathCommand{v\AM@letter}{\AM@letter}}
\DeclareBoldMathCommand{vnull}{0}
\DeclareBoldMathCommand{vone}{1}
\DeclareBoldMathCommand{valpha}{\alpha}
\DeclareBoldMathCommand{vbeta}{\beta}
\DeclareBoldMathCommand{vgamma}{\gamma}
\DeclareBoldMathCommand{vdelta}{\delta}
\DeclareBoldMathCommand{vepsilon}{\epsilon}
\DeclareBoldMathCommand{vvarepsilon}{\varepsilon}
\DeclareBoldMathCommand{vzeta}{\zeta}
\DeclareBoldMathCommand{veta}{\eta}
\DeclareBoldMathCommand{vtheta}{\theta}
\DeclareBoldMathCommand{vvartheta}{\vartheta}
\DeclareBoldMathCommand{viota}{\iota}
\DeclareBoldMathCommand{vkappa}{\kappa}
\DeclareBoldMathCommand{vlambda}{\lambda}
\DeclareBoldMathCommand{vmu}{\mu}
\DeclareBoldMathCommand{vnu}{\nu}
\DeclareBoldMathCommand{vpi}{\pi}
\DeclareBoldMathCommand{vxi}{\xi}
\DeclareBoldMathCommand{vrho}{\varrho}
\DeclareBoldMathCommand{vsigma}{\sigma}
\DeclareBoldMathCommand{vtau}{\tau}
\DeclareBoldMathCommand{vupsilon}{\upsilon}
\DeclareBoldMathCommand{vphi}{\varphi}
\DeclareBoldMathCommand{vchi}{\chi}
\DeclareBoldMathCommand{vpsi}{\psi}
\DeclareBoldMathCommand{vomega}{\omega}
\DeclareBoldMathCommand{vGamma}{\Gamma}
\DeclareBoldMathCommand{vDelta}{\Delta}
\DeclareBoldMathCommand{vTheta}{\Theta}
\DeclareBoldMathCommand{vLambda}{\Lambda}
\DeclareBoldMathCommand{vXi}{\Xi}
\DeclareBoldMathCommand{vPi}{\Pi}
\DeclareBoldMathCommand{vSigma}{\Sigma}
\DeclareBoldMathCommand{vUpsilon}{\Upsilon}
\DeclareBoldMathCommand{vPhi}{\Phi}
\DeclareBoldMathCommand{vPsi}{\Psi}
\DeclareBoldMathCommand{vOmega}{\Omega}

\newcommand\DeclareDiscreteBoldMathCommand[2]{%
  \protected@edef\@tempc{%
    \noexpand\DeclareRobustCommand{\csname #1\endcsname}{\boldsymbol{\mathrm{#2}}}}
  \@tempc}
\@tfor\AM@letter:=abcdefghijklmnopqrstuvwxyz%
\do{\DeclareDiscreteBoldMathCommand{vd\AM@letter}{\AM@letter}}
\@tfor\AM@letter:=ABCDEFGHIJKLMNOPQRSTUVWXYZ%
\do{\DeclareDiscreteBoldMathCommand{vd\AM@letter}{\AM@letter}}
\DeclareDiscreteBoldMathCommand{vdnull}{0}
\DeclareDiscreteBoldMathCommand{vdone}{1}
\DeclareDiscreteBoldMathCommand{vdalpha}{\alpha}
\DeclareDiscreteBoldMathCommand{vdbeta}{\beta}
\DeclareDiscreteBoldMathCommand{vdgamma}{\gamma}
\DeclareDiscreteBoldMathCommand{vddelta}{\delta}
\DeclareDiscreteBoldMathCommand{vdepsilon}{\epsilon}
\DeclareDiscreteBoldMathCommand{vdvarepsilon}{\varepsilon}
\DeclareDiscreteBoldMathCommand{vdzeta}{\zeta}
\DeclareDiscreteBoldMathCommand{vdeta}{\eta}
\DeclareDiscreteBoldMathCommand{vdtheta}{\theta}
\DeclareDiscreteBoldMathCommand{vdvartheta}{\vartheta}
\DeclareDiscreteBoldMathCommand{vdiota}{\iota}
\DeclareDiscreteBoldMathCommand{vdkappa}{\kappa}
\DeclareDiscreteBoldMathCommand{vdlambda}{\lambda}
\DeclareDiscreteBoldMathCommand{vdmu}{\mu}
\DeclareDiscreteBoldMathCommand{vdnu}{\nu}
\DeclareDiscreteBoldMathCommand{vdpi}{\pi}
\DeclareDiscreteBoldMathCommand{vdxi}{\xi}
\DeclareDiscreteBoldMathCommand{vdrho}{\varrho}
\DeclareDiscreteBoldMathCommand{vdsigma}{\sigma}
\DeclareDiscreteBoldMathCommand{vdtau}{\tau}
\DeclareDiscreteBoldMathCommand{vdupsilon}{\upsilon}
\DeclareDiscreteBoldMathCommand{vdphi}{\varphi}
\DeclareDiscreteBoldMathCommand{vdchi}{\chi}
\DeclareDiscreteBoldMathCommand{vdpsi}{\psi}
\DeclareDiscreteBoldMathCommand{vdomega}{\omega}
\DeclareDiscreteBoldMathCommand{vdGamma}{\Gamma}
\DeclareDiscreteBoldMathCommand{vdDelta}{\Delta}
\DeclareDiscreteBoldMathCommand{vdTheta}{\Theta}
\DeclareDiscreteBoldMathCommand{vdLambda}{\Lambda}
\DeclareDiscreteBoldMathCommand{vdXi}{\Xi}
\DeclareDiscreteBoldMathCommand{vdPi}{\Pi}
\DeclareDiscreteBoldMathCommand{vdSigma}{\Sigma}
\DeclareDiscreteBoldMathCommand{vdUpsilon}{\Upsilon}
\DeclareDiscreteBoldMathCommand{vdPhi}{\Phi}
\DeclareDiscreteBoldMathCommand{vdPsi}{\Psi}
\DeclareDiscreteBoldMathCommand{vdOmega}{\Omega}

\providecommand*{\dd}{%
  \@ifnextchar^{\@dd}{\@dd^{}}}
\def\@dd^#1{%
  \mathop{\mathrm{\mathstrut d}}%
  \nolimits^{#1}\dd@gobblespace}
\def\dd@gobblespace{%
  \futurelet\diffarg\dd@opspace}
\def\dd@opspace{%
  \let\dd@space\!%
  \ifx\diffarg(
\let\dd@space\relax%
\else%
\ifx\diffarg[
\let\dd@space\relax%
\else%
\ifx\diffarg\{%
\let\dd@space\relax%
\fi%
\fi%
\fi%
\dd@space}

\newcommand{\Frac}{%
  \@ifnextchar[
  {\Frac@i}
  {\Frac@ii}}
\newcommand{\Frac@i}{}
\def\Frac@i[#1]#2#3{%
  \genfrac{}{}{#1}{}{\displaystyle{#2}}{\displaystyle{#3}}}
\newcommand{\Frac@ii}[2]{\frac{\displaystyle{#1}}{\displaystyle{#2}}}
      \newcommand{\diff@diffspace}{\,}
\newcommand{\diff@mathfrac}[2]{\frac{#1}{#2}}
\newcommand{\diff@mathFrac}[2]{\Frac{#1}{#2}}
\newcommand{\diff@textfrac}[2]{%
  \bgroup #1\egroup\mkern-1mu/\mkern-1mu\bgroup #2\egroup}
\newcommand{\diff}{%
  \global\let\diff@diffop\dd
  \global\let\diff@frac\diff@mathfrac
  \@ifnextchar[
  {\diff@i}
  {\diff@ii}}
\newcommand{\Diff}{%
  \global\let\diff@diffop\dd
  \global\let\diff@frac\diff@mathFrac
  \@ifnextchar[
  {\diff@i}
  {\diff@ii}}
\newcommand{\tdiff}{%
  \global\let\diff@diffop\dd
  \global\let\diff@frac\diff@textfrac
  \@ifnextchar[
  {\diff@i}
  {\diff@ii}}
\newcommand{\pdiff}{%
  \global\let\diff@diffop\partial
  \global\let\diff@frac\diff@mathfrac
  \@ifnextchar[
  {\diff@i}
  {\diff@ii}}
\newcommand{\Pdiff}{%
  \global\let\diff@diffop\partial
  \global\let\diff@frac\diff@mathFrac
  \@ifnextchar[
  {\diff@i}
  {\diff@ii}}
\newcommand{\tpdiff}{%
  \global\let\diff@diffop\partial
  \global\let\diff@frac\diff@textfrac
  \@ifnextchar[
  {\diff@i}
  {\diff@ii}}

\newcommand*{\diff@i}{}
\def\diff@i[#1]#2#3{\eval{\diff@ii{#2}{#3}}_{#1}}

\newcommand*{\diff@ii}[2]{%
  \begingroup
  \toks0={}\count0=0
  \diff@degree #2\diff@degree
  \diff@frac{\diff@diffop\ifnum\count0>1^{\the\count0}\fi\diff@diffspace#1}%
  {\the\toks0}%
  \endgroup}

\newcommand*{\diff@degree}[1]{%
  \ifx #1\diff@degree \expandafter\diff@stopd
  \else \expandafter\diff@addd \fi #1^1$#1\diff@addd}
\newcommand{\diff@stopd}{}
\def\diff@stopd #1\diff@addd{}
\newcommand*{\diff@addd}{}
\def\diff@addd #1^#2#3$#4\diff@addd{%
  \advance\count0 #2
  \toks0=\expandafter{\the\toks0%
    {\diff@diffop\diff@diffspace #4}%
    \diff@diffspace}\diff@degree}

\def\rs#1{\@ifnextchar[
  {\@rs{#1}}{\@@rs{#1}}}
\def\@rs#1[#2]#3{\mathinner{%
    \setbox\@ne\hbox{$\displaystyle{\vphantom{#3}}#1{#3}\m@th$}%
    \setbox\tw@\hbox{$\displaystyle{#3}#2\m@th$}%
    \hskip\wd\@ne\hskip-\wd\tw@\mathord{\hskip\wd\tw@\hskip-\wd\@ne%
      {\vphantom{#3}}#1{#3}#2}}}
\def\@@rs#1#2{\mathinner{%
    \setbox\@ne\hbox{$\displaystyle{\vphantom{#2}}#1{#2}\m@th$}%
    \hskip\wd\@ne\mathord{\hskip-\wd\@ne%
      {\vphantom{#2}}#1{#2}}}}

\newcommand{\MR}{\mathalpha{\mathbb{R}}}

\newcommand*{\norm}[1]{\mathinner{\Vert#1\Vert}}

\definecolor{notecolor}{cmyk}{0,1,1,.2}
\newcommand*\AM@notesname{Notes}

\newenvironment{titleditemize}[1]{%
  \paragraph{#1}
  \begin{itemize}}
  {\end{itemize}}

\makeatother

\modulolinenumbers[5]

\graphicspath{{02_figures/}}

\journal{International Journal for Numerical Methods in Engineering}

\begin{document}

\begin{frontmatter}

\title{A Computational Model for Molecular Interactions Between Curved Slender Fibers Undergoing Large 3D Deformations With a Focus on Electrostatic, van der Waals and Repulsive Steric Forces}

\author{Maximilian J. Grill\corref{cor1}}
\ead{grill@lnm.mw.tum.de}
\author{Wolfgang A. Wall}
\author{Christoph Meier}

\address{Technical University of Munich, Institute for Computational Mechanics, Boltzmannstr.~15, 85748 Garching b.~M\"unchen, Germany}

\cortext[cor1]{Corresponding author}

\begin{abstract}
This contribution proposes the first 3D beam-to-beam interaction model for molecular interactions between curved slender fibers undergoing large deformations.
While the general model is not restricted to a specific beam formulation, in the present work it is combined with the geometrically exact beam theory and discretized via the finite element method.
A direct evaluation of the total interaction potential for general 3D bodies requires the integration of contributions from molecule or charge distributions over the volumes of the interaction partners, leading to a 6D integral (two nested 3D integrals) that has to be solved numerically.
Here, we propose a novel strategy to formulate reduced section-to-section interaction laws for the resultant interaction potential between a pair of cross-sections of two slender fibers such that only two 1D integrals along the fibers' length directions have to be solved numerically.
This section-to-section interaction potential (SSIP) approach yields a significant gain in efficiency, which is essential to enable the simulation of relevant time and length scales for many practical applications.
In a first step, the generic structure of SSIP laws, which is suitable for the most general interaction scenario (e.\,g.~fibers with arbitrary cross-section shape and inhomogeneous atomic/charge density within the cross-section) is presented.
Assuming circular, homogeneous cross-sections, in a next step, specific analytical expressions for SSIP laws describing short-range volume interactions (e.\,g.~van der Waals or steric interactions) and long-range surface interactions (e.\,g.~Coulomb interactions) are proposed.
Besides ready-to-use expressions for the total interaction potential, also the resulting virtual work contributions, its finite element discretizations as well as a suitable numerical regularization for the limit of zero separation are derived.
The validity of the SSIP laws as well as the accuracy and robustness of the general SSIP approach to beam-to-beam interactions is thoroughly verified by means of a set of numerical examples considering steric repulsion, electrostatic or van der Waals adhesion.
\end{abstract}


\end{frontmatter}

\section{Introduction}

Biopolymer fibers such as actin, collagen, cellulose and DNA, but also glass fibers or carbon nanotubes are ubiquitous examples for slender, deformable structures to be found on the scale of nano- to micrometers.
On these length scales, molecular interactions such as electrostatic or van der Waals (vdW) forces are of utmost importance for the formation and functionality of the complex fibrous systems they constitute~\cite{French2010,israel2011,parsegian2005}.
Biopolymer networks such as the cytoskeleton or the extracellular matrix, muscle fibers, Gecko spatulae or chromosomes are some of the most popular examples.
To foster the understanding of such systems, which in turn allows for innovations in several fields from medical treatment to novel synthetic materials, there is an urgent need for powerful simulation tools.
Finite element formulations based on the geometrically exact beam theory~\cite{jelenic1999,crisfield1999,Meier2017c} are known to model the transient (elastic) deformation of these slender structures in an accurate and efficient manner.
However, no corresponding numerical methods for above mentioned molecular interactions between deformable fibers have been published yet.
We thus aim to develop methods that both accurately as well as efficiently describe these molecular phenomena based on the geometrically exact beam theory in order to ultimately solve relevant practical problems on the scale of complex systems consisting of a large number of fibers in arbitrary arrangement.

A comprehensive review of the origin, characteristics and mathematical description of intermolecular forces can nowadays be found in (bio)physical textbooks~\cite{israel2011,parsegian2005}.
The critical point is to transfer the first principles formulated for the interaction between atoms or single molecules to the interaction between macromolecules such as slender fibers.
Here, the analytical approaches to be found in textbooks and also in recent contributions~\cite{Ohshima2009,Jaiswal2012,Stedman2014,Maeda2015} from the field of theoretical biophysics are (naturally) restricted to undeformable, rigid bodies with primitive geometries such as spheres, half spaces or, most relevant in our case, cylinders.
Some computational approaches can be found in the literature, but rather aim at including more complex phenomena such as retardation and solvent effects in vdW interactions \cite{Dryden2015}, still limited to rigid bodies.
All-atom simulation methods like molecular dynamics do not suffer from this restriction, but their computational cost is by orders of magnitude too high to be applied to the relevant, complex biological systems mentioned in the beginning and thus currently limited to time scales of nano- to microseconds~\cite{israel2011}.

On the other hand, studying the deformation of elastic, slender bodies has a long history in mechanics and today's geometrically exact finite element formulations for shear-deformable (Simo-Reissner) as well as shear-rigid (Kirchhoff-Love) beams have proven to be both highly accurate and efficient~\cite{jelenic1999,crisfield1999,Meier2017c}.
Moreover, contact interaction between beams has been considered in a number of publications, e.\,g.~\cite{wriggers1997,litewka2005,durville2010,Kulachenko2012,Chamekh2014,GayNeto2016a,Konyukhov2016,Weeger2017,meier2016,Meier2017a}.
However, all these methods are motivated by the macroscopic perspective of non-penetrating solid bodies rather than the microscopic view considering first principles of intermolecular repulsive forces.

The combination of elastic deformation of general 3D bodies and intermolecular forces has first been considered by Argento et.\,al.~\cite{Argento1997} for small deformations, by Sauer and Li~\cite{Sauer2007a} for large deformations and finally by Sauer and Wriggers~\cite{Sauer2009} also for three-dimensional problems.
In order to reduce the high computational cost associated with the required high-dimensional numerical integrals, a possible model reduction from body to surface interaction in case of sufficiently short-ranged interactions as e.\,g.~predominant in (adhesive) contact scenarios has already been addressed in these first publications and has been the focus of subsequent publications~\cite{Sauer2013,Fan2015}.
However, since these formulations aim to describe the interaction between 3D bodies of arbitrary shape, the inherent complexity of the problem still requires a four-dimensional integral over both surfaces in case of surface interactions and a six-dimensional integral over both volumes for volume interactions, respectively.
In contrast, beam theory describes a slender body as a 1D Cosserat continuum, such that a further reduction in the dimensionality and thus computational cost can be achieved.
So far, this idea has been applied to describing the interaction between a beam and an infinite half-space in 2D as a model for the adhesion of a Gecko spatula on a rigid surface~\cite{Sauer2009,Sauer2014} and later also for the interaction of a carbon nanotube with a Lennard-Jones wall in 3D~\cite{Schmidt2015}.
In both cases, the influence of the rigid half space can be evaluated analytically and formulated as a distributed load on the beam.

To the best of the authors' knowledge, no approach for describing molecular interactions between curved 3D beams for arbitrary configurations and large deformations has been proposed yet.
Notable previous approaches to similar problems have made simplifying assumptions.
Ahmadi and Menon~\cite{Ahmadi2014} study the clumping of fibers due to vdW adhesion by means of an analytical 2D beam method, yet only include vdW interaction between the hemispherical tips based on an analytical expression for the interaction of two spheres.
A numerical study of the influence of inter-fiber adhesion on the mechanical behavior of 2D fiber networks assumes an effective adhesion energy per unit length of perfectly parallel fiber segments and solves for the unknown contact length in a second, nested minimization algorithm~\cite{Negi2018}.

In this article, we propose the first model specifically for molecular interactions between arbitrarily curved and oriented slender fibers undergoing large deformations in 3D.
While the general model is not restricted to a specific beam formulation, in the present work it is combined with the geometrically exact beam theory and discretized via the finite element method.
This novel approach is based on reduced section-to-section interaction potential (SSIP) laws that describe the resulting interaction potential between a pair of cross-sections as a closed-form analytical expression.
Thus, the two-body interaction potential follows from two nested 1D integrals over this SSIP law along both fibers' axes, which are evaluated numerically.
In this way, the proposed, so-called SSIP approach significantly reduces the dimensionality of the required numerical integration from six to two, and hence the associated computational cost.
As compared to methods for 3D solid bodies or even to all-atom methods, this gain in efficiency opens up new fields of applications, e.\,g., the complex biological systems mentioned above.
Regarding the practicability of the SSIP approach, it is also important to emphasize that it can be seamlessly integrated into an existing finite element framework for solid mechanics.
In particular, it does neither depend on any specific beam formulation nor the applied spatial discretization scheme and in the context of the present work, it has exemplarily been used with geometrically exact Kirchhoff-Love as well as Simo-Reissner type beam finite elements.
Likewise, it is independent of the temporal discretization and we have used it along with static and (Lie group) Generalized-Alpha time stepping schemes as well as inside a Brownian dynamics framework.

For the proposed SSIP laws, which can either be derived analytically or postulated and fitted e.\,g.~to experimental data, we first present the most general form describing the interaction between arbitrarily shaped cross-sections with inhomogeneous distribution of the elementary interacting points (e.\,g.~atoms or charges).
Subsequently, we focus on homogeneous, circular cross-sections and propose specific, ready-to-use SSIP laws for vdW adhesion, steric repulsion and electrostatic interaction.
Based on the fundamental distinction into either short-range or long-range interactions, we present the required steps and theoretical considerations underlying the analytical derivation of the SSIP laws in a general manner, starting from first principles in form of a point pair interaction potential that is described by a power law with general exponent.
Besides the expressions for the total interaction potential, also the corresponding virtual work contributions, its finite element discretization and the consistent linearization are presented.
Due to the characteristic singularity of molecular interactions in the limit of zero separation, also a suitable numerical regularization of the SSIP laws will be proposed.

The remainder of this article is structured as follows.
\secref{sec::theoretical_foundation_molecular_interactions} briefly summarizes the fundamental concepts and theory of molecular interactions.
Along with the fundamentals of the geometrically exact beam theory to be introduced in \secref{sec::fundamentals_beams}, this forms the basis for the novel SSIP approach to be proposed in \secref{sec::method_pot_based_ia}.
This general approach will then be applied to specific types of physical interactions, namely vdW, steric and electrostatic interactions in \secref{sec::method_application_to_specific_types_of_interactions}.
In \secref{sec::FE_discretization_algorithmic_aspects}, we turn to the finite element discretization of the newly developed numerical methods and discuss some important algorithmic aspects such as the regularization of the reduced interaction laws and the algorithmic complexity of the reduced approach.
Finally, the accuracy of the proposed SSIP laws as well as the general SSIP approach to beam-to-beam interaction is validated by means of analytical as well as numerical reference solutions for academic test cases in~\secref{sec::verification_methods}.
In a series of numerical examples including steric repulsion, electrostatic or vdW adhesion, the effectiveness and robustness of the novel approach is verified in the remainder of~\secref{sec::numerical_results}.
We conclude the article in \secref{sec::summary_outlook} and provide an outlook to promising future enhancements of the novel approach.

\section{Fundamentals of Intermolecular Forces and Potentials}\label{sec::theoretical_foundation_molecular_interactions}
Interactions between molecules may result from various physical origins and are a complex and highly active field of research within the community of theoretical as well as experimental  physics.
The methods to be derived in this work make use of the most essential and well established findings as summarized e.\,g.~in the textbooks \cite{israel2011} and \cite{parsegian2005}.
This section briefly presents a selection of aspects relevant for this work.

\subsection{Characterization, terminology and disambiguation}\label{sec::molecular_interactions_classification_general_informations}
To begin with, a number of universal aspects characterizing molecular interactions, especially with regard to the numerical methods to be developed in this work, shall be presented.
A few simple facts about the examples mostly considered throughout this work, namely electrostatic and vdW interaction, are presented straight away, whereas the details on these and further types of molecular interactions are to be discussed in the subsequent \secref{sec::theory_molecular_interactions_pointpair}.
\begin{titleditemize}{A collection of characteristics of molecular interactions with high relevance for this work}
  \item
  \begin{description}
    \item[Type of elementary interaction partners]
    Interaction may originate from unit charges as in the case of electrostatics.
    Another popular example are vdW effects that are caused by fluctuating dipole interactions occurring in every molecule and hence are related to the molecular density of the material.
  \end{description}
  \item
  \begin{description}
    \item[Spatial distribution of elementary interaction partners]
    Thinking of the resulting interaction between two bodies as accumulation of all molecular interactions, the question for the locations of all elementary interaction partners arises.
    Charges can often be found on the bodies' surfaces whereas molecules relevant for vdW interactions spread over the entire volume of the bodies.
    This work focuses on solid bodies (i.\,e.~condensed matter) that are non-conducting such that interaction partners will not redistribute, i.\,e., change their position within a body.
  \end{description}
  \item
  \begin{description}
    \item[Distance-dependency of the fundamental potential law]
    Generally, the strength of molecular interactions decays with increasing distance.
    Most frequently, inverse power laws with different exponents or exponential decay can be identified.
  \end{description}
  \item
  \begin{description}
    \item[Range of interactions]
    As a result of the previous aspects, a range of significant strength of an interaction can be defined.
    Rather than an inherent property, the classification of long- versus short-ranged interactions is a theoretical concept to judge the perceptible impact in specific scenarios.
    Moreover, it is a decisive factor in the derivation of well-suited numerical methods.
  \end{description}
  \item
  \begin{description}
    \item[Additivity and higher-order contributions]
    Many approaches including the one presented in this work make use of superposition, i.\,e., accumulating all the individual contributions from elementary interaction partners to obtain the total effect of interaction.
    This assumes that the interactions behave additively, i.\,e., that the sum of all pair-wise interactions describes the overall interaction sufficiently well.
    More specifically, the presence of other elementary interaction partners in the surrounding must not have a pronounced effect as compared to an isolated system of an interacting pair.
    Otherwise, the sum of all pair-wise interactions would need to be extended by contributions from sets of three, four and more elementary interaction partners.
  \end{description}
\end{titleditemize}
As a matter of course, this list is not exhaustive but represents a selection of the most relevant aspects considered in the development of our methods throughout~\secref{sec::method_pot_based_ia} and \ref{sec::method_application_to_specific_types_of_interactions}.

\paragraph{Interaction potential and corresponding force}$\,$\\
An interaction potential~$\Phi(r)$, also known as (Gibbs) free energy of the interaction, is defined as the amount of energy required to approach the interaction partners starting from a reference configuration with zero energy at infinite separation.
Hence, the following relations between the interaction potential~$\Phi(r)$ and the magnitude of the force $f(r)$ acting upon each of the partners, each in terms of the distance between both interacting partners~$r$, hold true:
\begin{align}\label{eq::force_potential_relation}
  \Phi(r) = \int \limits_\infty^r -f(r) \dd r \qquad \leftrightarrow \qquad f(r) = - \diff{\Phi(r)}{r}
\end{align}
Although the final quantities of interest are often the resulting, vectorial forces on slender bodies, it is yet convenient and sensible to consider the scalar interaction potential throughout large parts of this work.
This is underlined by the fact that nonlinear finite element methods in the context of structural dynamics can be formulated on the basis of energy and work expressions.
Equation \eqref{eq::force_potential_relation} expresses the direct and inherent relation between force and potential.
Note that the forces emanating from such interaction potentials are conservative and the integral value in \eqref{eq::force_potential_relation} is path-independent.
Furthermore, the interaction is symmetric in a sense that the force acting upon the first interacting partner $\vf_1(r)$ has the same magnitude yet opposite direction as compared to the force acting on the second partner $\vf_2(r)$.
Using the partners' position vectors~$\vx_1, \vx_2 \in \MR^3$, we can formulate the vectorial equivalent of the formula above:
\begin{align}
 \vf_1(r) = - \diff{\Phi(r)}{r} \frac{\vx_1 - \vx_2}{\norm{\vx_1-\vx_2}} \qquad \text{and} \qquad \vf_2(r) = \diff{\Phi(r)}{r} \frac{\vx_1 - \vx_2}{\norm{\vx_1-\vx_2}} \qquad \text{with} \qquad r=\norm{\vx_1 - \vx_2}
\end{align}

\paragraph{Disambiguation}$\,$\\
In order to particularize the very general term \textit{molecular interactions}, we may note that we solely consider interactions between distinct, solid (macro-)molecules, i.\,e., no covalent or other chemical bonds, but rather what is sometimes referred to as~\textit{physical bonds}.
Thus, we restrict ourselves to \textit{intermolecular} forces as opposed to \textit{intramolecular} ones.

\subsection{Interactions between pairs of atoms, small molecules or point charges}\label{sec::theory_molecular_interactions_pointpair}
First principles describing molecular interactions are formulated for a pair of atoms, molecules or point charges.
In the following, all types of interactions to be considered in this work are thus first presented for a minimal system consisting of one pair of these elementary interaction partners.

\subsubsection{Electrostatics}\label{sec::theory_electrostatics_pointcharges}
Coulomb's law is one of the most fundamental laws in physics and describes the interaction of a pair of point charges under static conditions by
\begin{align}\label{eq::pot_ia_elstat_pointpair_Coulomb}
  \Phi_\text{elstat}(r) = \frac{Q_1 Q_2}{4\pi \varepsilon_0 \varepsilon} \frac{1}{r}, \qquad \norm{\vf_\text{elstat}(r)} = \frac{Q_1 Q_2}{4\pi \varepsilon_0 \varepsilon} \frac{1}{r^2}, \qquad \vf_{\text{elstat},1}(r) = \frac{Q_1 Q_2}{4\pi \varepsilon_0 \varepsilon} \frac{\vx_1 - \vx_2}{\norm{\vx_1-\vx_2}^3}
\end{align}
where~$\varepsilon_0$ and $\varepsilon$ are the vacuum and dielectric permittivity, respectively.
For the sake of brevity in any later usage, let us define the abbreviation~$C_\text{elstat} \defvariable \left( 4\pi \varepsilon_0 \varepsilon \right)^{-1}$.
Depending on the signs of~$Q_1$ and~$Q_2$, electrostatic forces may either be repulsive or attractive.
Besides a pair of point charges, Coulomb's law likewise holds for a pair of spherically symmetric charge distributions with resulting charges~$Q_1$ and~$Q_2$, respectively.
This is an important insight, since ultimately we are interested in interactions between two bodies with finite extension rather than points.
Furthermore, interactions between rigid spheres and rigid bodies are of interest for applications such as particle diffusion in hydrogels.
Throughout the entire work, no electrodynamic effects shall be considered.
This is a valid assumption as long as bodies are non-conductive and the motion of bodies carrying the attached charges happens on much larger time scales than relevant eigenfrequencies in electrodynamics.

Due to the inverse-first power law, the electrostatic potential has quite a long range, meaning that two point charges at a large distance still experience a considerable interaction force as compared to small distances.
This behavior is even more pronounced for the interaction of two extended bodies, where the whole lot of all distant point pairs dominates the total interaction energy as compared to the few closest point pairs.
This property is crucially different as compared e.\,g.~to vdW interactions considered in the next section.
We will account for and indeed make use of this important property in the development of the methods to be presented in this work.

\subsubsection{Van der Waals interactions}\label{sec::theory_vdW_pointpair}
\textit{Van der Waals} forces originate from charge fluctuations, thus being an electrodynamic effect caused by quantum-mechanical uncertainties in positions and orientations of charges.
Depending on the interaction partners, three subclasses can be distinguished as Keesom (two permanent dipoles), Debye (one permanent dipole, one induced dipole) and London dispersion interactions (two transient dipoles).
The ubiquitous nature of van der Waals interactions is due to the fact that the latter contribution even arises in neutral, nonpolar, yet polarizable matter that means basically every atom or molecule.
All three kinds of dipole interactions can be unified in that their interaction free energy follows an inverse-sixth power law in the separation \cite{parsegian2005}:
\begin{align}\label{eq::pot_ia_vdW_pointpair}
  \Phi_\text{vdW}(r) = - \frac{C_\text{vdW}}{r^6}
\end{align}
This is a pleasantly simple expression, yet intricate when it comes to transferring it to two-body interactions, as we will discuss in~\secref{sec::theory_molecular_interactions_twobody_vdW}.
In general, van der Waals forces are always attractive for two identical or similar molecules, yet may be repulsive for other material combinations.

\subsubsection{Steric exclusion}\label{sec::theory_steric_exclusion_pointpair}
Two approaching atoms or molecules will at some very small separation suddenly experience a seemingly infinite repulsive force.
This effect is attributed to the overlap of electron clouds and referred to as \textit{steric repulsion}, \textit{steric exclusion} or \textit{hard core repulsion}.
Without thorough theoretical foundation, several (almost) infinitely steep repulsive potential laws are empirically used to model this phenomenon.
The first option is a \textit{hard wall/core/sphere} potential which has a singularity at zero separation
\begin{equation}
  \lim_{r \to 0} \Phi_\text{c,hs}(r) = \infty \qquad \text{and} \qquad \Phi_\text{c,hs}(r) \equiv 0 \quad \text{for} \quad r>0.
\end{equation}
Other common choices include a \textit{power-law} potential with a large integer exponent~$n_\text{c,pow}$
\begin{equation}
  \Phi_\text{c,pow}(r) = C_\text{c,pow} \, r^{-n_\text{c,pow} }
\end{equation}
and finally an \textit{exponential} potential
\begin{equation}
  \Phi_\text{c,exp}(r) = C_\text{c,exp} \, e^{-r / r_\text{c,exp} }.
\end{equation}
Note that the former two coincide in the limit~$n_\text{c,pow} \rightarrow \infty$ and increase indefinitely for~$r\rightarrow 0$ while the exponential one does not.
Generally, this behavior of steric exclusion is in good agreement with our intuition based on macroscopic solid bodies coming into contact.

\subsubsection{Total molecular pair potentials and force fields}\label{sec::theory_total_point_pair_potential}
\begin{figure}[htp]%
  \centering
  \includegraphics[width=0.4\textwidth]{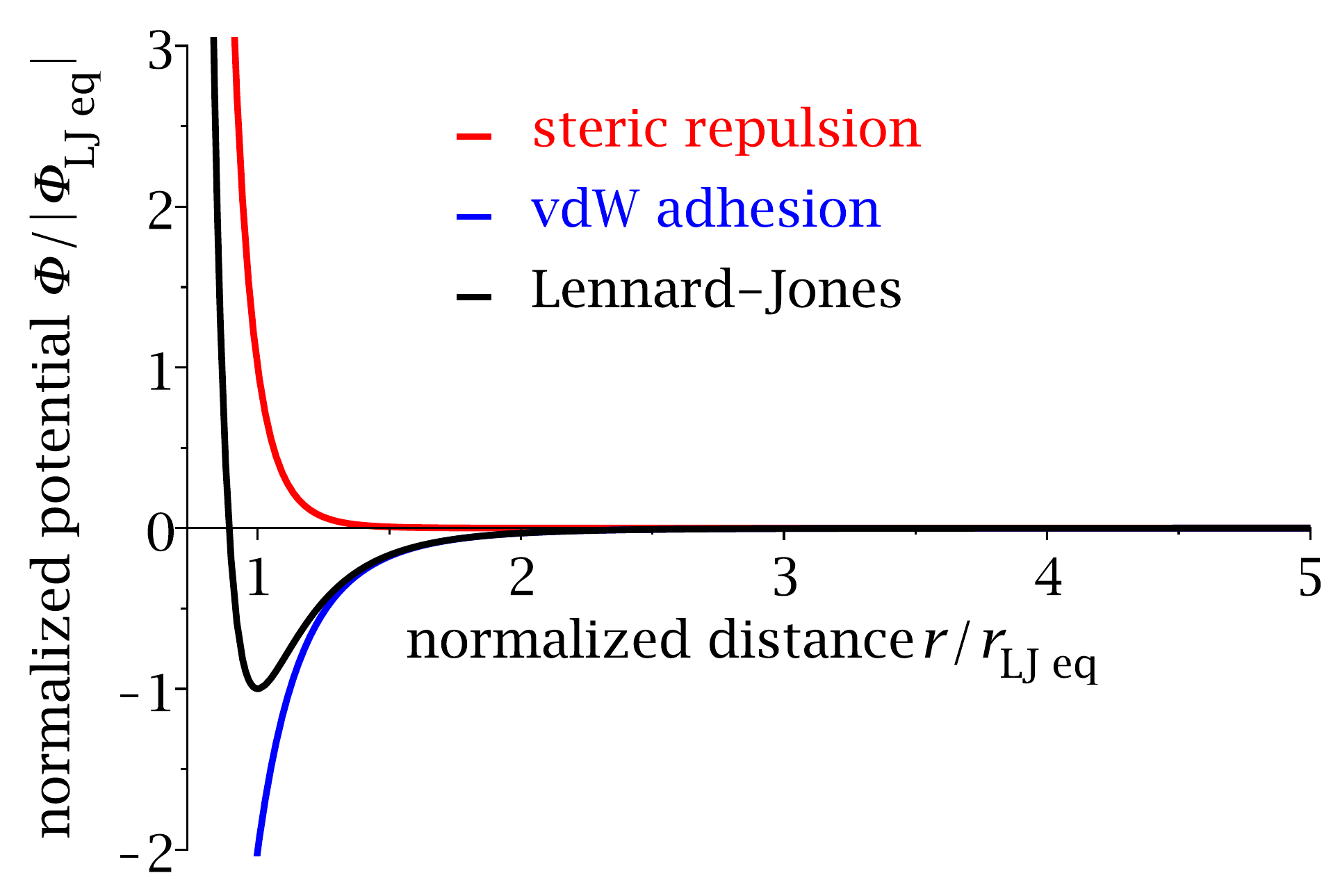}
  \caption{Lennard-Jones interaction potential for a pair of points, i.\,e.~atoms.}
  \label{fig::point-point_pot_LJ}
\end{figure}

In many systems of interest, any two or more of the aforementioned effects may be relevant at the same time such that a combination of the pair potentials is required.
This is typically done by summation of the individual potential contributions and leads to a \textit{total intermolecular pair potential}.
Among the large number of possible combinations\footnote{See \cite{israel2011}[p. 138] for a comprehensive list of combinations used as total pair potential laws.}, the \textit{Lennard-Jones} (LJ) potential is probably the most commonly used variant (see \figref{fig::point-point_pot_LJ}).
\begin{align}\label{eq::pot_ia_LJ_pointpair}
  \Phi_\text{LJ}(r) = k_{12} r^{-12} + k_6 r^{-6} = -\Phi_\text{LJ,eq} \left( \left( \frac{r_\text{LJ,eq}}{r} \right)^{12} - 2 \,\left( \frac{r_\text{LJ,eq}}{r} \right)^{6} \right)
\end{align}
It is a special case of the \textit{Mie potential}~$\Phi_\text{Mie}(r) = C_\text{Mie,m} \, r^{-m} - C_\text{Mie,n} \, r^{-n}$ with exponents being chosen to model the inverse-sixth van der Waals attraction on the one hand and a strong repulsion on the other hand.
The parameters can be identified as the minimal value~$\Phi_\text{LJ,eq}<0$ that the Lennard-Jones potential takes at equilibrium separation~$r_\text{LJ,eq}>0$, i.\,e., at the separation where the resulting force is zero.

Other important quantities characterizing the LJ force law
\begin{align}
  f_\text{LJ}(r) = -\frac{12\,\Phi_\text{LJ,eq}}{r_\text{LJ,eq}} \, \left( \left( \frac{r_\text{LJ,eq}}{r} \right)^{13} - \left( \frac{r_\text{LJ,eq}}{r} \right)^{7} \right)
\end{align}
are the minimal force value~$f_\text{LJ,min}$ and corresponding distance~$r_\text{LJ,f$_\text{min}$}$
\begin{align}
  f_\text{LJ,min} \approx \num{2.6899} \,\frac{\Phi_\text{LJ,eq}}{r_\text{LJ,eq}}
  \qquad \text{and} \qquad
  r_\text{LJ,f$_\text{min}$} = \left( \frac{13}{7} \right)^{1/6} \, r_\text{LJ,eq} \approx \num{1.1087} \, r_\text{LJ,eq}.
\end{align}
The minimal force, i.\,e., the maximal adhesive force, is commonly referred to as~\textit{pull-off force}.
Israelachvili~\cite{israel2011} also points out the chance of a fortunate cancellation of errors in total pair potentials, especially close to the limit~$r \rightarrow 0$.
In this regime, attractive forces tend to be underestimated by the simplified inverse-sixth term but likewise the steric repulsion is probably stronger than estimated from the power law.
Both errors cancel rather than accumulate and increase the model accuracy.

\begin{figure}[htp]%
  \centering
  \subfigure[]{
    \includegraphics[width=0.45\textwidth]{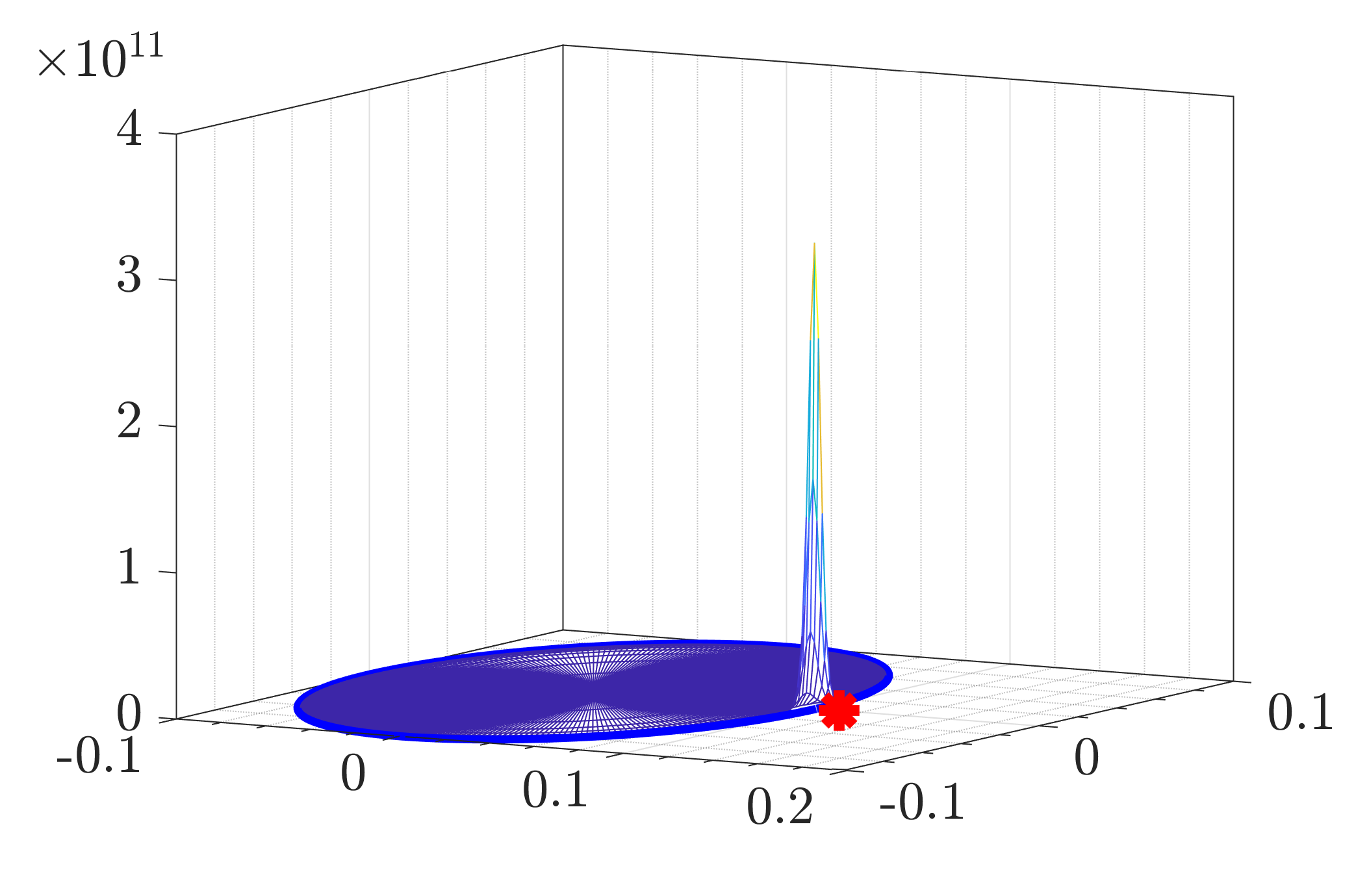}
    \label{fig::pot_ia_point_pair_r_exp-12_d1e-1}
  }
  \subfigure[]{
    \includegraphics[width=0.45\textwidth]{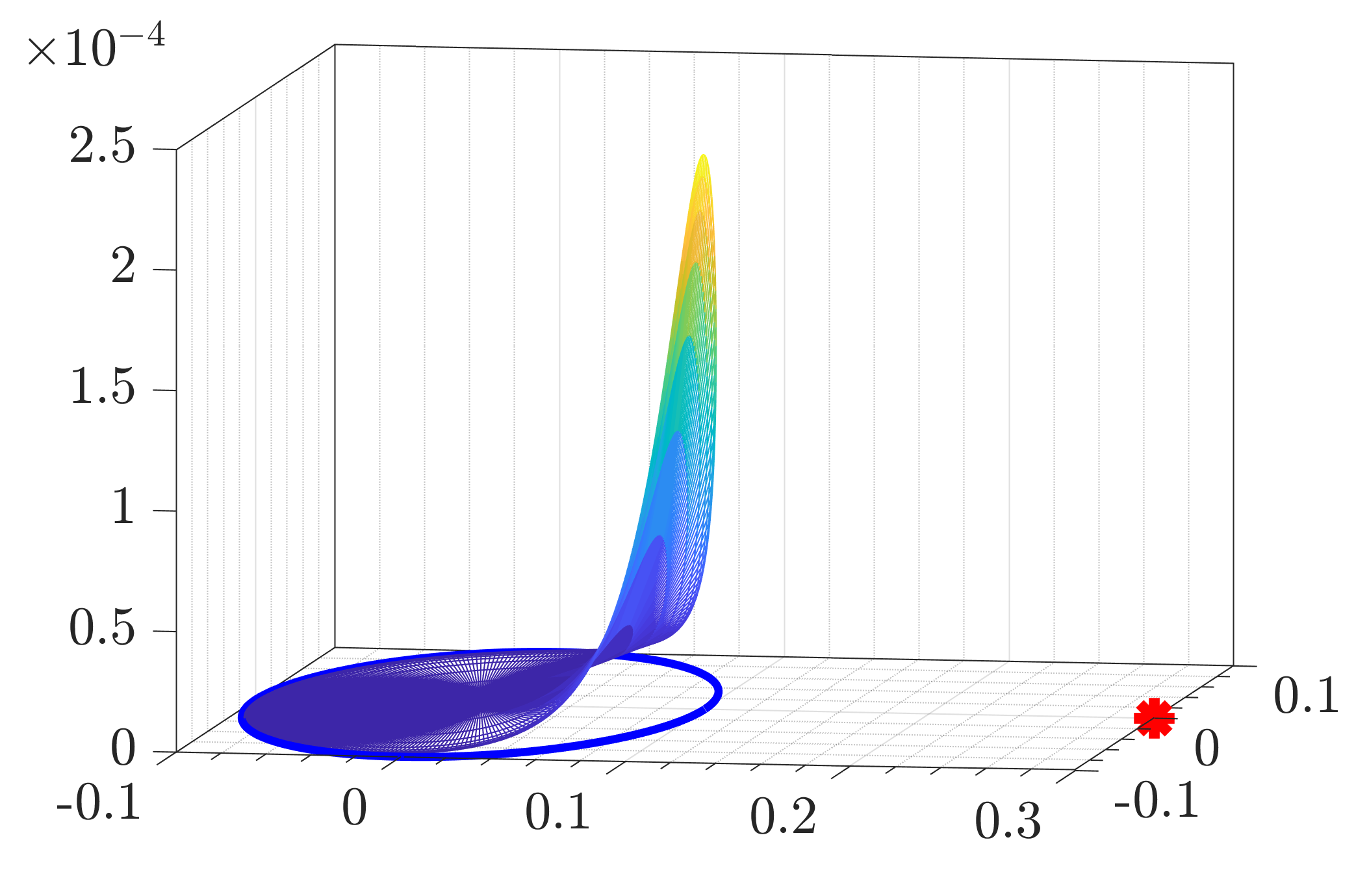}
    \label{fig::pot_ia_point_pair_r_exp-12_d2}
  }
  \caption{Example of a point-point interaction potential~$\Phi(r) \propto r^{-12}$ plotted over a circular domain (blue circle) with (a) small and (b) large distance to the point-like interaction partner (red dot). Note the huge difference in scales.}
  \label{fig::pot_ia_point_pair_r_exp-12}
\end{figure}

\paragraph{Remarks}
\begin{enumerate}
  \item Many of the presented point-point interaction potentials decay rapidly with the distance as shown exemplarily for a law~$\Phi(r) \propto r^{-12}$ in \figref{fig::pot_ia_point_pair_r_exp-12}.
  In anticipation of the numerical methods to be proposed in this work we can already state at this point that these extreme gradients are very challenging for numerical quadrature schemes that will therefore be discussed in a dedicated \secref{sec::method_numerical_integration}.

  \item In molecular dynamics, a \textit{force field} is typically used instead of the potential law to model the total interaction of a pair of atoms.
  Specific forms are being proposed for (coarse-grained) force fields modeling the interaction of macromolecules such as DNA instead of atoms.
  Since these all-atom approaches follow an entirely different approach as compared to the continuum model proposed here, we will not discuss force fields any further at this point.
\end{enumerate}
\subsection{Two-body interaction: Surface vs. volume interaction}\label{sec::theory_molecular_interactions_twobody}
In this section, we take the important step from interacting point pairs to interactions between two bodies with defined spatial extension containing many of the fundamental point-like interaction partners considered throughout the preceding \secref{sec::theory_molecular_interactions_pointpair}.
The obvious question for the spatial distribution of the interaction partners leads to the important distinction between \textit{surface} and \textit{volume interactions}.
As the name suggests, in the first case, the elementary interaction partners are distributed over the surface of the bodies but not in the interior.
The most important example from this category are electrostatic interactions between bodies where the charges sit on the surfaces and are not free to move around.
This applies to a large number of charged, non-conductive biopolymer fibers such as actin or DNA.
In the second case of volume interactions, the elementary interacting partners are distributed over the entire volume of the bodies.
The most important examples here are van der Waals interactions and steric exclusion.
As compared to surface interactions, this further increases the dimension of the problem making it more challenging to tackle, both by analytical as well as numerical means.
In terms of notation one may also find the expressions \textit{body forces} or \textit{bulk interaction} referring to this category of interactions.

\begin{figure}[htpb]%
  \centering
  \def\svgwidth{0.35\textwidth}
  \input{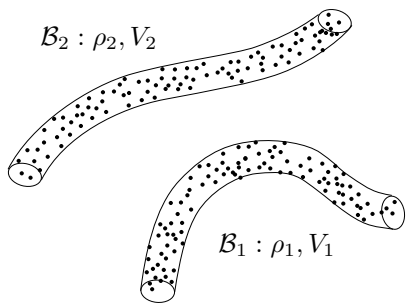}
  \vspace{-0.8cm}
  \caption{Two arbitrarily shaped, deformable bodies~$\mathcal{B}_1$ and~$\mathcal{B}_2$ with volumes~$V_1,V_2$ and continuous particle densities $\rho_1,\rho_2$.}
  \label{fig::beam_to_beam_interaction_particle_clouds}
\end{figure}

Let us briefly look at volume and surface interactions as an abstract concept, leaving aside the specifics of the underlying physical effects that are to be discussed in the subsequent \secref{sec::theory_molecular_interactions_twobody_surf} and \secref{sec::theory_molecular_interactions_twobody_vdW}.
Likewise, we assume additivity here and discuss the applicability later with the physical type of interaction.
Since volume interactions are the more general and challenging case, we will discuss most aspects and approaches first for volume interactions and later only point out the differences considering surface interactions throughout this article.
\figref{fig::beam_to_beam_interaction_particle_clouds} schematically visualizes the distribution of elementary interaction partners within two macromolecular or macroscopic bodies.

Assuming additivity, we apply \textit{pairwise summation} to arrive at the two-body interaction potential
\begin{equation}
 \Pi_\text{ia} = \sum_{i\in \mathcal{B}_1} \sum_{j\in \mathcal{B}_2} \Phi(r_{ij}).
\end{equation}
Further assuming a continuous atomic density $\rho_i$, $i={1,2}$, the total interaction potential can alternatively be rewritten as integral over the volumes $V_1, V_2$ of both bodies~$\mathcal{B}_1$ and~$\mathcal{B}_2$:
\begin{equation}\label{eq::pot_fullvolint}
 \Pi_\text{ia} = \iint_{V_1,V_2} \rho_1(\vx_1) \rho_2(\vx_2) \Phi(r) \dd V_2 \dd V_1 \qquad \text{with} \quad r = \norm{\vx_1-\vx_2}
\end{equation}
It can be shown that this continuum approach is the result of \textit{coarse-graining}, i.\,e., smearing-out the discrete positions of atoms in a system into a smooth atomic density function $\rho(\vx)$ \cite{Sauer2007a}.

In the case of surface interactions, the dimensionality of the problem reduces and summation or integration is carried out over both bodies' surfaces~$\partial \mathfrak{B}_1$, $\partial \mathfrak{B}_2$:
\begin{equation}\label{eq::pot_fullvolint_surface}
 \Pi_\text{ia} = \iint_{S_1,S_2} \sigma_1(\vx_1) \sigma_2(\vx_2) \Phi(r) \dd S_2 \dd S_1 \qquad \text{with} \quad r=\norm{\vx_1-\vx_2}
\end{equation}
Accordingly, surface densities~$\sigma_i(\vx_i)$, $i={1,2}$, replace the volume densities in this case.

\paragraph{The \textbf{range of two-body interaction forces} originating from point pair potentials}$\,$\\
Let us assume a general inverse power law $\Phi(r)=k r^{-m}$ for the point pair interaction potential.
It is obvious, that the potential becomes infinitely large if the separation of the two individual points~$r$ approaches zero and, on the other hand, that the potential rapidly decays with increasing distance.
Turning to two bodies of finite size, i.\,e., two clouds of points, things are more involved as the following theoretical considerations demonstrate.
In short, it can be shown that there is a fundamental difference between potentials with an exponent $m \leq 3$ on the one hand and $m > 3$ on the other hand.
Starting with the case~$m > 3$, e.\,g., vdW interactions, the two-body interaction potential goes to infinity if the bodies approach until their surfaces touch each other.
This can be illustrated by the simple example of two spheres of radius~$R$, where the vdW interaction potential scales with~$\Pi_\text{vdW} \propto g^{-1}$ (cf.~\cite[p.255]{israel2011}) with surface-to-surface separation or gap~$g=d-2R$ and the distance between the spheres' centers~$d$.
This singularity of the two-body interaction potential in the limit of zero separation~$g\to 0$ is due to the fact that potentials with $m > 3$ decay so rapidly that the few point pairs with smallest separation outweigh the potentially very large number of all other, distant point pairs in terms of their potential contributions.
Therefore, we can conclude that potentials with~$m > 3$ have no significant large distance contribution and the two-body interaction potential is governed by the separation of any two closest points (and their immediate surrounding).
Considering the example of two cylinders later on, we will also see that the vdW interaction potential of two perpendicular cylinders does not change perceptibly if the length of the cylinders is increased (cf.~eq.~\eqref{eq::pot_ia_vdW_cyl_cyl_perpendicular_smallseparation}), which can again be attributed to the short range of vdW interactions.

The situation is substantially different for potentials with $m\leq 3$, e.\,g., Coulombic interactions.
Here, the total contribution of all distant point pairs dominates over the few closest point pairs and the total interaction potential remains finite even if both bodies are in contact.
Looking once again at the simple example of two spheres, Coulomb's law (cf.~eq.~\eqref{eq::pot_ia_elstat_pointpair_Coulomb}) directly shows~$\Pi_\text{elstat} \propto d^{-1}$ and thus no singularity occurs for (nearly) contacting surfaces~$g\to 0$, i.\,e., $d \to 2R$.
Also, in contrast to the case of vdW mentioned above, the Coulomb interaction potential of two perpendicular cylinders would increase if their length is increased.
The underlying theoretical derivations revealing also the turning point, i.\,e., the exponent~$m=3$ were first noted by Newton and can be found e.\,g.~in \cite[p. 11]{israel2011}.
Due to this crucial difference, potentials with~$m>3$ will be denoted as \textit{short-range} interactions (e.\,g.~repulsive as well as attractive part of LJ) and potentials with~$m\leq3$ as \textit{long-range} interactions (e.\,g.~Coulomb) throughout this work.

\subsubsection{Electrostatics of non-conductive bodies: An example for long-range surface interactions}\label{sec::theory_molecular_interactions_twobody_surf}
The Coulomb interaction is additive such that the net force acting on an individual point charge in a system of point charges can be calculated from superposition of all pair-wise computed force contributions \cite{israel2011}.
Equivalently, the net interaction potential results from summation of all pair potentials.
A large body of literature deals with the problem of electrostatic multi-body interaction.
One concept of high relevance for the present work is a well-known strategy called multipole expansion, which aims to express the resultant electrostatic potential of a (continuous) charge distribution as an (infinite) series (see e.\,g.~\cite{Prytz2015} for details).
The individual terms of the series expansion generally are inverse power laws in the distance with increasing exponent and referred to as mono-, di-, quadru-, up to $n$-pole moments.
At points far from the location of the charge cloud, the series converges quickly and can thus be truncated in good approximation.
Regarding the total interaction potential of two charged bodies as formulated in~\eqref{eq::pot_fullvolint} or~\eqref{eq::pot_fullvolint_surface}, this already outlines the way how to determine~$\Pi_\text{elstat}$ for trivial geometries of the interacting bodies, where the integrals can potentially be solved analytically.
We will return to this concept in the context of deformable slender fibers when proposing the general SSIP approach in the beginning of~\secref{sec::method_pot_based_ia} and make use of a (truncated) multi-pole expansion of the charged cross-sections for the (simplified) SSIP law for long-range surface interactions to be proposed in~\secref{sec::ia_pot_double_length_specific_evaluation_elstat}.

\subsubsection{Van der Waals interaction: An example for short-range volume interactions}\label{sec::theory_molecular_interactions_twobody_vol}
\label{sec::theory_molecular_interactions_twobody_vdW}
Here, we want to discuss vdW interactions as one example of physically relevant volume interactions that is based on the inverse-sixth power law \eqref{eq::pot_ia_vdW_pointpair}.
However, very similar considerations and formulae apply to steric interactions as well as LJ interactions.

Today, we know that vdW interactions are generally non-additive.
The latest and most accurate models for two-body vdW interactions are based on Lifshitz theory and, among other effects, include retardation, anisotropy and differences in polarizability.
Nevertheless, a ``happy convergence'' of old, i.\,e.,~Hamaker's pairwise summation, and new, i.\,e.~Lifshitz, theory allows to determine the distance dependency from pairwise summation and then estimate the prefactor, i.\,e.,~Hamaker ``constant''~$A_\text{Ham}$, from more advanced modern theory.
This approach yielding a so-called Hamaker-Lifshitz hybrid form~\cite{parsegian2005},~\cite[p.~257]{israel2011} is what motivates us to use pairwise summation in the derivation of the numerical methods to be proposed in the present work.
Also, there are some special scenarios (negligible retardation, negligible difference in optical properties of the bodies, interaction in vacuum, ...), where additivity can be assumed as a good approximation even without adaption of the Hamaker constant.

Generally, even the simple approach of pairwise summation requires two nested 3D integrals over both bodies' volumes, i.\,e.,~six-dimensional integration.
Mainly due to this high dimensionality of the problem, unfortunately, (closed-form) analytical solutions can only be obtained for some simple special cases.
Still, careful considerations and selection allow us to exploit some of these analytical expressions in order to develop efficient, reduced methods in \secref{sec::method_double_length_specific_integral}.
To get a concise overview of all expressions relevant for the remainder of this work, we will provide a collection of closed-form analytical solutions in the following.

First, we want to look at two cylinders representing the simplest model for two interacting straight, rigid fibers with circular cross-section.
A number of publications consider this scenario and due to the simplicity of the geometry they were able to derive analytical solutions for some special cases.
The resulting expressions are summarized in Table~\ref{tab::pot_ia_vdW_cylinders_formulae} and will be used for verification purposes in \secref{sec::verification_methods}.
A second, highly relevant scenario is the one considering two disks.
These analytical expressions, summarized in Table~\ref{tab::pot_ia_vdW_disks_formulae}, will be beneficial, and in fact provide the main ingredient, for the SSIP approach to describe molecular interactions between \textit{deformable} fibers modeled as 1D Cosserat continua.

\paragraph{Two Cylinders}$\,$\\
To begin with, we consider the cases of parallel and perpendicular cylinders.
Generally, the cylinders are assumed to be infinitely long, such that the boundary effects at its ends may be neglected.
As the interaction potential for parallel cylinders would be infinite, one typically considers a length-specific interaction potential~$\tilde \pi_\text{vdW,cyl$\parallel$cyl}$ with dimensions of energy per unit length.
This quantity thus describes the interaction of one infinitely long cylinder with a section of unit length of the other infinitely long cylinder.
For perpendicular orientation (and all other mutual angles apart from $\alpha=0$) on the other hand, the total interaction potential~$\Pi_\text{vdW,cyl$\perp$cyl}$ remains finite.

Even for this simple case of two cylinders, no closed-form analytical solution for the vdW interaction energy can be found for all mutual angles and all separations.
One thus resorts to the consideration of the limits of small and large surface-to-surface separations for which the general solution, an infinite series, converges to the expressions presented in the following Table~\ref{tab::pot_ia_vdW_cylinders_formulae}.
\begin{table}[htpb]
  \begin{center}
    \begin{tabular}{|c|c|c|}\hline
      &&\\
      & Limit of \textit{small} separations & Limit of \textit{large} separations\\
      & $g \ll R_1,R_2$ & $g,d \gg R_1,R_2$\\
      &&\\\hline
      &&\\
      parallel&
      $\tilde \pi_\text{vdW,cyl$\parallel$cyl,ss} = - \frac{ A_\text{Ham} }{ 24 } \, \sqrt{ \frac{ 2 R_1 R_2 }{ R_1 + R_2 } } \, g^{ -\frac{3}{2}}$
      \refstepcounter{equation}(\theequation)\label{eq::pot_ia_vdW_cyl_cyl_parallel_smallseparation}
      &
      $  \tilde \pi_\text{vdW,cyl$\parallel$cyl,ls} = - \frac{ 3 \pi }{ 8 } \, A_\text{Ham} \, R_1^2 R_2^2 \, d^{-5}$
      \refstepcounter{equation}(\theequation)\label{eq::pot_ia_vdW_cyl_cyl_parallel_largeseparation}\\
      $\left[ \frac{\text{energy}}{\text{length}} \right]$
      & see \cite{israel2011}[p.~255],\cite{parsegian2005}[p.~172] & see \cite{parsegian2005}[p.~16, p.~172]\\
      &&\\\hline
      &&\\
      perp.
      &
      $ \Pi_\text{vdW,cyl$\perp$cyl,ss} = - \frac{ A_\text{Ham} }{ 6 } \, \sqrt{ R_1 R_2 } \, g^{-1} $
      \refstepcounter{equation}(\theequation)\label{eq::pot_ia_vdW_cyl_cyl_perpendicular_smallseparation}
      &
      $\Pi_\text{vdW,cyl$\perp$cyl,ls} = - \frac{ \pi }{ 2 } \, A_\text{Ham} \, R_1^2 R_2^2 \, d^{ -4}$
      \refstepcounter{equation}(\theequation)\label{eq::pot_ia_vdW_cyl_cyl_perpendicular_largeseparation}\\
      $\left[ \text{energy} \right]$
      & see \cite{israel2011}[p.~255] & see \cite{parsegian2005}[p.~16]\\
      &&\\\hline
    \end{tabular}
  \end{center}
  \caption{A collection of analytical solutions for the cylinder-cylinder interaction potential derived via pairwise summation. Here, $R_i$ denotes the cylinder radii, $d$ denotes the closest distance between the cylinder axes, $g$ denotes the surface-to-surface separation, i.\,e., gap, $A_\text{Ham} := \pi^2 \rho_1 \rho_2 C_\text{vdW}$ is the commonly used abbreviation known as Hamaker constant where $\rho_i$ denotes the particle densities and $C_\text{vdW}$ denotes the constant prefactor of the point-pair potential law (see eq.~\eqref{eq::pot_ia_vdW_pointpair}).}
  \label{tab::pot_ia_vdW_cylinders_formulae}
\end{table}

Despite the different dimensions of the quantities for parallel and perpendicular cylinders, we can still compare these expressions as becomes clear by the following thought experiment.
Considering two ``sufficiently long'' cylinders of length~$L$ in parallel orientation, the total interaction potential is well described by~$\Pi_\text{vdW,cyl$\parallel$cyl} = \tilde \pi_\text{vdW,cyl$\parallel$cyl} \cdot L$ and thus shows the same scaling behavior in the separation as \eqref{eq::pot_ia_vdW_cyl_cyl_parallel_smallseparation} and \eqref{eq::pot_ia_vdW_cyl_cyl_parallel_largeseparation}.
In addition, \eqref{eq::pot_ia_vdW_cyl_cyl_perpendicular_smallseparation} and \eqref{eq::pot_ia_vdW_cyl_cyl_perpendicular_largeseparation} are also a good approximation for the perpendicular orientation of these cylinders of finite length~$L$ since the difference in the distant point pairs is negligible.

We would like to point out just a few interesting aspects of these equations.
First, it is remarkable how the expressions differ in the exponent of the power law describing the distance dependency of the potential.
This relates to a diverse and highly nonlinear behavior already for this simplest model system of fiber-fiber interactions composed of two cylinders.
Second, the parallel orientation is a very special orientation that gives rise to the strongest possible adhesive forces between two cylinders and at the same time is the only stable equilibrium configuration.
Third, the distance scaling behavior of two parallel cylinders at small separations $\tilde \pi_{\text{vdW,cyl}\parallel\text{cyl,ss}} \propto g^{ -\frac{3}{2}}$ lies between the fundamental solutions known for two infinite half spaces $\tilde{\tilde \pi} \propto g^{-2}$ (double tilde indicates a potential per unit area) and two spheres $\Pi \propto g^{-1}$.
Note that again multiplication of these laws by a length or area does not alter the scaling law in the distance.
Looking at the equations for large separations, we see similar relations, once again with a stronger distance scaling behavior in the parallel case.
Generally, the solutions for large separations are expressed more naturally in the inter-axis separation~$d$ rather than the surface-to-surface separation~$g$.

\paragraph{Two Disks}$\,$\\
This problem has been studied in literature on vdW interaction of straight, rigid cylinders of infinite length \cite{langbein1972}.
In analogy to the cylinder-cylinder scenario, it turns out that not even in the simplest case of parallel oriented disks, i.\,e., two disks with parallel normal vectors, a closed analytical solution can be found for all separations.
Instead, two expressions for the limit of small and large separations of the disks $g$ as compared to their radii $R_1,R_2$ are presented in the following Table~\ref{tab::pot_ia_vdW_disks_formulae}, respectively.

\begin{table}[htpb]
  \begin{center}
    \begin{tabular}{|c|c|c|}\hline
      &&\\
      & Limit of \textit{small} separations & Limit of \textit{large} separations\\
      & $g \ll R_1,R_2$ & $g,d \gg R_1,R_2$\\
      &&\\\hline
      &&\\
      parallel
      &
      $\tilde{ \tilde{ \pi}}_\text{vdW,disk$\parallel$disk,ss} = - \frac{ 3 A_\text{Ham} }{ 256 } \, \sqrt{ \frac{ 2 R_1 R_2 }{ R_1 + R_2 } } \, g^{ -\frac{5}{2}}$
      \refstepcounter{equation}(\theequation)\label{eq::pot_ia_vdW_disk_disk_parallel_smallseparation}
      &
      $\tilde{ \tilde{ \pi}}_\text{vdW,disk$\parallel$disk,ls} =  - A_\text{Ham} \,R_1^2 \, R_2^2 \, d^{-6}$
      \refstepcounter{equation}(\theequation)\label{eq::pot_ia_vdW_disk_disk_parallel_largeseparation}\\
      $\left[ \frac{\text{energy}}{\text{length}^2} \right]$
      & see \cite{langbein1972} & see \cite{langbein1972}\\
      &&\\\hline
    \end{tabular}
  \end{center}
  \caption{A collection of analytical solutions for the disk-disk interaction potential derived via pairwise summation. Here, $R_i$ denotes the disk radii, $d$ denotes the closest distance between the disk midpoints, $g$ denotes the surface-to-surface separation, i.\,e., gap, $A_\text{Ham} := \pi^2 \rho_1 \rho_2 C_\text{vdW}$ is the commonly used abbreviation known as Hamaker constant where $\rho_i$ denotes the particle densities and $C_\text{vdW}$ denotes the constant prefactor of the point-pair potential law (see eq.~\eqref{eq::pot_ia_vdW_pointpair}).}
  \label{tab::pot_ia_vdW_disks_formulae}
\end{table}
To summarize, a closed-form expression for the two-body vdW interaction potential is only known for some rare special cases and the ones relevant for fiber-fiber interactions have been identified in the voluminous literature on this topic and presented here in a brief and concise manner.

We would like to conclude this section on two-body vdW interactions with a note on the analogy to steric exclusion, i.\,e., contact interactions, as already discussed for point pairs in~\secref{sec::theory_steric_exclusion_pointpair}.
This class of physical interactions shares the two central properties of being extremely short in range and being volume interactions.
Starting from an inverse-twelve power law as in the repulsive part of the LJ interaction law, one may apply very similar solution strategies and finally obtain very similar expressions as the ones presented in this section.
For the sake of brevity, we refer to the derivations in \ref{sec::formulae_two_body_LJ_interaction} and the analysis of the resulting total LJ interaction that will also be used for the regularization of the reduced potential laws in \secref{sec::regularization}.

\section{Fundamentals of Geometrically Exact 3D Beam Theory}\label{sec::fundamentals_beams}
\label{sec::beam_theory}
This section aims to provide a brief and concise introduction to well-known concepts of beam theory to be used in the remainder of this article.
As commonly used in engineering mechanics, we refer to the term \textit{beam} as a mathematical model for a three-dimensional, slender, deformable body for which the following assumption can be made.
The much larger extent of the body in its axial direction as compared to all transverse directions often justifies the Bernoulli hypothesis of rigid and therefore undeformable cross-sections.
This in turn allows for a reduced dimensional description as a 1D Cosserat continuum embedded in the 3D Euclidian space.

The so-called Simo-Reissner beam theory dates back to the works of Reissner~\cite{reissner1981}, Simo~\cite{simo1985}, and Simo and Vu-Quoc~\cite{simo1986}, who generalized the linear Timoshenko beam theory~\cite{Timoshenko1921} to the geometrically nonlinear regime.
Since the Simo-Reissner model, which accounts for the deformation modes of axial tension, bending, torsion and shear deformation, is the most general representative of geometrically exact beam theories, we choose it as the one to be used exemplarily throughout this work.
Nevertheless, the novel approach to be proposed is not restricted to a specific beam formulation.
We have likewise applied it to formulations of Kirchhoff-Love type, which are known to be advantageous in the regime of high slenderness ratios where the underlying assumption of negligible shear deformation is met~\cite{Meier2017c,Meier2017b}.
Refer to~\secref{sec::method_pot_based_ia_FE_discretization} for more details.

\paragraph{Geometry representation}$\,$\\
A certain configuration of the 1D Cosserat continuum is uniquely defined by the centroid position and the orientation of the cross-section at every point of the continuum.
The set of all centroid positions is referred to as \textit{centerline} or \textit{neutral axis} and expressed by the curve
\begin{equation}
  s,t \mapsto \vr(s,t) \in \MR^3
\end{equation}
in space and time~$t \in \MR$.
Each  material point along the centerline is represented by a corresponding value of the arc-length parameter~$s \in \left[ 0, l_0 \right] =: \Omega_l \subset \MR$.
Note that this arc-length parameter~$s$ is defined in the stress-free, initial configuration of the centerline curve~\mbox{$\vr_0(s) = \vr(s,t\!=\!0)$}.
Thus, the norm of the initial centerline tangent vector yields
\begin{equation}
  \norm{ \vr_0'(s) } := \norm{ \pdiff{ \vr_0(s) }{ s } } \equiv 1,
\end{equation}
but generally, in presence of axial tension, $\norm{ \vr'(s,t) } = \norm{ \tpdiff{\vr(s,t)}{s} } \neq 1$.

\begin{figure}[htpb]%
  \centering
  \def\svgwidth{0.6\textwidth}
  \input{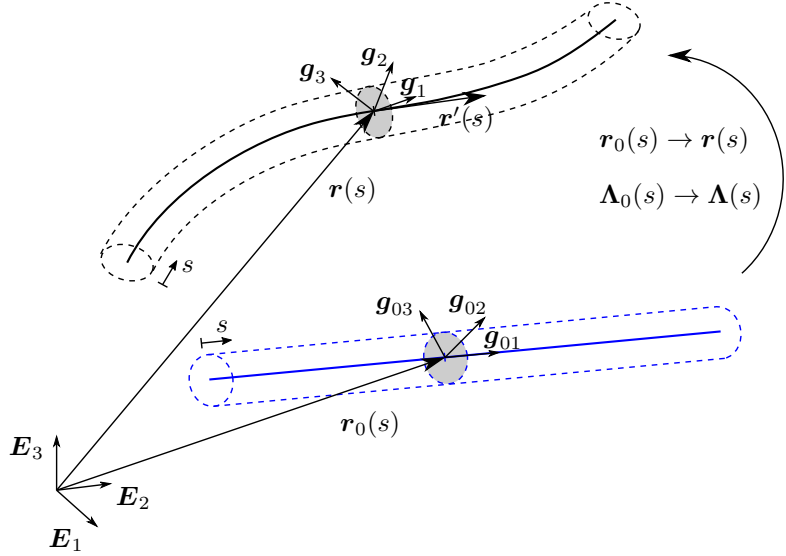}
  \caption{Geometry description and kinematics of the Cosserat continuum formulation of a beam: Initial, i.\,e.,~stress-free (blue) and deformed (black) configuration.
           Straight configuration in initial state is chosen exemplarily here without loss of generality.}
  \label{fig::beam_kinematics}
\end{figure}

Furthermore, the cross-section orientation at each of these material points is expressed by a right-handed orthonormal frame often denoted as \textit{material triad}:
\begin{equation}
  s,t \mapsto \vLambda(s,t) := \left[ \vg_1(s,t),\, \vg_2(s,t),\, \vg_3(s,t) \right] \in SO^3
\end{equation}
The second and third base vector follow those material fibers representing the principal axes of inertia of area.
Such a triad can equivalently be interpreted as rotation tensor transforming the base vectors of a global Cartesian frame~$\vE_i \in \MR^3, i \in {1,2,3}$ into the base vectors of the material triad $\vg_i \in \MR^3, i \in {1,2,3}$ via
\begin{equation}
  \vg_i(s,t) = \vLambda(s,t) \, \vE_i(s,t).
\end{equation}
In summary, a beam's configuration may be uniquely described by a field of centroid positions~$\vr(s,t)$ and a field of associated material triads~$\vLambda(s,t)$, altogether constituting a 1D Cosserat continuum (see \figref{fig::beam_kinematics}).
According to this concept of geometry representation, the position~$\vx$ of an arbitrary material point~$P$ of the slender body is obtained from
\begin{equation} \label{eq::position_material_point_Cosserat}
  \vx_\text{P}(s,s_2,s_3,t) = \vr(s,t) + s_2 \, \vg_2(s,t) + s_3 \, \vg_3(s,t).
\end{equation}
Here, the additional convective coordinates~$s_2$ and $s_3$ specify the location of P within the cross-section, i.\,e., as linear combination of the unit direction vectors~$\vg_2$ and $\vg_3$.
For a minimal parameterization of the triad, e.\,g.~the three-component rotation pseudo-vector~$\vpsi$ may be used such that we end up with six independent degrees of freedom~$(\vr,\vpsi)$ to define the position of each material point in the body by means of~\eqref{eq::position_material_point_Cosserat}.

\paragraph{Remark on notation}
Unless otherwise specified, all vector and matrix quantities are expressed in the global Cartesian basis~$\vE_i$. Differing bases as e.\,g.~the material frame are indicated by a subscript $\left[.\right]_{\vg_i}$.
Quantities evaluated at time~$t\!=\!0$, i.\,e., the initial stress-free configuration, are indicated by a subscript $0$ as e.\,g.~in~$\vr_0(s)$.
Differentiation with respect to the arc-length coordinate~$s$ is indicated by a prime, e.\,g., for the centerline tangent vector~$\vr'(s,t)=\tdiff{\vr(s,t)}{s}$.
Differentiation with respect to time~$t$ is indicated by a dot, e.\,g., for the centerline velocity vector~$\dot\vr(s,t) = \tdiff{\vr(s,t)}{t}$.
For the sake of brevity, the arguments~$s,t$ will often be omitted in the following.

\paragraph{Remark on finite 3D rotations}
To a large extent, the challenges and complexity in the numerical treatment of the geometrically exact beam theory can be traced back to the presence of large rotations.
In contrast to common \textit{vector spaces}, the rotation group $SO(3)$ is a \textit{nonlinear manifold} (with Lie group structure) and lacks essential properties such as additivity and commutativity, which makes standard procedures such as interpolation or update of configurations much more involved.
While \secref{sec::method_pot_based_ia} introduces the concept of section-to-section interaction laws in the most general manner, in \secref{sec::method_application_to_specific_types_of_interactions}, some additional (practically relevant) assumptions are made that allow to formulate the interaction laws as pure function of the beam centerline configuration.
In turn, this strategy will allow to avoid the handling of finite rotations and to achieve simpler and more compact numerical formulations.

\paragraph{Kinematics, deformation measures and potential energy of the internal, elastic forces}$\,$\\
\figref{fig::beam_kinematics} summarizes the kinematics of geometrically exact beam theory.
Based on these kinematic quantities, deformation measures as well as constitutive laws can be defined.
Finally, the potential of the internal (elastic) forces and moments~$\Pi_\text{int}$ is expressed uniquely by means of the set of six degrees of freedom~$(\vr,\vpsi)$ at each point of the 1D Cosserat continuum.
See e.\,g.~\cite{jelenic1999,crisfield1999,Meier2017c} for a detailed presentation of these steps.

\section{The Section-to-Section Interaction Potential (SSIP) Approach}\label{sec::method_pot_based_ia}
Based on the fundamentals of molecular interactions (\secref{sec::theoretical_foundation_molecular_interactions}) as well as geometrically exact beam theory (\secref{sec::fundamentals_beams}), this section will propose the novel SSIP approach to model various types of molecular interactions between deformable fibers undergoing large deflections in 3D.

\subsection{Problem statement}\label{sec::problem_statement_general_strategy}
For a classical conservative system, the total potential energy of the system can be stated taking into account the internal and external energy~$\Pi_{int}$ and~$\Pi_{ext}$.
The additional contribution from molecular interaction potentials~$\Pi_\text{ia}$ is simply added to the total potential energy as follows.
\begin{equation}
 \Pi_\text{TPE}=\Pi_\text{int}-\Pi_\text{ext}+\Pi_\text{ia} \defeq \text{min.}
\end{equation}
Note that the existing parts remain unchanged from the additional contribution.
One noteworthy difference is that internal and external energy are summed over all bodies in the system whereas the total interaction free energy is summed over all pairs of interacting bodies.

According to the \textit{principle of minimum of total potential energy}, the weak form of the equilibrium equations is derived by means of variational calculus.
The very same equation may alternatively be derived by means of the \textit{principle of virtual work} which also holds for non-conservative systems:
\begin{align}\label{eq::total_virtual_work_is_zero}
 \delta \Pi_\text{int} - \delta \Pi_\text{ext} + \delta \Pi_\text{ia} = 0
\end{align}
Clearly, the evaluation of the interaction potential~$\Pi_{ia}$, or rather its variation~$\delta \Pi_{ia}$, is the crucial step here.
Recall~\eqref{eq::pot_fullvolint} to realize that it generally requires the evaluation of two nested 3D integrals%
\footnote{It is important to mention that, assuming additivity of the involved potentials, systems with more than two bodies can be handled by superposition of all pair-wise two-body interaction potentials.
It is thus sufficient to consider one pair of beams in the following.
The same reasoning applies to more than one type of physical interaction, i.\,e., potential contribution.}.
The direct approach using 6D numerical quadrature turns out to be extremely costly and in fact inhibits any application to (biologically) relevant multi-body systems.
See \secref{sec::method_numerical_integration} for more details on the complexity and the cost of this naive, direct approach as well as the novel SSIP approach to be proposed in the following.

\subsection{The key to dimensional reduction from 6D to 2D}\label{sec::method_double_length_specific_integral}
We propose a split of the integral in the length dimensions $l_1, l_2$ on the one hand and the cross-sectional dimensions $A_1,A_2$ on the other hand:
\begin{equation}\label{eq::pot_split_int2D_int4D}
 \Pi_\text{ia} = \iint_{l_1,l_2} \; \underbrace{\iint_{A_1,A_2} \rho_1(\vx_1) \rho_2(\vx_2) \Phi(r) \dd A_2 \dd A_1}_{=: \; \tilde{\tilde{\pi}}(\vr_{1-2},\vpsi_{1-2})} \; \dd s_2 \dd s_1 \qquad \text{with} \quad r=\norm{\vx_1-\vx_2}.
\end{equation}
Exploiting the characteristic slenderness of beams, the 4D integration over both undeformable cross-sections shall be tackled analytically and only the remaining two nested 1D integrals along the centerline curves shall be evaluated numerically to allow for arbitrarily deformed configurations.
Generally speaking, we follow the key idea of reduced dimensionality from beam theory and thus aim to express the relevant information about the cross-sectional dimensions by the point-wise six degrees of freedom~$(\vr_i,\vpsi_i)$ of the 1D Cosserat continua without loss of significant information.
To this end we need to consider the resulting interaction between all the elementary interaction partners within two cross-sections expressed by an interaction potential~$\tilde{\tilde{\pi}}(\vr_{1-2},\vpsi_{1-2})$ that depends on the separation~$\vr_{1-2}$ of the centroid positions and the relative rotation~$\vpsi_{1-2}$ between both material frames attached to the cross-sections.
For this reason, the novel approach is referred to as the \textit{section-to-section interaction potential} (SSIP) approach.
The SSIP~$\tilde{\tilde{\pi}}$ is a double length-specific quantity in the way that it measures an energy per unit length of beam~$1$ per unit length of beam~$2$, which is indicated here by the double tilde.
The sought-after total interaction potential~$\Pi_\text{ia}$ of two slender deformable bodies thus results from double numerical integration of the double length-specific SSIP along both centerline curves:
\begin{align}\label{eq::ia_pot_double_integration}
 \Pi_\text{ia} = \int \limits_0^{l_1} \int \limits_0^{l_2} \tilde{\tilde{\pi}}(\vr_{1-2},\vpsi_{1-2}) \dd s_2 \dd s_1
\end{align}
This relation suggests another, alternative interpretation of the SSIP~$\tilde{\tilde{\pi}}$.
In analogy to the term~\textit{inter-surface potential}, introduced by~\cite{Argento1997}, $\tilde{\tilde{\pi}}$ can be understood as an \textit{inter-axis potential}, i.\,e., describing the interaction of two spatial curves (with attached material frames).

\begin{figure}[htpb]%
  \centering
  \def\svgwidth{0.5\textwidth}
  \input{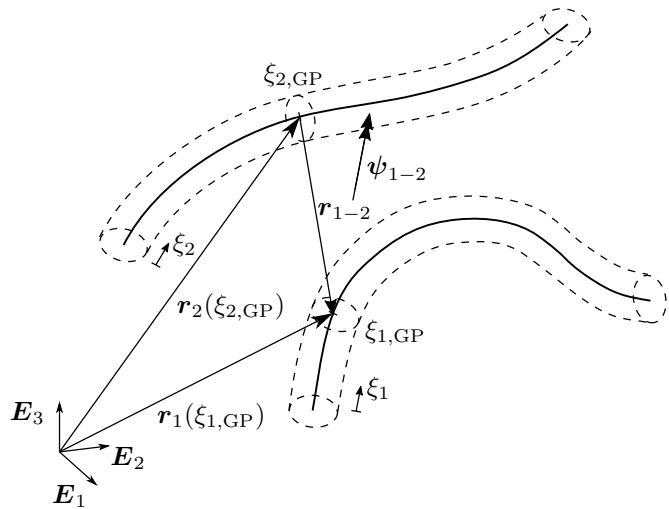}
  \caption{Illustration of the novel SSIP approach: Two cross-sections at integration points~$\xi_{1/2,\text{GP}}$ of beam~$1$ and~$2$, respectively, their separation~$\vr_{1-2}$ and relative rotation~$\vpsi_{1-2}$.}
  \label{fig::beam_to_beam_interaction_pot_double_integration}
\end{figure}

To further illustrate this novel concept, a simple, demonstrative example is shown in~\figref{fig::beam_to_beam_interaction_pot_double_integration}.
In this scenario of two beams with circular cross-section, the SSIP~$\tilde{\tilde{\pi}}(\vr_{1-2},\vpsi_{1-2})$ describes the interaction of two circular disks at arbitrary mutual distance and orientation.
To evaluate the two nested 1D integrals along the beam axes numerically, the SSIP needs to be evaluated for all combinations of integration points (denoted here as Gaussian quadrature points (GP), without loss of generality).
For one of these pairs~($\xi_{1,\text{GP}}$, $\xi_{2,\text{GP}}$), the geometrical quantities are shown exemplarily.

While analytical integration of the inner 4D integral of \eqref{eq::pot_split_int2D_int4D} has already been suggested above as one way to find a closed-form expression for the SSIP~$\tilde{\tilde{\pi}}$, we would like to stress the generality of the SSIP approach at this point.
The question of how to find~$\tilde{\tilde{\pi}}$ is independent of the strategy to determine the interaction energy~$\Pi_\text{ia}$ of two slender bodies via numerical double integration as proposed in this section.
This is important to understand because the SSIP~$\tilde{\tilde{\pi}}$ will obviously depend on the type of interaction as well as the cross-section shape and a number of other factors and there might also be cases where no analytical solution can be obtained and one wants to resort to relations fitted to experimental data.
In the scope of this work, several specific expressions of~$\tilde{\tilde{\pi}}$, e.\,g., for vdW as well as electrostatic interactions, will be derived analytically in the following \secref{sec::method_application_to_specific_types_of_interactions}.

In its most general form $\tilde{\tilde{\pi}}$ will be a function of the relative displacement~$\vr_{1-2}$ and the relative rotation~$\vpsi_{1-2}$ between both cross-sections, i.\,e., three translational and three rotational degrees of freedom.
This becomes clear if one recalls that the position~$\vx_\text{P}$ of every material point in a slender body can be uniquely described by the six degrees of freedom of a 1D Cosserat continuum (cf.\,eq.~\eqref{eq::position_material_point_Cosserat}).
Thus, keeping one cross-section fixed, the position~$\vx_{\text{P}1-\text{P}2}$ of every material point in the second cross-section relative to the (centroid position and material frame of the) first cross-section is again uniquely described by six degrees of freedom~$(\vr_{1-2},\vpsi_{1-2})$.
This insight naturally leads to the interesting question under which conditions the SSIP~$\tilde{\tilde{\pi}}$ can be described by a smaller set of degrees of freedom, thus simplifying the expressions.
Rotational symmetry of the interacting cross-sections is one common example where the SSIP would be invariant under rotations around the cross-section's normal axis.
We will return to this topic in~\secref{sec::assumptions_simplifications} as a preparation for the following derivation of specific expressions for the SSIP.

\paragraph{Remark on the included special case of surface interactions} \label{sec::method_surface_pot_ia}
It is very convenient that the practically highly relevant case of surface potentials is already included as a simpler, special case in the proposed SSIP approach to model molecular interactions between the entire volume of flexible fibers.
In simple words, it is sufficient to omit one spatial dimension of analytical integration on each interacting body in the analytical derivation of the required SSIP~$\tilde{\tilde{\pi}}$.
More specifically, this means that~$\tilde{\tilde{\pi}}$ may be obtained from solving analytically two nested 1D integrals along both, e.\,g.~ring-shaped, contour lines of the fiber cross-sections.

\section{Application of the General SSIP Approach to Specific Types of Interactions}\label{sec::method_application_to_specific_types_of_interactions}
At this point we would like to return to the fact that the SSIP approach proposed in \secref{sec::method_pot_based_ia} is general in the sense that it does not depend on the specific type of physical interaction.
This section provides the necessary information and formulae to apply the newly proposed, generally valid approach from \secref{sec::method_double_length_specific_integral} to certain types of real-world, physical interactions such as electrostatics or vdW.
As mentioned above, the approach requires a closed-form expression for the SSIP~$\tilde{\tilde{\pi}}$.
We basically see two alternative promising ways to arrive at such a reduced interaction law~$\tilde{\tilde{\pi}}$:
\begin{enumerate}
  \item analytical integration, e.\,g., as presented in \secref{sec::ia_pot_double_length_specific_evaluation_vdW}
  \item postulate~$\tilde{\tilde{\pi}}$ as a general function of separation~$\vr_{1-2}$ and mutual orientation~$\vpsi_{1-2}$ and determine the free parameters via fitting to
  \begin{enumerate}
    \item experimental data for specific section-to-section configurations, i.\,e., discrete values of separation~$\vr_{1-2}$ and mutual orientation~$\vpsi_{1-2}$
    \item data from (one-time) numerical 4D integration for specific section-to-section configurations, i.\,e., discrete values of~$\vr_{1-2}$ and~$\vpsi_{1-2}$
    \item experimental data for the global system response, e.\,g., of the entire fiber pair or a fiber network
  \end{enumerate}
\end{enumerate}
As a starting point we will restrict ourselves to the first option based on analytical integration throughout the remainder of this work.
See~\secref{sec::ia_pot_double_length_specific_evaluation_vdW} for an example of the further steps required to derive the final, ready-to-use expressions in case of vdW interactions.
To give but one example for a recent experimental work, which could serve as the basis for the second option listed above, we refer to \cite{Hilitski2015} measuring cohesive interactions between a single pair of microtubules.
Postulating an SSIP and studying the global system response in numerical simulations could also be used as a verification of theoretical predictions for the system behavior.
To give an example we refer to the work of theoretical biophysicists studying the structural polymorphism of the cytoskeleton resulting from molecular rod-rod interactions \cite{Borukhov2005}, which is based on a postulated model potential ``that captures the main features of any realistic potential''.
In summary, we see a large number of promising future use cases for the proposed SSIP approach.

\subsection{Additional assumptions and possible simplification of the most general form of SSIP laws}\label{sec::assumptions_simplifications}
Recall that the most general form of the SSIP is uniquely described by a set of six degrees of freedom, three for the relative displacement and three for the relative orientation of the two interacting cross-sections, as presented in the preceding~\secref{sec::method_double_length_specific_integral}.
The following assumptions turn out to significantly simplify this most general form of the SSIP law by reducing the number of relevant degrees of freedom from six to four, two or even one.
This in turn eases the desirable derivation of analytical closed-form solutions of the SSIP~$\tilde{\tilde{\pi}}$ based on the point pair potentials~$\Phi(r)$ presented in~\secref{sec::theory_molecular_interactions_pointpair}.
Specifically, these assumptions are:
\begin{enumerate}
  \item undeformable cross-sections
  \item circular cross-section shapes
  \item homogeneous (or, more generally, rotationally symmetric) particle densities~$\rho_1,\rho_2$ in the cross-sections or surface charge densities~$\sigma_1,\sigma_2$ over the circumference
\end{enumerate}
The first assumption is typical for geometrically exact beam theory and the second and third assumption are reasonable regarding our applications to biopolymer fibers such as actin or DNA that can often be modeled as homogeneous fibers with circular cross-sections.
Based on these three assumptions, we can conclude that the interaction between two cross-sections is geometrically equivalent to the interaction of two homogeneous, circular disks (or rings in case of surface interactions).
The rotational symmetry of the circular disks then implies that the interaction potential is invariant under rotations around their own axes and thus reduces the number of degrees of freedom to four.
The relative importance of the remaining degrees of freedom, i.\,e., modes, will be the crucial point in the following discussion, where we turn to the interaction of two slender bodies, i.\,e., consider the entirety of all cross-section pairs.
At this point, recall the fundamental distinction between either short-range or long-range interactions as outlined in~\secref{sec::theory_molecular_interactions_twobody_vol}.

\begin{figure}[htb]
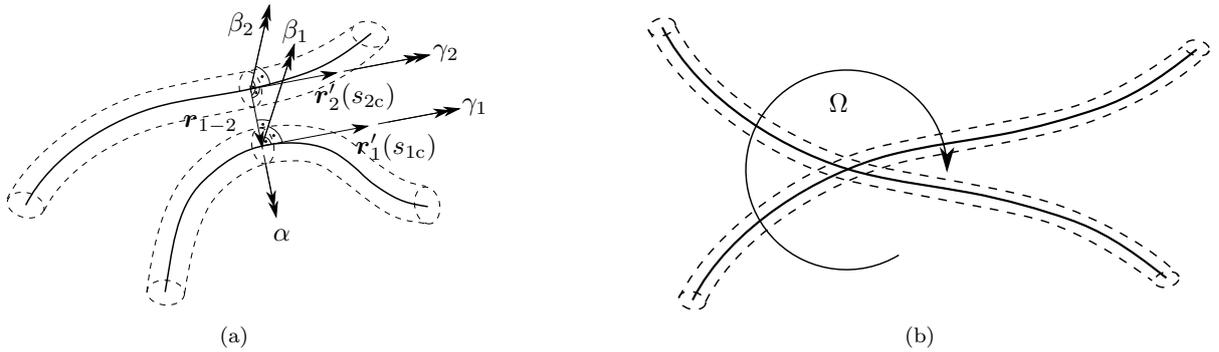
%
  \centering
  \subfigure[]{
    \def\svgwidth{0.41\textwidth}
    \input{beam_to_beam_interaction_sketch_assumptions_short-range.pdf_tex}
    \label{fig::beam_to_beam_interaction_sketch_assumptions_short_range}
  }
  \hfill
  \subfigure[]{
    \def\svgwidth{0.45\textwidth}
    \input{beam_to_beam_interaction_sketch_assumptions_long-range.pdf_tex}
    \label{fig::beam_to_beam_interaction_sketch_assumptions_long_range}
  }
  \caption{Sketches to illustrate the simplifications resulting for (a) short- and (b) long-range interactions.}
  \label{fig::beam_to_beam_interaction_sketch_assumptions}
\end{figure}
In the case of short-range interactions, the cross-section pairs in the immediate vicinity of the mutual closest points of the slender bodies dominate the total interaction.
As is known from macroscopic beam contact formulations~\cite{wriggers1997,meier2016,Meier2017a}, the criterion for the closest point is that the distance vector~$\vr_{1-2}$ is perpendicular to both centerline tangent vectors~$\vr'_i$, i.\,e., (assuming small shear angles) the normal vectors of the disks (see~\figref{fig::beam_to_beam_interaction_sketch_assumptions_short_range}).
Since only cross-section pairs in the direct vicinity of the closest points are relevant, arbitrary relative configurations (i.\,e.~separations and relative rotations) between those cross-sections shall in the following be discussed on the basis of six alternative degrees of freedom as illustrated in~\figref{fig::beam_to_beam_interaction_sketch_assumptions_short_range}.
By considering the cross-sections~$A_{1\text{c}}$ and $A_{2\text{c}}$ at the closest points as reference, the relative configuration between cross-sections in the direct vicinity of $A_{1\text{c}}$ and $A_{2\text{c}}$ can be described via (small) rotations of $A_{1\text{c}}$ around the axis $\vr_1'$ (angle~$\gamma_1$) and $\vr_{1-2} \times \vr_1'$ (angle~$\beta_1$), (small) rotations of $A_{2\text{c}}$ around the axes $\vr_2'$ (angle~$\gamma_2$) and $\vr_{1-2} \times \vr_2'$ (angle~$\beta_2$), (small) relative rotations between $A_{1\text{c}}$ and $A_{2\text{c}}$ around the axis $\vr_{1-2}$ (angle~$\alpha$), and (small) changes in the (scalar) distance $d=\norm{\vr_{1-2}}$.
As a consequence of assumptions 1--3 discussed above, the considered interaction potentials are invariant under rotations $\gamma_1$ and $\gamma_2$.
From the remaining four degrees of freedom, the scalar distance~$d$ clearly has the most significant influence on the interaction potential because changes in~$d$ directly affect the mutual distance~$r$ of all point pairs in the body and, most importantly, the smallest surface separation~$g$ between both bodies.
The second most significant influence is expected for the scalar relative rotation angle~$\alpha$ between the cross-section normal vectors, i.\,e., $\cos(\alpha) = \vn_1 \cdot \vn_2 \approx \vr_1' \cdot \vr_2' / (\norm{\vr_1'} \norm{\vr_2'})$.
A change in~$\alpha$ does not alter the gap~$g$, but influences the distance of all next nearest point pairs in the immediate vicinity of the closest surface point pair.
For the remaining two relative rotations~$\beta_1$ and $\beta_2$, arguments for both sides, either significant or rather irrelevant influence on the total interaction potential, can be found at this point.
On the one hand, the orthogonality conditions~$\vr_i' \cdot \vr_{1-2} = 0$, $i=1,2$ at the closest points are fulfilled in good approximation also for cross-sections in the direct vicinity of the closest points, such that the influence of~$\beta_{1/2}$ could be considered negligible.
On the other hand, even small rotations~$\beta_{1/2}$ change the smallest separation of any two point pairs in the immediate neighborhood of the closest point pair as soon as the centroid distance vector $\vr_{1-2}$ rotates out of the two cross-section planes.
Therefore, it seems hard to draw a final conclusion with respect to the influence and thus importance of~$\beta_{1/2}$ based on the qualitative theoretical considerations of this section.

To summarize, the scalar distance~$d$ between the cross-section centroids, the scalar relative rotation angle~$\alpha$ between the cross-section normal vectors and possibly also the relative rotation components~$\beta_{1/2}$ are supposed to have a perceptible influence on the short-range interaction between slender beams fulfilling assumptions 1--3, with a relative importance which decreases in this order.
In favor of the simplest possible model, we will therefore assume at this point that the effect of this scalar relative rotations~$\alpha, \beta_1$ and $\beta_2$ is negligible as compared to the effect of the scalar separation~$d$.
This allows us to directly use the analytical, closed-form expression for the disk-disk interaction potential as presented in~\secref{sec::theory_molecular_interactions_twobody_vdW}.
The error for arbitrary configurations associated with this model assumption will be thoroughly analyzed in \secref{sec::verif_approx}.
In this context, it is a noteworthy fact, that the first published method for 2D beam-to-rigid half space LJ interaction \cite{Sauer2009} likewise neglects the effect of cross-section orientation.
In the subsequent publication \cite{Sauer2014}, the effect of cross-section rotation, i.\,e., interaction moments, has finally been included and a quantitative analysis considering a peeling experiment of a Gecko spatula revealed that the differences in the resulting maximum peeling force and bending moment are below~$8\%$ and~$2\%$, respectively.
However, it is unclear whether this assessment also holds for beam-to-beam interactions modeled via the proposed SSIP approach.
Including the orientation of the cross-sections thus is a work in progress and will be addressed in a subsequent publication.
Finally, it is emphasized that by the discussed assumptions the SSIP law~$\tilde{\tilde{\pi}}$ as well as the total two-body interaction potential~$\Pi_\text{ia}$ can be formulated as pure function of the beam centerlines~$\vr_1$ and $\vr_2$ without the necessity to consider cross-section orientations via rotational degrees of freedom.
This is considered a significant simplification of the most general case of the SSIP approach and thus facilitates both the remaining derivations in the present work as well as potential future applications.

\paragraph{Remark on configurations with non-unique closest points}
It is well-known from the literature on macroscopic beam contact that the location of the closest points is non-unique for certain configurations of two interacting beams, e.\,g., the trivial case of two straight beams, where an infinite number of closest point pairs exists (see e.\,g.~\cite{meier2016}).
Note however that the reasoning presented above also holds in these cases, since the cross-section pairs in either one or several of these regions will dominate the total interaction potential.\\

In the case of long-range interactions, the situation is fundamentally different.
Recall that here the large number of cross-section pairs with large separation~$d\gg R$ outweighs the contributions from those few pairs in the vicinity of the closest point and dominates the total interaction.
Thus, the regime of large separations is decisive in this case and it has already been shown in the literature considering disk-disk interaction (see the brief summary in~\ref{sec::derivation_pot_ia_powerlaw_disks}) that in this regime the exact orientation of the disks can be neglected as compared to the centroid separation~$d$.
In simple terms, this holds because the distance~$\vx_{\text{P}1-\text{P}2}$ between any point in disk~$1$ and any point in disk~$2$ may be approximated by the centroid separation~$d$, if~$d$ is much larger than the disk radii~$R_i$, which - again - holds for the large majority of all possible cross-section pairs.
The validity of this assumption will be thoroughly verified by means of numerical reference solutions in~\secref{sec::verif_SSIP_disks_cyls_elstat}.

\paragraph{Remark}
The following, similar reasoning from the perspective of slender continua comes to the same conclusion.
As visualized in ~\figref{fig::beam_to_beam_interaction_sketch_assumptions_long_range}, even pure (rigid body) rotations of slender bodies always entail large displacements of the centerline%
\footnote{Disregarding rotations around its own axis, which are irrelevant here due to rotational symmetry, as mentioned above.}
in the region far away from the center of rotation.
The displacement of any material point due to cross-section rotation will be in the order of~$\Omega R$, where~$\Omega$ is the angle of rotation and~$R$ denotes the cross-section radius, whereas the displacement due to centerline displacement will be in the order of~$\Omega L$, where~$L$ is the distance from the center of rotation and thus in the order of the beam length~$l$.
Due to the high slenderness~$l/R\gg 1$ of beams, the displacement from translation of the centroid will dominate in the region of large separations with~$L\gg R$, which is the decisive one here, because it includes the large majority of all possible cross-section pairs, as outlined above.
The original, analogous reasoning has been applied to the relative importance of translational versus rotational contributions to the mass inertia of beams.\\

To conclude, we have discussed the possibility of defining and using SSIP laws~$\tilde{\tilde{\pi}}$ as a function of the scalar separation of the centroids~$d$ instead of the six degrees of freedom in the most general form.
This significantly simplifies the theory because the analytical solutions for the planar disk-disk interaction from literature can directly be used and the complex treatment of large rotations is avoided.
Having considered the additional assumptions above in the context of short-range interactions, the relative importance of cross-section rotations still needs to be verified in the subsequent quantitative analysis of~\secref{sec::verif_approx}.
In the case of long-range interactions between slender bodies, we have argued that the application of such simple SSIP laws~$\tilde{\tilde{\pi}}(d)$ is expected to be a good approximation which will be confirmed by the quantitative analysis of~\secref{sec::verif_SSIP_disks_cyls_elstat}.

\subsection{Short-range volume interactions such as van der Waals and steric repulsion}\label{sec::ia_pot_double_length_specific_evaluation_vdW}
In the following, a generic short-range volume interaction described by the point-pair potential law
\begin{equation}\label{eq::pot_general_powerlaw_pointpair}
  \Phi_\text{m}(r)= k_\text{m} \, r^{-m}, \, m>3
\end{equation}
will be considered, because it includes vdW interaction for exponent~$m=6$ (cf.~eq.~\eqref{eq::pot_ia_vdW_pointpair}) as well as steric repulsion as modeled by LJ for exponent~$m=12$ (cf.~eq.~\eqref{eq::pot_ia_LJ_pointpair}).
As outlined already in the preceding section, only the regime of small separations is practically relevant in this case of short-range interactions and we neglect the effect of cross-section rotations throughout this article.
At this point, we can thus return to the results for the disk-disk scenario obtained in literature on vdW interactions \cite{langbein1972} and summarized in Table~\ref{tab::pot_ia_vdW_disks_formulae}.
In particular, we make use of expression~\eqref{eq::pot_ia_vdW_disk_disk_parallel_smallseparation} or rather the more general form \eqref{eq::approx_small_sep}.
The latter is valid for all power-law point pair interaction potentials with a general exponent~$m>7/2$, i.\,e., all interactions where the strength decays ``fast enough''.

First, let us introduce the following abbreviation containing all constants in the lengthy expression:
\begin{equation}\label{eq::vdW_small_sep_def_constant}
 c_\text{m,ss} \defvariable k_\text{m}\rho_1 \rho_2 \frac{2 \pi}{(m-2)^2} \quad \sqrt{ \frac{2 R_1 R_2}{R_1+R_2} } \quad \frac{\Gamma(m-\tfrac{7}{2}) \Gamma(\tfrac{m-1}{2})}{\Gamma(m-2) \Gamma(\tfrac{m}{2}-1)}
\end{equation}
Using eq.~\eqref{eq::approx_small_sep} in combination with the general SSIP approach \eqref{eq::ia_pot_double_integration} from \secref{sec::method_double_length_specific_integral}, we directly obtain an expression for the total interaction potential of two deformable fibers in the case of short-range interactions:
\begin{align}\label{eq::iapot_small_sep}
 \Pi_\text{m,ss} &= \int_0^{l_1} \int_0^{l_2} c_\text{m,ss} \, g^{-m+\tfrac{7}{2}} \, \dd s_2 \dd s_1 \qquad \text{for} \quad m>\tfrac{7}{2}\\
 & \text{with} \quad g(s_1,s_2) = \norm{ \vr_1(s_1) - \vr_2(s_2) } - R_1 - R_2\label{eq::gap}
\end{align}
Here, the so-called gap $g$ is the (scalar) surface-to-surface separation, i.\,e., the beams' centerline curves $\vr_1(s_1)$ and $\vr_2(s_2)$ minus the two radii~$R_i$, as visualized in Figure~\ref{fig::beam_to_beam_interaction_pot_double_integration}.
In general, the particle densities $\rho_{1/2}$ may depend on the curve parameters $s_{1/2}$, i.\,e., vary along the fiber, without introducing any additional complexity at this point.
For the sake of brevity, these arguments~$s_{1/2}$ will be omitted in the remainder of this section.

The variation of the interaction potential required to solve eq.~\eqref{eq::total_virtual_work_is_zero} finally reads
\begin{align}\label{eq::var_iapot_small_sep}
 \delta \Pi_\text{m,ss} = (-m+\tfrac{7}{2}) \int_0^{l_1} \int_0^{l_2} c_\text{m,ss} \left( \delta \vr_1^T - \delta \vr_2^T \right) \frac{\vr_1 - \vr_2}{d} g^{-m+\tfrac{5}{2}} \dd s_2 \dd s_1 \qquad \text{for} \quad m>\tfrac{7}{2}.
\end{align}
Here, we used the variation of the gap $\delta g$, which is a well-known expression from the literature on macroscopic beam contact \cite{wriggers1997} and is identical to the variation of the separation of the beams' centerlines $\delta d$ to be used in \eqref{eq::var_iapot_large_sep_surface}, because the cross-sections are assumed to be undeformable:
\begin{equation}
 \delta g = \delta d = \left( \delta \vr_1^T - \delta \vr_2^T \right) \frac{\vr_1 - \vr_2}{d}
\end{equation}
Solving eq.~\eqref{eq::total_virtual_work_is_zero} generally requires two further steps of discretization and subsequent linearization of this additional contribution~$\delta \Pi_\text{m,ss}$ to the total virtual work.
The resulting expressions will be presented in \secref{sec::method_vdW_FE_discretization} and \ref{sec::method_vdW_linearization}, respectively.
As discussed along with the general SSIP approach in \secref{sec::method_double_length_specific_integral}, the remaining two nested~1D integrals are evaluated numerically, e.\,g., by means of Gaussian quadrature.
See \secref{sec::method_numerical_integration} for details on this algorithmic aspect.

\paragraph{Remark on the regularization of the integrand}
The inverse power law in the integrand of eq.~\eqref{eq::var_iapot_small_sep} has a singularity for the limit of zero surface-to-surface separation~$g\rightarrow0$.
Consequently, a so-called \textit{regularization} of the potential law is needed to numerically handle (the integration of) this term robustly as well as sufficiently accurate.
This approach is well-known e.\,g.~from (beam) contact mechanics (see e.\,g.~\cite{durville2007,Sauer2013,meier2016}) and will be further discussed and elaborated in \secref{sec::regularization}.\\

At the end of this section, we can conclude that we found specific, ready-to-use expressions for the interaction free energy as well as virtual work of generic short-range interactions described via the SSIP approach.
Thus, vdW interaction or steric exclusion of slender, deformable continua can now be modeled in an efficient manner, reducing the numerical integral to be evaluated from six to two dimensions.
A detailed quantitative study of the approximation quality with regard to the assumptions discussed in the preceding~\secref{sec::assumptions_simplifications} is content of \secref{sec::verif_approx}.

\subsection{Long-range surface interactions such as electrostatics}\label{sec::ia_pot_double_length_specific_evaluation_elstat}
Having discussed short-range volume interactions, we now want to consider one example of \textit{long-range surface potentials}.
Since electrostatic interaction is the prime example of surface potential interaction and at the same time of high interest for the application to biopolymers we have in mind, we will focus on this case throughout the following section and mostly speak of \textit{point charges} as the elementary interaction partners.
However, the required steps and formulae will be presented as general as possible in order to allow for a smooth future transfer to other applications.

Especially in this context, it is important to stress again that within this model the elementary interaction partners, i.\,e., charges must not redistribute within the bodies.
Hence, only non-conducting materials can be modeled with the SSIP approach.
This however covers our main purpose to model electrostatic interactions between bio-macromolecules such as protein filaments and DNA because charges are not free to move therein.

According to the SSIP approach proposed in~\secref{sec::method_double_length_specific_integral}, we aim to use analytical expressions for the two inner integrals over the cross-section circumferences, while the integration along the two beam centerlines will be evaluated numerically (cf.~eq.~\eqref{eq::pot_split_int2D_int4D} in combination with the remark on surface interactions at the end of the corresponding section).
As discussed in~\secref{sec::assumptions_simplifications}, the regime of large separations is the decisive one for beam-to-beam interactions in this case of long-range interactions and the SSIP law can be simplified in good approximation to depend only on the centroid separation~$d$, which will be confirmed numerically in~\secref{sec::verif_SSIP_disks_cyls_elstat}.

At this point, we again return to the expressions for the disk-disk interaction based on a generic point pair potential~$\Phi_m(r)=k \, r^{-m}$, as derived in the literature on vdW interactions~\cite{langbein1972} and summarized in~\ref{sec::derivation_pot_ia_powerlaw_disks}.
In particular, the relation~\eqref{eq::approx_large_sep} will be used, which is the same approximation used to derive eq.~\eqref{eq::pot_ia_vdW_disk_disk_parallel_largeseparation} that describes the practically rather irrelevant scenario of short-range vdW interactions in the regime of large separations.
Note that in the context of electrostatics, this result is well-known as the first term, i.\,e., zeroth pole or \textit{monopole} of the multipole expansion of the ring-shaped charge distribution on each of the disks' circumference, which represents the effect of the net charge of a (continuous) charge distribution and has no angular dependence (see~\secref{sec::theory_molecular_interactions_twobody_surf}).
In simple terms, this monopole-monopole interaction means that the point pair interaction potential~$\Phi(r)$ is evaluated only once for the distance between the centers of the distributions~$r=d=\norm{ \vr_1 - \vr_2 }$ and weighted with the number of all point charges on the two circumferences of the circular cross-sections.
The expression for the SSIP law to be used throughout this work would thus be exact for the scenario of the net charge of each cross-section concentrated at the centroid position (or distributed spherically symmetric around the centroid position).
If the accuracy of the SSIP approach needs to be improved beyond the level resulting from this simplified SSIP law (see \secref{sec::verif_SSIP_disks_cyls_elstat} for the analysis), one could simply include more terms from the multipole expansion of the (ring-shaped) charge distributions to the SSIP law, which would take the relative rotation of the cross-sections into account.
Throughout this work and for the applications we have in mind, the simplified SSIP law, which is based on the monopole-monopole interaction of cross-sections, turns out to be an excellent approximation for the true electrostatic interaction law and we thus restrict ourselves to this variant.

Two nested 1D integrals over the beams' length dimensions then yield the two-body interaction potential for two fibers with arbitrary centerline shapes
\begin{align}\label{eq::iapot_large_sep_surface}
 \Pi_\text{ia,ls} =\int_0^{l_1} \int_0^{l_2} 2\pi R_1 \sigma_1 \, 2\pi R_2 \sigma_2 \, \Phi(r=d) \dd s_2 \dd s_1 \qquad \text{with} \qquad d(s_1,s_2) = \norm{ \vr_1(s_1) - \vr_2(s_2)}.
\end{align}
The surface (charge) densities~$\sigma_j$, $j=1,2$ have already been introduced in eq.~\eqref{eq::pot_fullvolint_surface}.
Particularly for the case of electrostatics, the surface charge per unit length can be identified as $\lambda_j = 2\pi R_j \sigma_j$, $j=1,2$, and is commonly referred to as \textit{linear charge density}.
Note however that eq.~\eqref{eq::iapot_large_sep_surface} holds for all long-range point pair potential laws $\Phi(r)$, e.\,g., all power laws~$\Phi_m(r)=k \, r^{-m}$ with $m\leq3$.
In order to obtain the weak form of the continuous problem, the variation of this total interaction energy needs to be derived.
This variational form can immediately be stated as
\begin{align}\label{eq::var_iapot_large_sep_surface}
 \delta \Pi_\text{ia,ls} &=  \int_0^{l_1} \int_0^{l_2} \lambda_1 \lambda_2 \, \pdiff{{\Phi(r=d)}}{d} \, \delta d \, \dd s_2 \dd s_1
 \quad \text{with} \quad \delta d = \left( \delta \vr_1^T - \delta \vr_2^T \right) \frac{\vr_1 - \vr_2}{d}
\end{align}
as the consistent variation of the separation of the beams' centerlines~$d$, which is well-known from macroscopic beam contact formulations \cite{wriggers1997}.
By inserting the generic (long-range) power law
\begin{equation}
 \Phi_m(r) = k \, r^{-m}, \, m \leq 3
\end{equation}
into \eqref{eq::var_iapot_large_sep_surface}, we obtain the final expression for the variation of the two-body interaction energy of two deformable slender bodies
\begin{equation}\label{eq::var_pot_ia_powerlaw_large_sep_surface}
 \delta \Pi_\text{m,ls} = \int_0^{l_1} \int_0^{l_2} \, \underbrace{k m \, \lambda_1 \lambda_2}_{=:c_\text{m,ls}} \, \left( - \delta \vr_1^T + \delta \vr_2^T \right) \, \frac{\vr_1 - \vr_2}{d^{m+2}}  \, \dd s_2 \dd s_1.
\end{equation}
The specific case of Coulombic surface interactions follows directly for~$m=1$ and~$k=C_\text{elstat}$ (cf.~eq.~\eqref{eq::pot_ia_elstat_pointpair_Coulomb}).
At this point, we have once again arrived at the sought-after contribution to the weak form~\eqref{eq::total_virtual_work_is_zero} of the space-continuous problem.
The steps of finite element discretization and linearization will again be presented later, in \secref{sec::method_elstat_FE_discretization} and \ref{sec::method_elstat_linearization}, respectively.

\paragraph{Remark on volume interactions}
Note that there is no conceptual difference if long-range volume interactions were considered instead of the long-range surface interactions presented exemplarily in this section.
The only difference lies in the constant prefactor~$c_\text{m,ls}$, which would read~$c_\text{m,ls}=k m A_1 A_2 \rho_1 \rho_2$ instead.
Rather than the spatial distribution of the elementary interaction points in the volume or on the surface, it is the long-ranged nature of the interactions, which is important for the derivations in this section and allows the use of approximations for large separations (refer to the extensive discussion in~\secref{sec::assumptions_simplifications}).

\paragraph{Remark on intra- versus inter-body interactions}
The electrostatic interaction of point charges on the same slender body may cause unexpected effects.
Assuming equal charges along the beam length leads to repulsive forces which in turn cause tensile axial forces in the beam.
At the start of a dynamic simulation, a simply supported beam will undergo axial strain oscillations before eventually an equilibrium state is found.
Alternatively, these interactions of charges within the same body may be included in the constitutive model used for the continuous body and to this end be modeled by an increased effective stiffness as has e.\,g.~been suggested by \cite{Sauer2007a,Cyron2013a}.
The latter approach has been applied in the numerical examples of \secref{sec::numerical_results}.

\section{Finite Element Discretization and Selected Algorithmic Aspects}\label{sec::FE_discretization_algorithmic_aspects}
Having discussed the space-continuous theory in \secref{sec::method_pot_based_ia} and \ref{sec::method_application_to_specific_types_of_interactions}, we now turn to the step of spatial discretization by means of finite elements.
Subsequently, the most important aspects of the required algorithmic framework will be presented briefly and discussed specifically in the light of the novel SSIP approach.
This includes the applied regularization technique, multi-dimensional numerical integration, an analysis of the algorithmic complexity as well as the topics of search for interaction partners and parallel computing.

\subsection{Spatial discretization based on beam finite elements}\label{sec::method_pot_based_ia_FE_discretization}
As presented in \secref{sec::beam_theory}, the centerline position~$\vr$ and the triad~$\vLambda$ arise as the two primary fields of unknowns.
Within Simo-Reissner beam theory, both fields are uncorrelated and their discretization can hence be considered independently as follows.
The Simo-Reissner finite beam element used throughout this work originates from \cite{jelenic1999,crisfield1999}, although we apply a different centerline interpolation scheme here.
We employ cubic Hermite polynomials based on nodal position vectors~$\hat\vdd^1, \hat\vdd^2$ and tangent vectors~$\hat\vdt^1, \hat\vdt^2$ as the primary variables.
See \cite{Meier2014} for a detailed discussion of Hermite centerline interpolation in the context of geometrically exact (Kirchhoff) beams and \cite{Meier2017b} for the details on the Hermitian Simo-Reissner beam element that is used within this article.
Applying this interpolation scheme results in the following discretized centerline geometry and variation:
\begin{align}
\begin{split}\label{eq::centerline_discretization}
  \vr(\xi) \approx \vr_\text{h}(\xi) &= \sum_{i=1}^2 H_d^i(\xi) \, \hat \vdd^i + \frac{l}{2} \sum_{i=1}^2 H_t^i(\xi) \, \hat \vdt^i =: \vdH \, \hat \vdd, \\
  \delta \vr(\xi) \approx \delta \vr_\text{h}(\xi) &= \sum_{i=1}^2 H_d^i(\xi) \, \delta \hat \vdd^i + \frac{l}{2} \sum_{i=1}^2 H_t^i(\xi) \, \delta \hat \vdt^i =: \vdH \, \delta \hat \vdd
\end{split}
\end{align}
Here, all the degrees of freedom of one element relevant for the centerline interpolation, i.\,e., nodal positions~$\hat\vdd^i$ and tangents~$\hat\vdt^i$, $i=1,2$ are collected in one vector~$\hat\vdd$ and~$\vdH$ is the accordingly assembled matrix of shape functions, i.\,e., Hermite polynomials~$H_d^i$ and $H_t^i$.
The newly introduced element-local parameter~$\xi \in [-1;1]$ is biuniquely related to the arc-length parameter~$s \in [s_\text{ele,min}; s_\text{ele,max}]$ describing the very same physical domain of the beam as follows and the scalar factor defining this mapping between both infinite length measures is called the element \textit{Jacobian}~$J(\xi)$:
\begin{equation}
  \dd s = \diff{ s }{ \xi } \dd \xi =: J(\xi) \dd \xi \qquad \text{with} \qquad J(\xi) := \norm{ \diff { \vr_{0,\text{h}}(\xi) }{ \xi } }.
\end{equation}
Our motivation to use Hermite interpolation is that it ensures $C_1$-continuity, i.\,e., a smooth geometry representation even across element boundaries.
This property turned out to be crucial for the robustness of simulations in the context of macroscopic beam contact methods \cite{Meier2017b}, and is just as important if we include molecular interactions as proposed in this article.
See \cite{Sauer2011} for a comprehensive discussion of (non-)smooth geometries and adhesive, molecular interactions using 2D solid elements.
Note however that neither the SSIP approach proposed in \secref{sec::method_pot_based_ia} nor the specific expressions for the interaction free energy and the virtual work are limited to this Hermite interpolation scheme.
In fact, all of the following discrete expressions will be equally valid for a large number of other beam formulations, where the discrete centerline geometry is defined by polynomial interpolation, which can generally be expressed in terms of the generic shape function matrix~$\vdH$ introduced above.

Recall also, that the proposed SSIP laws from \secref{sec::method_application_to_specific_types_of_interactions} solely depend on the centerline curve description, i.\,e., the rotation field does not appear in the additional contributions and hence its discretization is not relevant in the context of this work.
It is therefore sufficient to apply the discretization scheme for the centerline field stated in eq.~\eqref{eq::centerline_discretization} to the expressions for the virtual work contributions~$\delta \Pi_\text{ia}$ presented in \secref{sec::method_application_to_specific_types_of_interactions} and finally end up with the discrete element residual vectors~$\vdr_{\text{ia},1/2}$.
The latter need to be assembled into the global residual vector~$\vdR$ as it is standard in the (nonlinear) finite element method.
Note that the linearization of all the expressions presented in this~\secref{sec::method_pot_based_ia_FE_discretization} is provided in~\ref{sec::linearization}.

\subsubsection{Short-range volume interactions such as van der Waals and steric repulsion}\label{sec::method_vdW_FE_discretization}

Discretization of the centerline curves according to \eqref{eq::centerline_discretization}, i.\,e., $\vr_j \approx \vr_{\text{h},j} = \vdH_j \, \hat \vd_j$ and $\delta \vr_j^\text{T} \approx \delta \vr_{\text{h},j}^\text{T} = \delta \hat \vd_j^\text{T} \, \vdH_j^\text{T}$, for both beam elements~$j={1,2}$ turns the space-continuous form~\eqref{eq::var_iapot_small_sep} of the two-body virtual work contribution from molecular interactions~$\delta \Pi_\text{m,ss}$ into its discrete counterpart
\begin{align}\label{eq::var_discrete_iapot_small_sep}
 \delta \Pi_\text{m,ss,h} = -(m-\tfrac{7}{2}) \int_0^{l_1} \int_0^{l_2} c_\text{m,ss} \left( \delta \hat \vd_1^\text{T} \vdH_1^\text{T} - \delta \hat \vd_2^\text{T} \vdH_2^\text{T} \right) \frac{\vr_{h,1} - \vr_{h,2}}{d_{h}} g_h^{-m+\tfrac{5}{2}} \dd s_2 \dd s_1.
\end{align}
Refer to~\eqref{eq::vdW_small_sep_def_constant} for the definition of the constant~$c_\text{m,ss}$.
Note that \eqref{eq::var_discrete_iapot_small_sep} only contributes to those scalar residua associated with the centerline, i.\,e., translational degrees of freedom~$\hat \vdd$.
This is a logical consequence of the fact that the SSIP law solely depends on the beams' centerline curves, as discussed in detail in \secref{sec::method_application_to_specific_types_of_interactions}.
For the sake of brevity, the index 'h', indicating all discrete quantities, will be omitted from here on since all following quantities are considered discrete.
In eq.~\eqref{eq::var_discrete_iapot_small_sep}, the discrete element residual vectors of the two interacting elements~$j=1,2$ can finally be identified as
\begin{align}\label{eq::res_ia_pot_smallsep_ele1}
  \vdr_{\text{m,ss},1} &=  - (m-\tfrac{7}{2}) \int_0^{l_1} \int_0^{l_2} c_\text{m,ss} \, \vdH_1^\text{T} \frac{ \left( \vr_{1} - \vr_{2} \right)}{d} \, g^{-m+\tfrac{5}{2}} \, \dd s_2 \dd s_1 \quad \text{and}\\
  \vdr_{\text{m,ss},2} &=  (m-\tfrac{7}{2}) \int_0^{l_1} \int_0^{l_2} c_\text{m,ss} \, \vdH_2^\text{T} \frac{ \left( \vr_{1} - \vr_{2} \right)}{d} \, g^{-m+\tfrac{5}{2}} \, \dd s_2 \dd s_1 \label{eq::res_ia_pot_smallsep_ele2}.
\end{align}
See \secref{sec::method_numerical_integration} for details on the numerical quadrature required to evaluate these expressions.

\subsubsection{Long-range surface interactions such as electrostatics}\label{sec::method_elstat_FE_discretization}
In analogy to the previous section, we discretize~\eqref{eq::var_pot_ia_powerlaw_large_sep_surface} and obtain the discrete element residual vectors
\begin{align}\label{eq::res_ia_pot_largesep}
 \vdr_{\text{m,ls},1} &= - \int_0^{l_1} \int_0^{l_2} \, c_\text{m,ls} \, \vdH_1^\text{T} \frac{ \left( \vr_{1} - \vr_{2} \right)}{d^{m+2}} \, \dd s_2 \dd s_1 \quad \text{and} \quad
 \vdr_{\text{m,ls},2} &= \int_0^{l_1} \int_0^{l_2} \, c_\text{m,ls} \, \vdH_2^\text{T} \frac{ \left( \vr_{1} - \vr_{2} \right)}{d^{m+2}} \, \dd s_2 \dd s_1.
\end{align}
As mentioned already in~\secref{sec::ia_pot_double_length_specific_evaluation_elstat}, the discrete element residual vectors in the specific case of Coulombic interactions follow directly for~$m=1$ and~$c_\text{m,ls} = C_\text{elstat} \lambda_1 \lambda_2$.
See \secref{sec::theory_electrostatics_pointcharges} for the definition of~$C_\text{elstat}$ and~\secref{sec::ia_pot_double_length_specific_evaluation_elstat} for the definition of the linear charge densities~$\lambda_i$.
Again, as mentioned in~\secref{sec::ia_pot_double_length_specific_evaluation_elstat}, the case of long-range \textit{volume} interactions only requires to adapt the constant prefactor via~$c_\text{m,ls}=k m A_1 A_2 \rho_1 \rho_2$.

\subsection{Objectivity and conservation properties}\label{sec::conservation_properties}
It can be shown that the proposed SSIP approach from~\secref{sec::method_pot_based_ia} in combination with the SSIP laws from~\secref{sec::method_application_to_specific_types_of_interactions} fulfills the essential mechanical properties of objectivity, global conservation of linear and angular momentum as well as global conservation of energy.
Due to the equivalent structure of the resulting space-discrete contributions, e.\,g., equation~\eqref{eq::var_discrete_iapot_small_sep}, as compared to the terms obtained in macroscopic beam contact formulations, we refer to the proof and detailed discussion of these important aspects in~\cite[Appendix B]{Meier2017a}.
The fulfillment of conservation properties will furthermore be verified by means of the numerical examples in~\secref{sec::example_contact_repusive_LJpot} and~\secref{sec::num_ex_twocrossedbeams_elstat_snapintocontact}.

\subsection{Regularization of SSIP laws in the limit of zero separation}\label{sec::regularization}
The singularity of inverse power laws for zero separation is a well-known pitfall when dealing with this kind of interaction laws.
See e.\,g.~\cite[p.137]{israel2011} for a discussion of this topic in the context of point-point LJ interaction as compared to a hard-sphere model.
In numerical methods, one therefore typically applies a regularization that cures the singularity and ensures the robustness of the method.
Sauer \cite{Sauer2011} gives an example for a regularized LJ force law between two half-spaces, where the force is linearly extrapolated below a certain separation, which is chosen as~$1.05$ times the equilibrium spacing of the two half spaces.
Also, existing, macroscopic beam contact formulations rely on the regularization of the seemingly instantaneous and infinite jump in the contact force when two macroscopic beams come into contact (see e.\,g.~\cite{Meier2017a,durville2012}).

However, the SSIP laws derived for disk-disk vdW or LJ interaction from~\secref{sec::ia_pot_double_length_specific_evaluation_vdW} have not yet been considered in literature.
Note that LJ is the most general and challenging case considered in this work, since strong adhesive forces compete with even stronger repulsive forces whenever two fibers are about to come into contact.
To be more precise, it is not only the strength of these competing forces, but also the high gradients in the force-distance relation that lead to a very stiff behavior of the governing partial differential equations.
This alone places high demands on the nonlinear solver, which in combination with the already mentioned singularity at zero separation~$g=0$, and the fact that LJ interaction laws are not defined for configurations~$g<0$ where both fibers penetrate each other, makes it extremely demanding to solve the problem numerically.

The results and conclusions discussed throughout this section are mainly based on the extensive numerical peeling and pull-off experiment with two adhesive fibers, which is presented in the authors' recent contribution~\cite{GrillPeelingPulloff}.
In absence of a regularization, only the pragmatic yet effective approach of applying a very restrictive upper bound of the displacement increment per nonlinear iteration (see~\ref{sec::algorithm_implementation_aspects} for details) proved successful to solve for the quasi-static equilibrium configurations without occurrence of any invalid configuration~$g\leq0$ for any integration point pair in any nonlinear iteration.
It must be emphasized that even a single occurrence of the latter is fatal and aborts the simulation, such that the mentioned approach is the only way to compute a solution for the full LJ interaction law, which can in turn serve as a reference solution during the validation of the regularization to be proposed and applied.
However, the mentioned approach severely deteriorates the convergence behavior and leads to a large number of nonlinear iterations per time step.
Thus, the regularization to be proposed in this section is superior in two respects: it guarantees the avoidance of singular/undefined values and saves a factor of five in the number of iterations of the nonlinear solver.

\begin{figure}[htp]%
  \centering
  \includegraphics[width=0.4\textwidth]{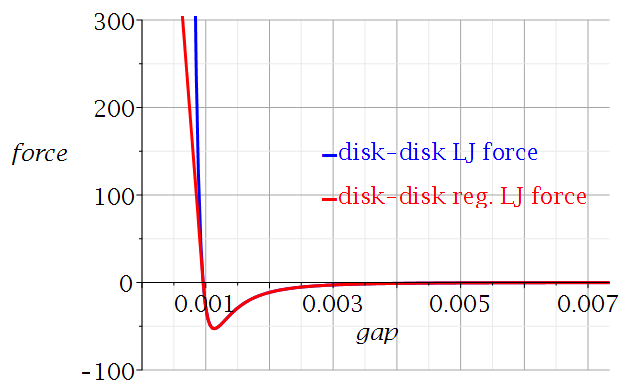}
  \caption[Comparison of regularized and full LJ disk-disk force law.]{Comparison of regularized (red) and full (blue) LJ disk-disk force law. Here, $g_\text{reg,LJ}=g_\text{LJ,eq,disk$\parallel$disk}$ is shown exemplarily\footnotemark.}
  \label{fig::force_disk-disk_LJ_vs_reg_LJ_linextpol}
\end{figure}

Specifically, we apply a linear extrapolation of the total LJ force law below a certain separation~$g_\text{reg,LJ}$ in a manner very similar to \cite{Sauer2011} with the only difference that it is applied to the length-specific disk-disk force law instead of the force law between two half spaces.
Linear extrapolation means that the original expression~$(m-7/2)\, c_\text{m,ss} \, g^{-m+\frac{5}{2}}$ in~\eqref{eq::res_ia_pot_smallsep_ele1} and \eqref{eq::res_ia_pot_smallsep_ele2} is replaced by a linear equation~$a\,g+b$ in the gap~$g$ for all~$g<g_\text{reg,LJ}$.
The two constants~$a$ and $b$ are determined from the requirements that the force value as well as the first derivative of the original and the linear expression are identical for the regularization separation~$g=g_\text{reg,LJ}$.
\figref{fig::force_disk-disk_LJ_vs_reg_LJ_linextpol} shows both the original (blue) and the regularized (red) LJ disk-disk force law as a function of the smallest surface separation~$g$.
\footnotetext{See eq.~\eqref{eq::equilibrium_spacing_LJ_disks_parallel_smallsep} for an analytical expression of~$g_\text{LJ,eq,disk$\parallel$disk}$.}

The numerical experiment of adhesive fibers studied in~\cite{GrillPeelingPulloff} reveals that this regularization yields the already mentioned great enhancement in terms of robustness as well as efficiency without any change in the system response.
As shown in the comparison of the force-displacement curves therein, the results obtained with the full LJ and with the regularized LJ force law do indeed coincide down to machine precision.
This is reasonable and expected, because we chose a regularization parameter~$g_\text{reg,LJ} \leq g_\text{LJ,eq,cyl$\parallel$cyl}$ that is smaller than any separation value~$g$ occurring anywhere in the system in any converged equilibrium state.
Thus, the solution never ``sees'' the modification to the vdW force law in the interval~$g<g_\text{reg,LJ}$ and the results are identical.
However, since during the nonlinear iterations also non-equilibrium configurations with~$g<g_\text{reg,LJ}$ occur, the nonlinear solution procedure is influenced in an extremely positive way, leading to an overall saving of a factor of five in the number of nonlinear iterations as compared to the full LJ interaction without any regularization.
For more details on the comparison including all parameter values we kindly refer the reader to~\cite{GrillPeelingPulloff}.

\subsection{Numerical evaluation of n-dimensional integrals of intermolecular potential laws}\label{sec::method_numerical_integration}
Generally, we use $n$ nested loops of a 1D Gauss-Legendre quadrature scheme which is the well-established and de-facto standard method in nonlinear finite element frameworks and has been used also in previous publications in the context of molecular interactions \cite{Argento1997,Sauer2013}.
Due to the strong nonlinearity, i.\,e., high gradients of the power laws, a large number of quadrature points is required in each dimension to achieve sufficient accuracy.
This effect is most critical for high exponents of the potential law, i.\,e., vdW and steric interactions, and small separations of the interacting bodies.
We thus implemented the possibility to subdivide the domain of a finite element into~$n_\text{IS}$ integration segments and apply an~$n_\text{GP}$-point Gauss rule on each of them in order to achieve sufficient density of quadrature points in every case.

\subsection{Algorithm complexity}\label{sec::algorithm_complexity}
Multi-dimensional numerical integration of the intermolecular potential laws as discussed above turns out to be the crucial factor in terms of efficiency.
For the following analysis of efficiency, we consider the associated algorithmic complexity.
Generally, all possible pairs of elements need to be evaluated, which has~$\bigO \left( n_\text{ele}^2 \right)$ complexity.
Let us assume we apply a total of~$n_\text{GP,tot,ele-length}$ integration points along the element length and~$n_\text{GP,tot,transverse}$ integration points in the transversal, i.\,e., cross-sectional in-plane directions.
Thus, the complexity of an approach based on full 6D numerical integration over the 3D volumes of the two interacting bodies (cf.~\eqref{eq::pot_fullvolint}) can be stated as
\begin{equation}
  \bigO \left( n_\text{ele}^2 \cdot n_\text{GP,tot,ele-length}^2 \cdot n_\text{GP,tot,transverse}^4 \right).
\end{equation}
In contrast to that, the novel SSIP approach proposed in~\secref{sec::method_double_length_specific_integral} reduces the dimensionality of numerical integration from six to two (cf.~\eqref{eq::ia_pot_double_integration}) and thus yields
\begin{equation}
  \bigO \left( n_\text{ele}^2 \cdot n_\text{GP,tot,ele-length}^2 \right)
\end{equation}
complexity.
The resulting difference between both clearly depends on the problem size, type of interaction and other factors.
To get an impression, typical numbers for the total number of quadrature points in transverse dimensions based on the numerical examples of~\secref{sec::numerical_results} are given as $n_\text{GP,tot,transverse} = 10 \ldots 100$.
The gain in efficiency thus easily exceeds a factor of~$10^4$ and can be as large as a factor of~$10^8$.
In addition to this tremendous saving from the inherent algorithmic complexity, the power law integrand has a smaller exponent due to the preliminary analytical integration in case of the SSIP approach.
This in turn allows for a smaller number of integration points~$n_\text{ele} \cdot n_\text{GP,tot,ele-length}$ for each of the two nested 1D integrations along the centerline, given the same level of accuracy.
To give an example, the vdW interaction force scales with an exponent of~$-7$ if formulated for two points (cf.~\eqref{eq::pot_ia_vdW_pointpair}) as compared to an exponent of~$-7/2$ for two circular cross-sections (cf.~\eqref{eq::var_iapot_small_sep} for~$m=6$).
This makes another significant difference, especially if very small separations - as typically observed for contacting bodies - are considered.
The combination of high dimensionality and strong nonlinearity of the integrand renders the direct approach of six-dimensional numerical quadrature to evaluate eq.~\eqref{eq::pot_fullvolint} unfeasible for basically any problem of practical relevance.
In fact, even a single evaluation of the vdW potential of two straight cylinders to serve as a reference solution turned out to be too computationally costly below some critical, small separation. See~\secref{sec::verif_approx} for details on this numerical example.
Note that although there might be more elaborate numerical quadrature schemes for these challenging integrands consisting of rational functions (see e.\,g.~\cite{Gautschi2001}), the basic problem and the conclusions drawn from this comparison of algorithmic complexities remain the same.

These cost estimates based on theoretical algorithm complexity and the experience from rather small academic examples considered in~\secref{sec::numerical_results} show that the SSIP approach indeed makes the difference between feasible and intractable computational problems.
This directly translates to the applicability to complex biopolymer as well as synthetic fibrous systems that we have in mind and thus significantly extends the range of (research) questions that are accessible by means of numerical simulation.

\subsection{Search algorithm and parallel computing}\label{sec::search_parallel_computing}
In order to find the relevant pairs of interaction partners, the same search algorithms as in the case of macroscopic contact (between beams or 3D solids) may be applied, however, the obvious difference lies in the search radius.
For contact algorithms, a very small search radius covering the immediate surrounding of a considered body is sufficient, whereas for molecular interactions the search radius depends on the type of interaction and must be at least as large as the so-called cut-off radius.
Only at separations beyond the cut-off radius, the energy contributions from a particular interaction are assumed to be small enough to neglect them.
Depending on the interaction potential and partners, the range and thus cut-off radius can be considerably large which underlines the importance of an efficient search algorithm.
In the scope of this work, a so-called bucket search strategy has been used, that divides the simulation domain uniformly into a number of cells or buckets and assigns all nodes and elements to these cells to later determine spatially proximate pairs of elements based on the content of neighboring cells.
This leads to an algorithmic complexity of~$\bigO (n_\text{ele})$ and the search thus turned out to be insignificant in terms of computational cost as compared to the evaluation of pair interactions as discussed in the preceding section.
See \cite{Wriggers2006} for an overview of search algorithms in the context of computational contact mechanics.

To speed up simulations of large systems, parallel computing is a well-established strategy of ever increasing importance.
Key to this concept is the ability to partition the problem such that an independent and thus simultaneous computation on several processors is enabled.
In our framework, this partitioning is based on the same bucket strategy that handles the search for interaction partners.
Regarding the evaluation of interaction forces, a pair (or set) of interacting beam elements is assigned to the processor which owns and thus already evaluates the internal and external force contribution of the involved elements.
At processor boundaries, i.\,e., if the two interacting elements are owned by different processors, one processor is chosen to evaluate the interaction forces and the required information such as the element state vector of the element owned by the other processor is communicated beforehand.
Upon successful evaluation of the element pair interaction, the resulting contribution to the element residual vector and stiffness matrix is again communicated for the element whose owning processor was not responsible for the pair evaluation.

\section{Numerical Examples}\label{sec::numerical_results}
The set of numerical examples studied in this section aims to verify the effectiveness, accuracy and robustness of the proposed SSIP approach and the corresponding SSIP laws as a computational model for either steric repulsion, electrostatic or vdW adhesion and also a combination of those.
Supplementary information on the code framework and the algorithms used for the simulations is provided in~\ref{sec::algorithm_implementation_aspects}.

\subsection{Verification of the simplified SSIP laws using the examples of two disks and two cylinders}\label{sec::verification_methods}
As a follow-up to the general discussion of using simplified SSIP laws in~\secref{sec::assumptions_simplifications} and the proposal of specific closed-form analytic expressions in~\secref{sec::ia_pot_double_length_specific_evaluation_vdW} and \secref{sec::ia_pot_double_length_specific_evaluation_elstat}, this section aims to analyze the accuracy in a quantitative manner.
The minimal examples of two disks or two cylinders are considered in order to allow for a clear and sound analysis of either the isolated SSIP laws or its use within the general SSIP approach to modeling beam-to-beam interactions, respectively.

\subsubsection{Verification for short-range volume interactions such as van der Waals and steric repulsion}\label{sec::verif_approx}
Throughout this section, we consider the example of vdW interaction, but analogous results are expected for steric interaction or any other short-range volume interaction.
Specifically, we will focus on the approximation quality of the proposed SSIP law from~\secref{sec::ia_pot_double_length_specific_evaluation_vdW}, which is based on the assumptions and resulting simplifications discussed in-depth in~\secref{sec::assumptions_simplifications}.
Recall that, beside the obviously most important surface-to-surface separation~$g$, the rotation of the cross-sections around the closest point~$\alpha$ (quantified by the angle enclosed by their tangent vectors) and potentially also the rotation components~$\beta_1,\beta_2$ (see \figref{fig::beam_to_beam_interaction_sketch_assumptions_short_range}) have been identified as relevant degrees of freedom, yet are neglected in the simplified SSIP law proposed in~\secref{sec::ia_pot_double_length_specific_evaluation_vdW}.
The influence of these factors, separation and rotation, on the approximation quality will thus be analyzed numerically in the following.

Recall also from the discussions in~\secref{sec::assumptions_simplifications} and~\ref{sec::ia_pot_double_length_specific_evaluation_vdW} that only the regime of small separations will be of practical relevance in the case of short-range interactions considered here.
However, we include the regime of large separations in the following analyses, mainly because it will be interesting to see the transition from small to large separations and confirm that potential values indeed drop by several orders of magnitude as compared to the regime of small separations.
Moreover, it is a question of theoretical interest and has been considered in literature on vdW interactions~\cite{langbein1972}.
This regime of large separations can be treated without any additional effort as described for the case of long-range interactions in~\secref{sec::ia_pot_double_length_specific_evaluation_elstat} (where this regime is the decisive one) if we take into account the corresponding remark on volume interactions.

As presented in \secref{sec::theory_molecular_interactions_twobody_vdW}, analytical solutions for the special cases of parallel and perpendicular cylinders, for the regime of small and large separations, respectively, can be found in the literature \cite{israel2011,parsegian2005} and thus serve as reference solutions in this section.
To the best of the authors' knowledge, no analytical reference solution has yet been reported for the intermediate regime in between the limits of large and small separations.
Another source for reference solutions is the full numerical integration of the point pair potential over the volume of the interacting bodies, however it is limited due to the tremendous computational cost.
Only a combination of both analytical and numerical reference solutions thus allows for a sound verification of the novel SSIP approach and the proposed SSIP laws.

In the following analyses, either the SSIP, i.\,e., the interaction potential per unit length squared~$\tilde{\tilde{\pi}}$ of a pair of circular cross-sections, the interaction potential per unit length~$\tilde{\pi}$ of a pair of parallel cylinders or the interaction potential~$\Pi$ of a pair of perpendicular cylinders will be plotted as a function of the dimensionless surface-to-surface separation $g/R$, respectively.
For simplicity, the radii of the beams are set to~$R_1=R_2=:R=1$.

\begin{figure}[htp]%
  \centering
  \hfill
  \subfigure[]{
    \includegraphics[width=0.49\textwidth]{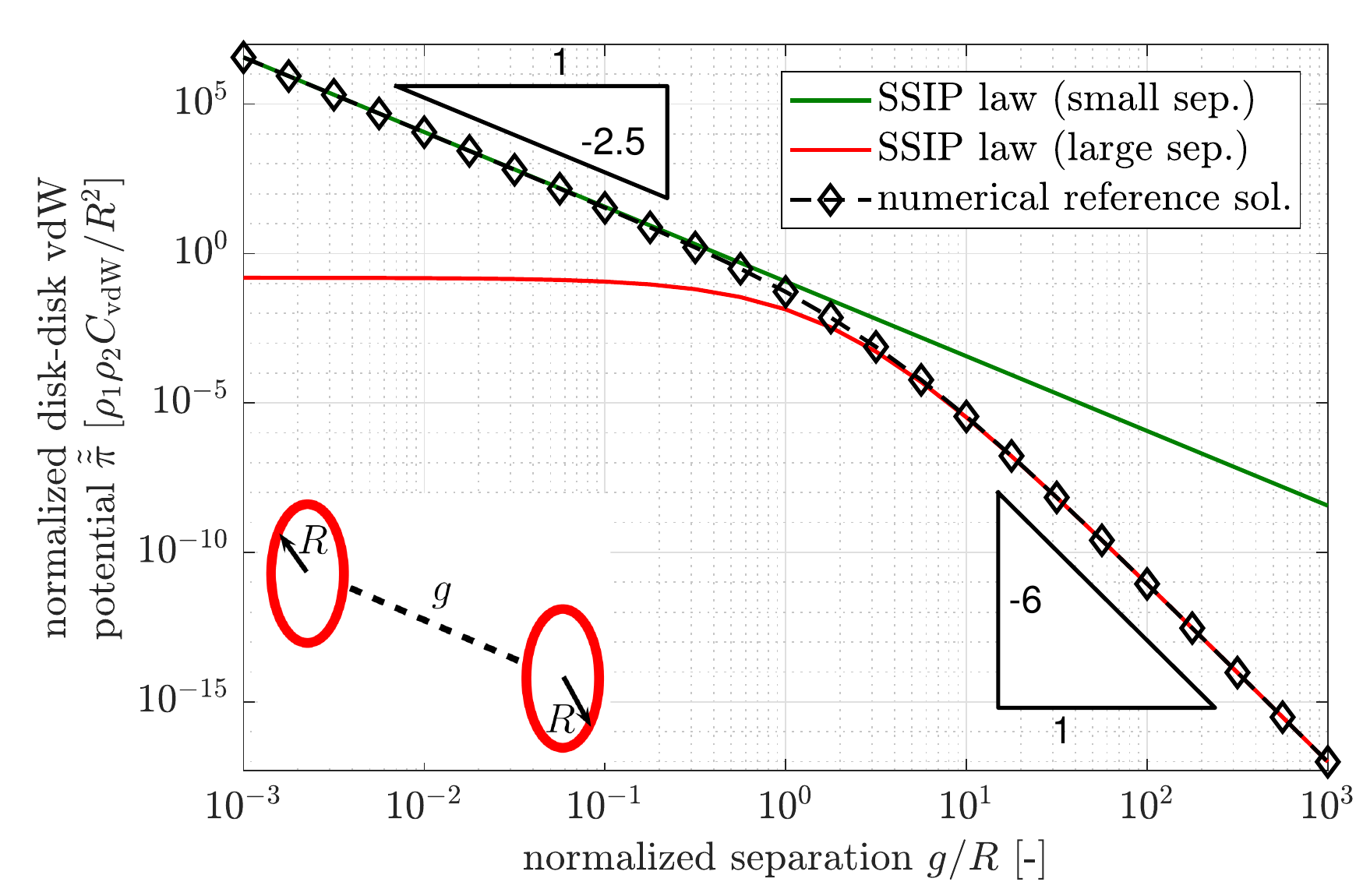}
    \label{fig::vdW_pot_over_gap_disks_parallel}
  }
  \hfill
  \subfigure[]{
    \includegraphics[width=0.15\textwidth]{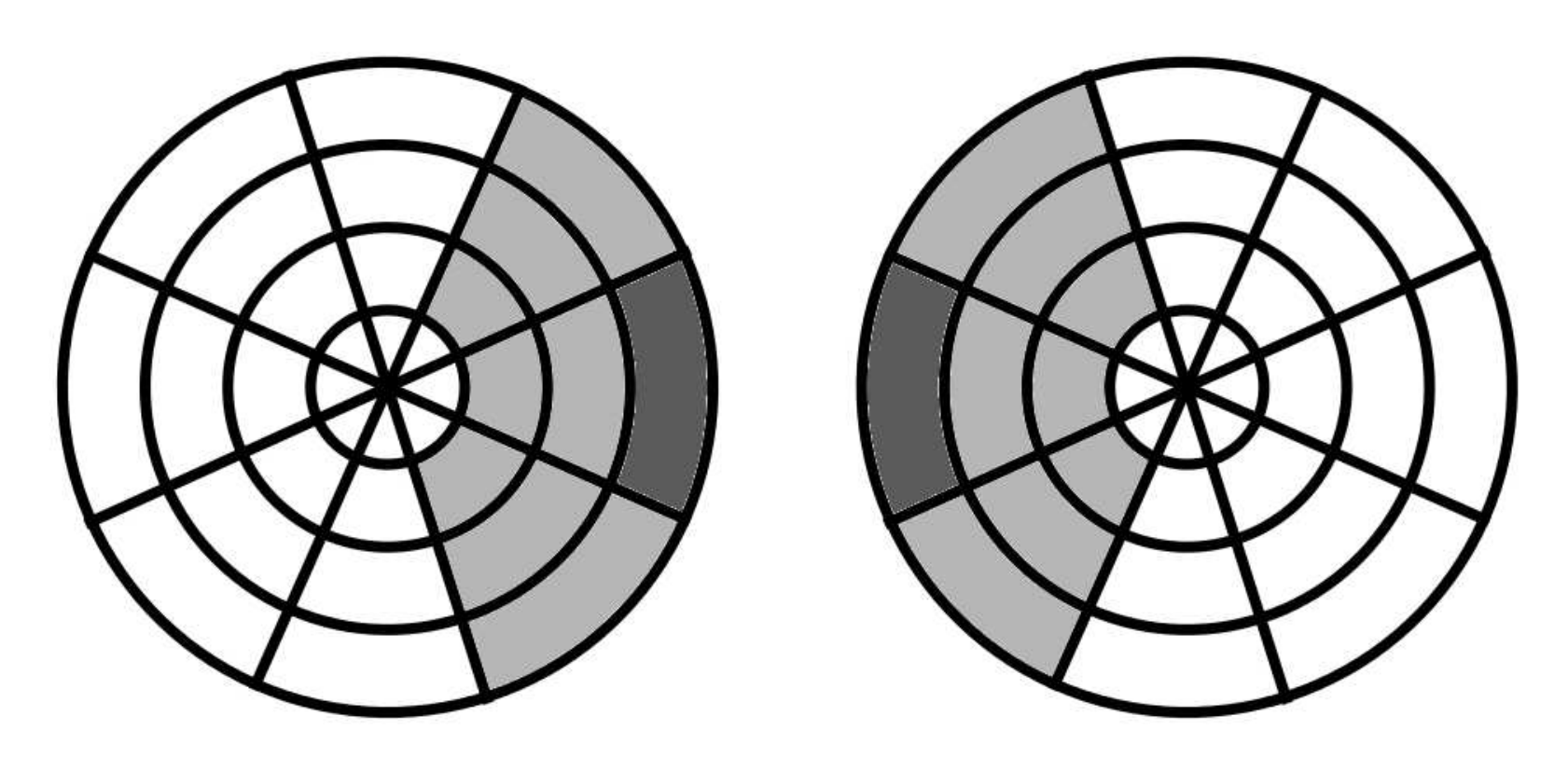}
    \label{fig::cross_sec_integration_sectors}
  }
  \hspace{2.5cm}
  \caption{(a) VdW interaction potential per unit length squared~$\tilde{\tilde{\pi}}$ of two disks in parallel orientation over normalized surface separation~$g/R$. The analytical expressions~\eqref{eq::pot_ia_vdW_disk_disk_parallel_smallseparation} (green line) and~\eqref{eq::pot_ia_vdW_disk_disk_parallel_largeseparation} (red line) used as SSIP laws throughout this work are verified by means of a numerical reference solution (black dashed line with diamonds). (b) Subdivision of circular cross-sections into integration sectors used to compute numerical reference solution. For rapidly decaying potentials, only the areas highlighted in dark and light gray considerably contribute to the total interaction potential.}
\end{figure}

\paragraph{Parallel disks and cylinders}$\;$\\
\figref{fig::vdW_pot_over_gap_disks_parallel} shows the SSIP~$\tilde{\tilde{\pi}}$ of two disks in parallel orientation, i.\,e., their normal vectors are parallel with mutual angle~$\alpha=0$, as a function of the normalized separation~$g/R$.
This is the simplest geometrical configuration and forms the basis of the proposed SSIP laws from~\secref{sec::method_application_to_specific_types_of_interactions}.
We thus begin our analysis with the verification of the used analytical solutions in the limit of small (green line, cf.~eq.~\eqref{eq::pot_ia_vdW_disk_disk_parallel_smallseparation}) and large (red line, cf.~eq.~\eqref{eq::pot_ia_vdW_disk_disk_parallel_largeseparation}) separations by means of a numerical reference solution (black dashed line with diamonds) obtained from 4D numerical integration of the point pair potential law~\eqref{eq::pot_ia_vdW_pointpair}.

\figref{fig::vdW_pot_over_gap_disks_parallel} confirms that both analytical expressions match the numerical reference solution perfectly well in the limit of large and small separations, respectively.
As predicted, the interaction potential of two circular disks follows a power law with (negative) exponent $2.5$ for small and $6$ for large separations%
\footnote{Note that in the double logarithmic plot, a power law with exponent $m$ is a linear function with slope $m$.}.
Note that all plots in this section are normalized with respect to the length scale $R$ and the energy scale $\rho_1\rho_2 C_\text{vdW}$.
It is remarkable that the obtained values span several orders of magnitude which illustrates the numerical challenges associated with power laws, especially in the context of numerical integration schemes as discussed already in \secref{sec::method_numerical_integration}.
Moreover, it underlines that the regime of large separations is practically irrelevant in the case of short-range interactions, because the potential values are basically zero as compared to those obtained in the small separation regime.

Regarding the full range of separations, one may ask where either of the two expressions may be used given a maximal tolerable error threshold.
As can be concluded from \figref{fig::vdW_pot_over_gap_disks_parallel}, the resulting error is small for separations $g/R<0.1$ with a relative error below $8\%$ and $g/R>10$ with a relative error below $7\%$.
In the region of intermediate separations, the analytical solution for small separations seems to yield an upper bound, whereas the one for large separations seems to yield a lower bound for the interaction potential.

Let us have a look at the efficiency gain from using the analytical solutions.
The numerical reference solution requires the evaluation of a 4D integral over both cross-sectional areas for a given separation~$g$ and has been carried out in polar coordinates.
Assuming Gaussian quadrature with the same number of Gauss points~$n_\text{GP,tot,transverse}$ in radial and circumferential dimension and for both cross-sections, this requires a total of $(n_\text{GP,tot,transverse})^4$ function evaluations.
In contrast, the analytical expressions for the large and small separation limit, respectively, require only one function evaluation.
This significant gain in efficiency is most pronounced for small separations, where the number of required Gauss points increases drastically due to the high gradient of the power law that needs to be resolved (see \secref{sec::method_numerical_integration} for details).
If the number of Gauss points is not sufficient, this leads to so-called underintegration and we observed that the obtained curve of the numerical reference solution erroneously flattens (because the contribution of the closest-point pair is not captured) or becomes steeper (because the contribution of the closest-point pair is overrated).

For these reasons, the computation of an accurate numerical reference solution shown in \figref{fig::vdW_pot_over_gap_disks_parallel} requires quite some effort.
The integration domains were subdivided into integration sectors (see \figref{fig::cross_sec_integration_sectors}) in order to further increase the Gauss point density.
But even in this planar disk-disk scenario requiring only 4D integration, we reached a minimal separation of~$g/R \approx \num{5e-3}$, below which the affordable number of Gauss points was not sufficient to correctly evaluate the SSIP~$\tilde{\tilde{\pi}}$ by means of full numerical integration%
\footnote{The maximum number of~$n_\text{GP,tot,transverse}=8\times 32=256$ considered in the scope of this work led to several hours of computation time on a desktop PC for the evaluation of~$\tilde{\tilde{\pi}}$ as a numerical reference solution for~\figref{fig::vdW_pot_over_gap_disks_parallel}.}.
For these very small separations, only the exact analytical dimensional reduction from 4D to 2D according to Langbein (cf.~\cite{langbein1972} and eq. \eqref{eq::dimred_langbein}) allowed to compute an accurate numerical reference solution.
The analytical solutions for the disk-disk interaction potential~\eqref{eq::pot_ia_vdW_disk_disk_parallel_smallseparation} (and~\eqref{eq::pot_ia_vdW_disk_disk_parallel_largeseparation}), used as SSIP law in eq.~\eqref{eq::var_iapot_small_sep} (and~\eqref{eq::var_iapot_large_sep_surface}), thus realize a significant increase in efficiency and indeed only enable the accurate evaluation of the interaction potential in the regime of very small separations.
Note that such small separations are highly relevant if we consider fibers in contact since surface separations are expected to lie on atomic length scale in this case.
For instance, the work of Argento et al.~\cite{Argento1997} mentions~$g=\SI{0.2}{\nano\meter}$ to be a typical value for contacting solid bodies and states that accurate numerical integration thus is the most challenging and in fact limiting factor for the numerical methods based on inter-surface potentials.
As a reference value for the applications we have in mind, the fiber radius~$R$ varies from several~$\SI{}{\nano\meter}$ for DNA to $\SI{}{\milli\meter}$ for synthetic polymer fibers, resulting in a potentially very small normalized separation~$g/R$.
An example for the simulation of adhesive fibers in contact can be found in the authors' recent contribution~\cite{GrillPeelingPulloff}, which studies the peeling and pull-off behavior of two fibers attracting each other either via vdW or electrostatic forces.

As a next step, the interaction potential per unit length~$\tilde{\pi}$ of two parallel straight beams is considered.
The length of the beams is chosen sufficiently high such that it has no perceptible influence on the results and meets the assumption of infinitely long cylinders made to derive the analytical reference solution from \cite{langbein1972}.
Accordingly, a slenderness ratio $\zeta=l/R=50$ is used in the regime of small separations, whereas $\zeta=l/R=1000$ is used for large separations.
\begin{figure}[htp]%
  \centering
    \includegraphics[width=0.49\textwidth]{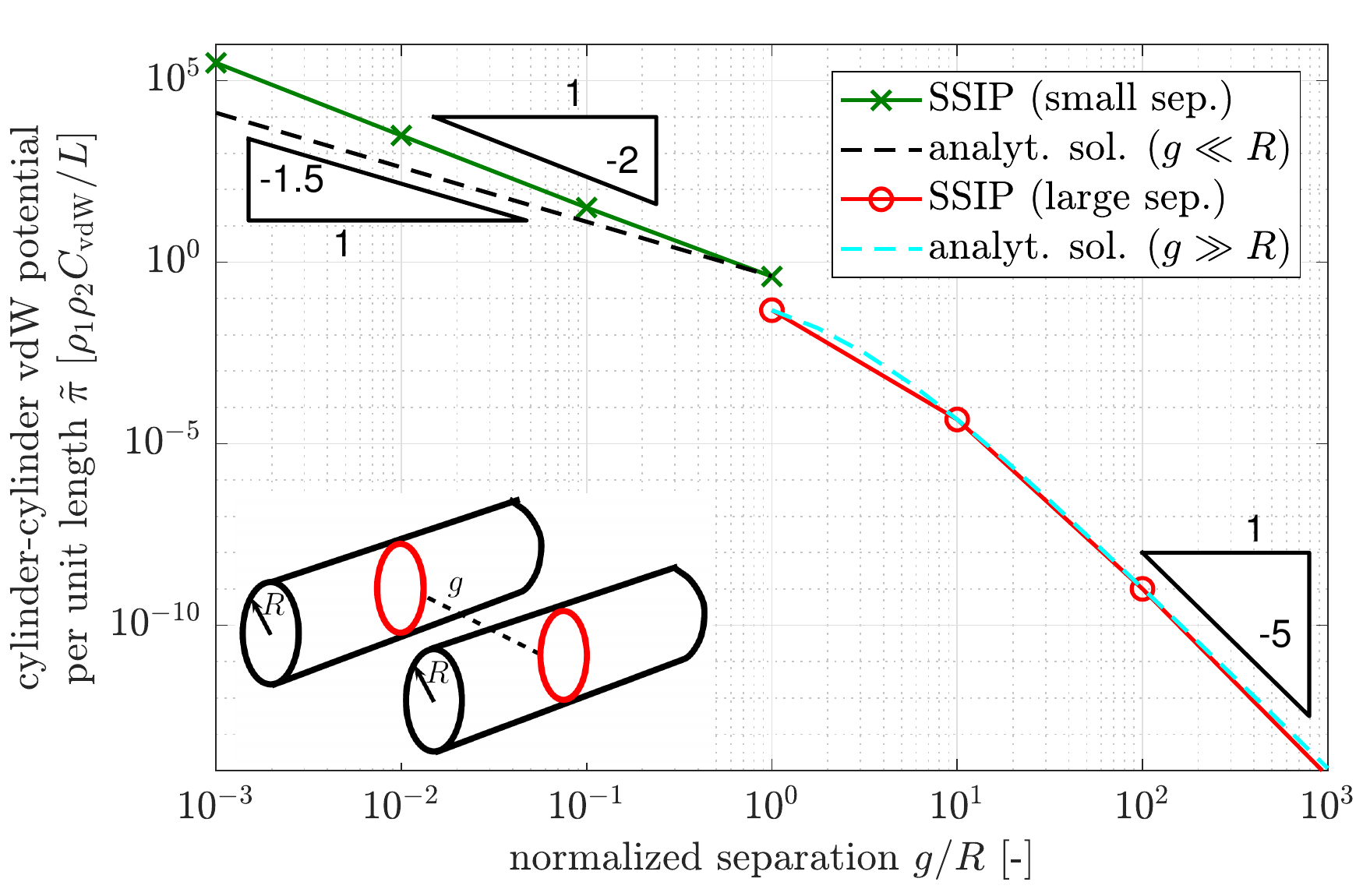}
  \caption{VdW interaction potential per unit length~$\tilde{\pi}$ of two parallel cylinders over normalized surface separation~$g/R$.}
  \label{fig::vdW_pot_perunitlength_over_gap_cylinders_parallel}
\end{figure}%
Based on the experience from the disk-disk scenario, it is not surprising that the full 6D numerical integration in this case exceeds the affordable computational resources by orders of magnitude and thus can not serve as a reliable reference solution.
In fact, we were not able to reproduce the theoretically predicted power law scaling in the regime of small separations despite using a number of Gauss points that led to computation times of several days.
However, instead of the numerical reference solution, the analytical solution for infinitely long cylinders in the limit of very small (black dashed line, cf.~eq.~\eqref{eq::pot_ia_vdW_cyl_cyl_parallel_smallseparation}) and very large separations (blue dashed line, cf.~eq.~\eqref{eq::pot_ia_vdW_cyl_cyl_parallel_largeseparation}) serves as a reference in~\figref{fig::vdW_pot_perunitlength_over_gap_cylinders_parallel}.
Note that as compared to the case of two circular disks the exponent of the power laws and thus the slope of the curves drops by one due to the integration over both cylinders' length dimension.

Interestingly, the SSIP approach using the simplified SSIP law from~\secref{sec::ia_pot_double_length_specific_evaluation_vdW} (green line with crosses) does not yield the correct scaling behavior even in this case of parallel cylinders.
This confirms the concerns from~\secref{sec::assumptions_simplifications} that the simplified SSIP law neglecting any relative rotations of the cross-sections deteriorates the accuracy of the approach in the case of short-ranged interactions in the regime of small separations.
Due to this specific scenario of parallel cylinders, this deterioration can be attributed solely to the rotation components~$\beta_{1/2}$ (see~\figref{fig::beam_to_beam_interaction_sketch_assumptions_short_range}) since the included angle of the cross-section normal vectors~$\alpha$ is zero for every of the infinitely many pairs of cross-sections.
Despite the correct trend of the resulting interaction potential as an inverse power law of the surface separation, it must thus be stated that the simplified SSIP law overestimates the strength of interaction and that the error increases with decreasing separation
\footnote{
Note that the numerical integration error has been ruled out as cause for this behavior by choosing a high number of Gauss points~$n_\text{GP,tot,ele-length}=2\times50=100$ for each of the $64$ elements used to discretize each cylinder.
A further increase of~$n_\text{GP}$ by a factor of five does not change the results using double precision.
}.
In the regime of large separations, however, the results for the SSIP approach (red line with circles) perfectly match the analytical reference solution (blue dashed line).
This confirms the hypothesis from~\secref{sec::assumptions_simplifications} that the relative rotation of cross-sections is negligible in this regime and a high accuracy can be achieved with the simplified SSIP law.
Although being of little practical importance here due to the negligible absolute values, this is a first numerical evidence for the validity of the SSIP approach in general and its high accuracy even in combination with simplified SSIP laws in the particular case of long-range interactions to be considered in the following~\secref{sec::verif_SSIP_disks_cyls_elstat}.

\paragraph{Perpendicular disks and cylinders}$\;$\\
Up to now, we have only discussed the situation of parallel orientation of disks and cylinders.
In the following, the accuracy of the simplified SSIP laws as well as the SSIP approach for twisted configurations will be analyzed by considering the most extreme configuration of perpendicular disks and cylinders.
Again, computing a reference solution by means of full numerical integration was only affordable for the 4D case of two disks.
\begin{figure}[htpb]%
  \centering
  \hspace{-0.3cm}
  \subfigure[]{
    \includegraphics[width=0.49\textwidth]{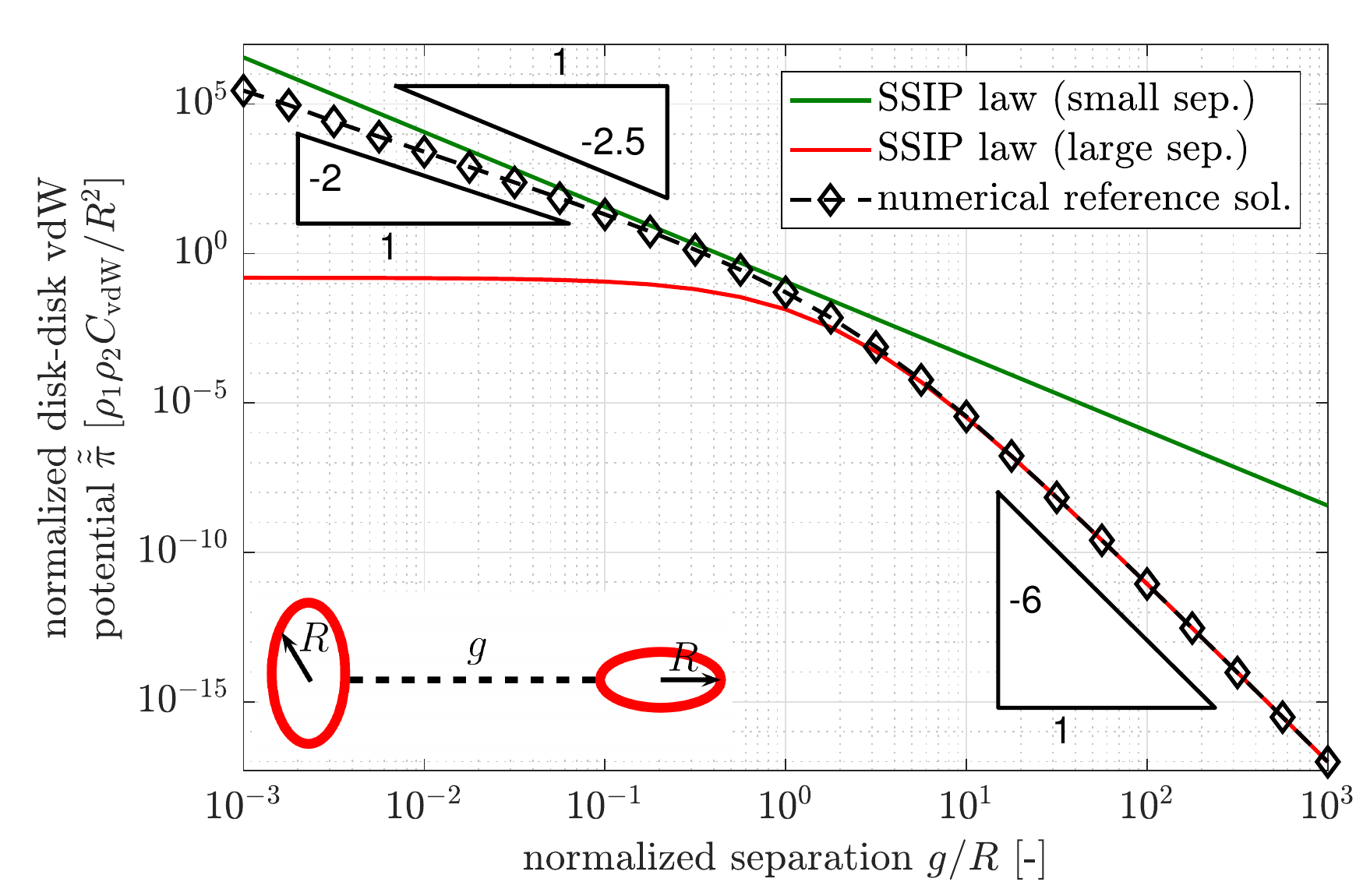}
    \label{fig::vdW_pot_over_gap_disks_perpendicular}
  }
  \hspace{-0.2cm}
  \subfigure[]{
    \includegraphics[width=0.49\textwidth]{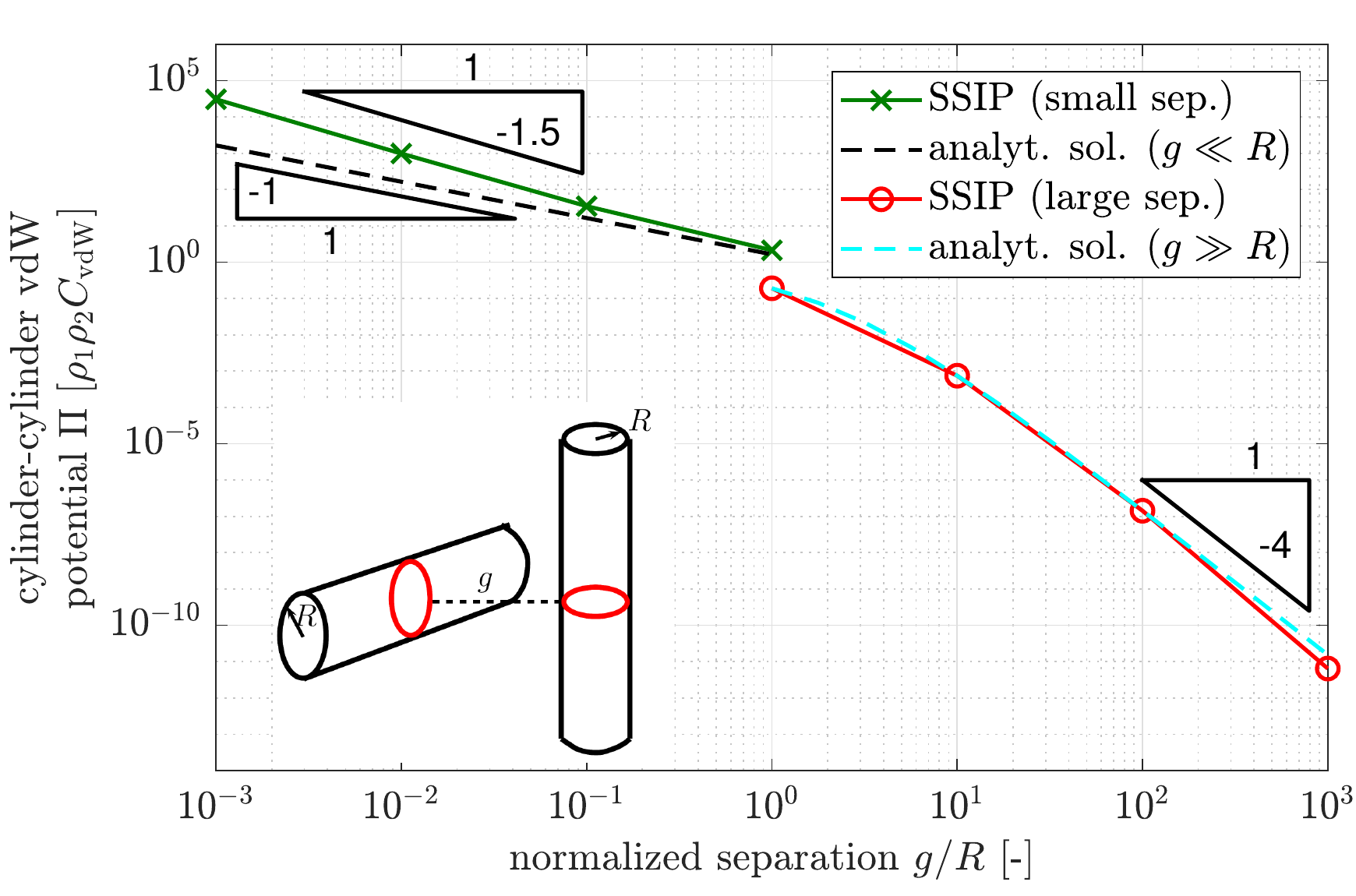}
    \label{fig::vdW_pot_over_gap_cylinders_perpendicular}
  }
  \caption{(a) VdW interaction potential per unit length squared~$\tilde{\tilde{\pi}}$ of two perpendicular disks and (b) vdW interaction potential~$\Pi$ of two perpendicular cylinders, plotted over the normalized surface separation~$g/R$, respectively.}
\end{figure}%
The results for perpendicular disks shown in \figref{fig::vdW_pot_over_gap_disks_perpendicular} confirm that there is no difference between perpendicular and parallel orientation for large separations and the scaling behavior of the numerical reference solution (black dashed line with diamonds) with exponent~$6$ is met by the simplified SSIP law (red line, cf.~eq.~\eqref{eq::pot_ia_vdW_disk_disk_parallel_largeseparation}).
On the other hand, there is a remarkable difference in the scaling behavior for small separations.
While the interaction potential of two parallel disks, which is the underlying assumption of the proposed SSIP law (green line, cf.~eq.~\eqref{eq::pot_ia_vdW_disk_disk_parallel_smallseparation}), follows an inverse~$2.5$ power law, the numerical reference solution (black dashed line with diamonds) suggests that this behavior changes for perpendicular disks to an inverse~$2$ power law.
This time, the difference in results can be attributed to the relative rotation~$\alpha$, i.\,e., the angle included by the cross-section normal vectors and again the error of the proposed simplified SSIP law increases with decreasing separation.

Finally, the scenario of perpendicular cylinders is considered and \figref{fig::vdW_pot_over_gap_cylinders_perpendicular} shows the total interaction potential~$\Pi$ as a function of the normalized smallest surface separation~$g/R$.
As discussed before, the computational cost of the full 6D numerical integration is too high to compute a reliable reference solution in the case of two 3D bodies and we resort to the analytical solutions for the limits of very small and very large separations, respectively.
Note that in contrast to the case of infinitely long \textit{parallel} cylinders (cf.~\figref{fig::vdW_pot_perunitlength_over_gap_cylinders_parallel}) the total interaction potential of infinitely long perpendicular cylinders is finite and the result thus has dimensions of energy instead of energy per length.
Perpendicular cylinders are worth to consider because they trigger both of the sources of errors that have been analyzed individually so far - neglecting relative rotations~$\alpha$ as well as~$\beta_{1/2}$ in the simplified SSIP law.
In short, the resulting accuracy is similar as for either perpendicular disks or parallel cylinders.
In the decisive regime of small separations, the SSIP approach based on the simplified SSIP law (green line with crosses) fails to reproduce the correct scaling behavior of the analytical reference solution (black dashed line, cf.~eq.~\eqref{eq::pot_ia_vdW_cyl_cyl_perpendicular_smallseparation}), whereas in the regime of large separations, the SSIP approach based on the simplified SSIP law (red line with circles) perfectly matches the analytical reference solution (blue dashed line, cf.~eq.~\eqref{eq::pot_ia_vdW_cyl_cyl_perpendicular_largeseparation}).

\paragraph{Conclusions}$\;$\\
First, this section reveals that full 6D numerical integration to compute the total interaction potential of slender continua is by orders of magnitude too expensive and can not reasonably be used as a numerical reference solution even in minimal examples of one pair of cylinders.
At most, 4D numerical integration required for disk-disk interactions allows to compute numerical reference solutions for the intermediate regime of separations where no analytical solutions are known.
This underlines the importance of reducing the dimensionality of numerical integration to 2D as achieved by the proposed SSIP approach in order to enable the simulation of large systems as well as a large number of time steps.

Second, the thorough analysis of the accuracy resulting from using the proposed simplified SSIP law, neglecting the cross-section rotations, reveals that one has to distinguish between the regime of small and large separations.
In the decisive regime of small separations, we find that the scaling behavior deviates from the analytical prediction for perpendicular disks and parallel as well as perpendicular cylinders and that the resulting error increases with decreasing separation.
A remedy of this limitation could be a calibration, i.e. a scaling of the prefactor k in the simple SSIP law, to fit a given reference solution within a small range of separations (e.g. around the equilibrium distance of the LJ potential).
In the authors' recent contribution~\cite{GrillPeelingPulloff}, this pragmatic procedure is shown to reproduce the global system response very well.
Still, it would be valuable to include the relative rotations of the cross-sections in the applied SSIP law to obtain the correct asymptotic scaling behavior.
To the best of the authors' knowledge, no analytical closed-form expression has been published yet and the like is far from trivial to derive.
Therefore, we leave this as a promising enhancement of the novel approach, which we are currently working on and will address in a future publication.
As mentioned before, the regime of large separations is of little practical relevance in the case of short-range volume interactions, however it is of some theoretical interest and the corresponding findings and conclusions of this regime will hold true also for long-range interactions such as electrostatics to be considered in the following section.
Here, the results are in excellent agreement with the theoretically predicted power laws for parallel as well as perpendicular disks and cylinders.

\subsubsection{Verification for long-range surface interactions such as electrostatics}\label{sec::verif_SSIP_disks_cyls_elstat}
Turning to long-range interactions, again parallel and perpendicular disks and cylinders will be considered in order to analyze the accuracy of the simplified SSIP law from~\secref{sec::ia_pot_double_length_specific_evaluation_elstat} both individually as well as applied within the general SSIP approach proposed in~\secref{sec::method_double_length_specific_integral}.
As before, Coulombic surface interactions are chosen as specific example, however the conclusions are expected to hold true also for other types of long-range interactions.
As compared to the preceding section, the computation of a numerical reference solution simplifies mainly due to the reduction from volume to surface interactions but also due to the smaller gradient values, which need to be resolved in the regime of small separations thus requiring less integration points.
This allows for a verification by means of a numerical reference solution also in the case of cylinder-cylinder interaction.

\begin{figure}[htpb]%
  \centering
  \hspace{-0.3cm}
  \subfigure[]{
    \includegraphics[width=0.49\textwidth]{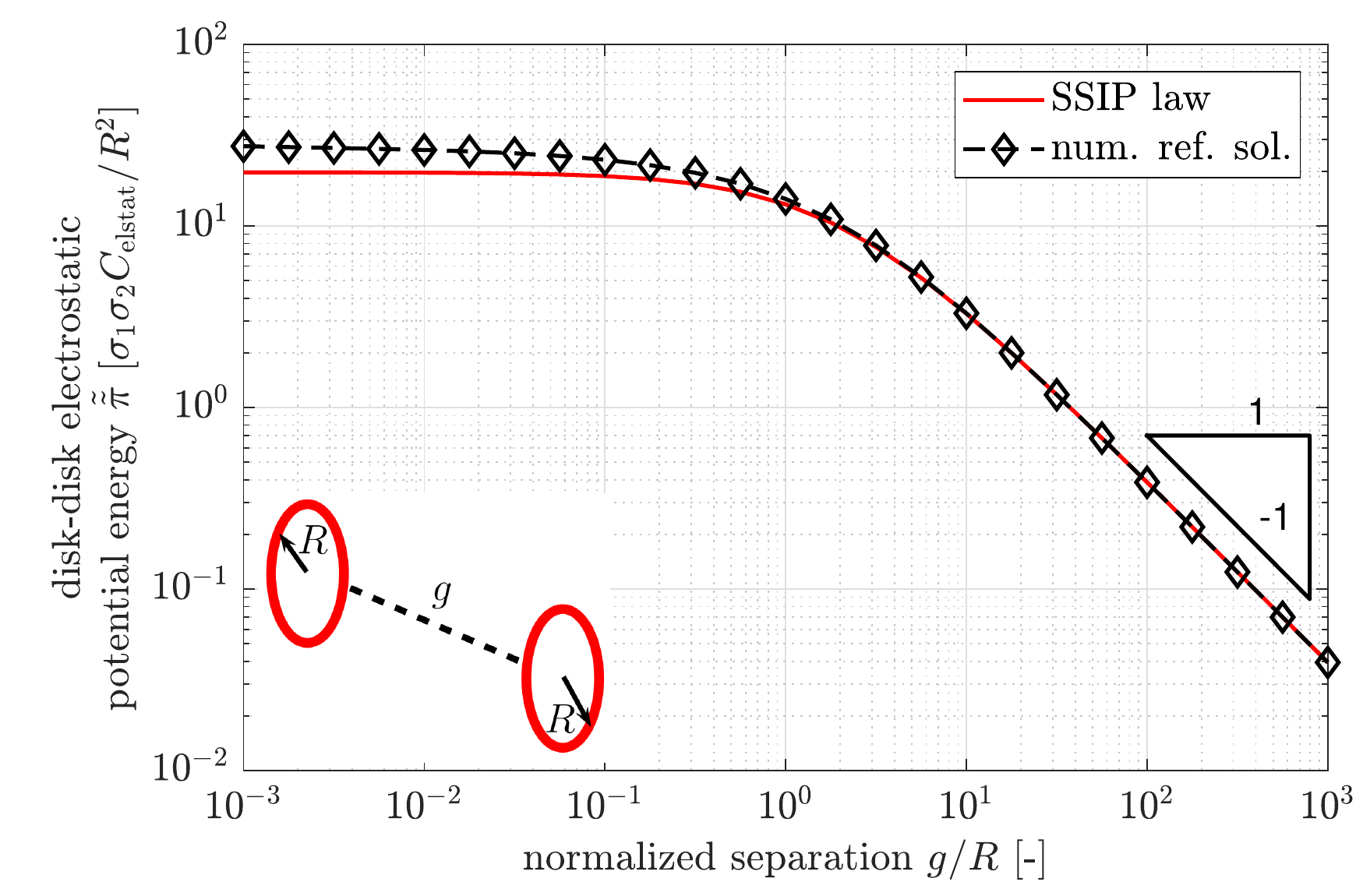}
    \label{fig::elstat_pot_over_gap_disks_parallel}
  }
  \hspace{-0.2cm}
  \subfigure[]{
    \includegraphics[width=0.49\textwidth]{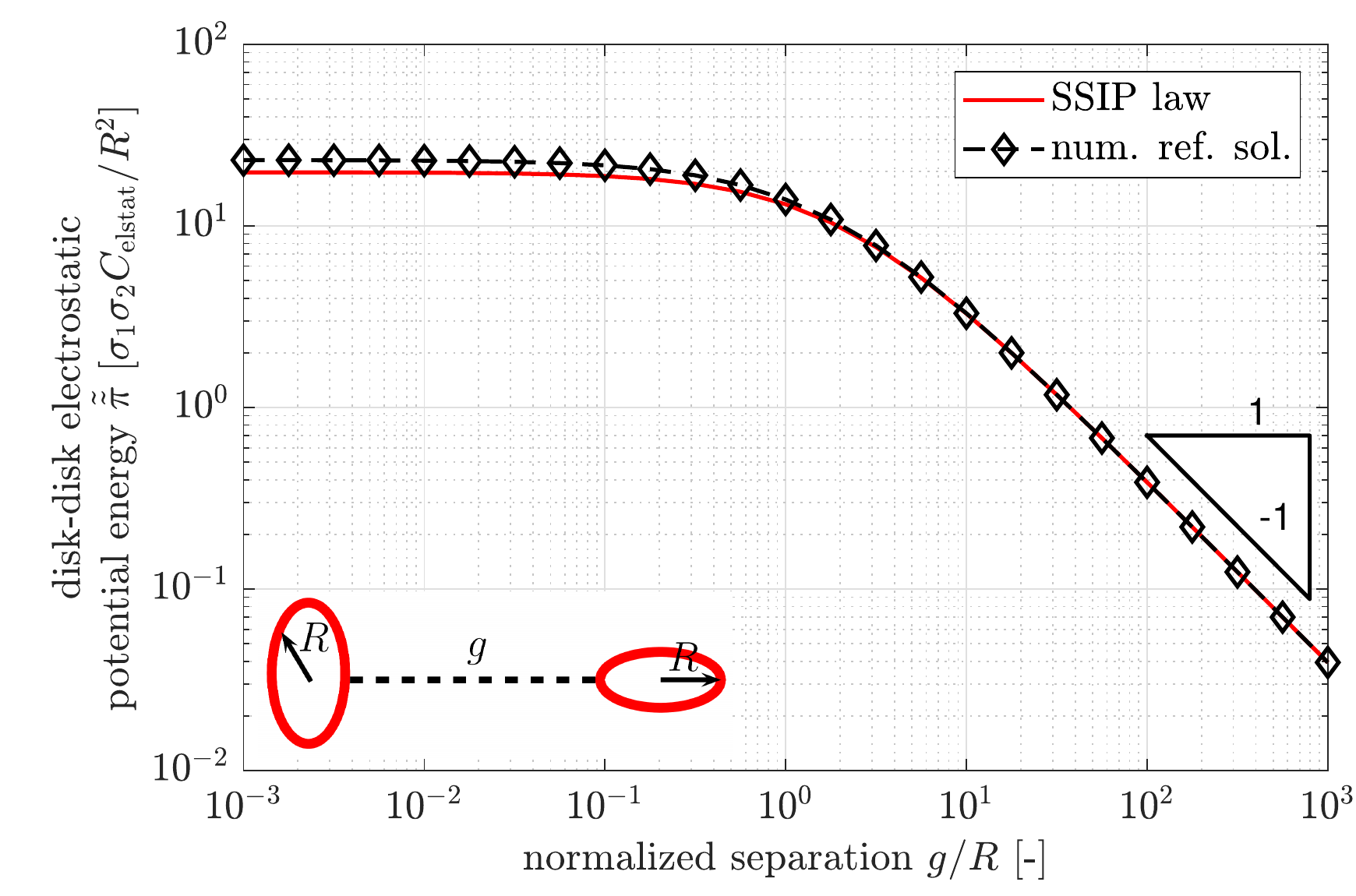}
    \label{fig::elstat_pot_over_gap_disks_perpendicular}
  }
  \caption{Electrostatic interaction potential per length squared~$\tilde{\tilde{\pi}}$ of (a) two parallel disks and (b) two perpendicular disks, plotted over the normalized surface separation~$g/R$, respectively.}
  \label{fig::elstat_pot_over_gap_disks}
\end{figure}%

\begin{figure}[htpb]%
  \centering
  \hspace{-0.3cm}
  \subfigure[]{
    \includegraphics[width=0.49\textwidth]{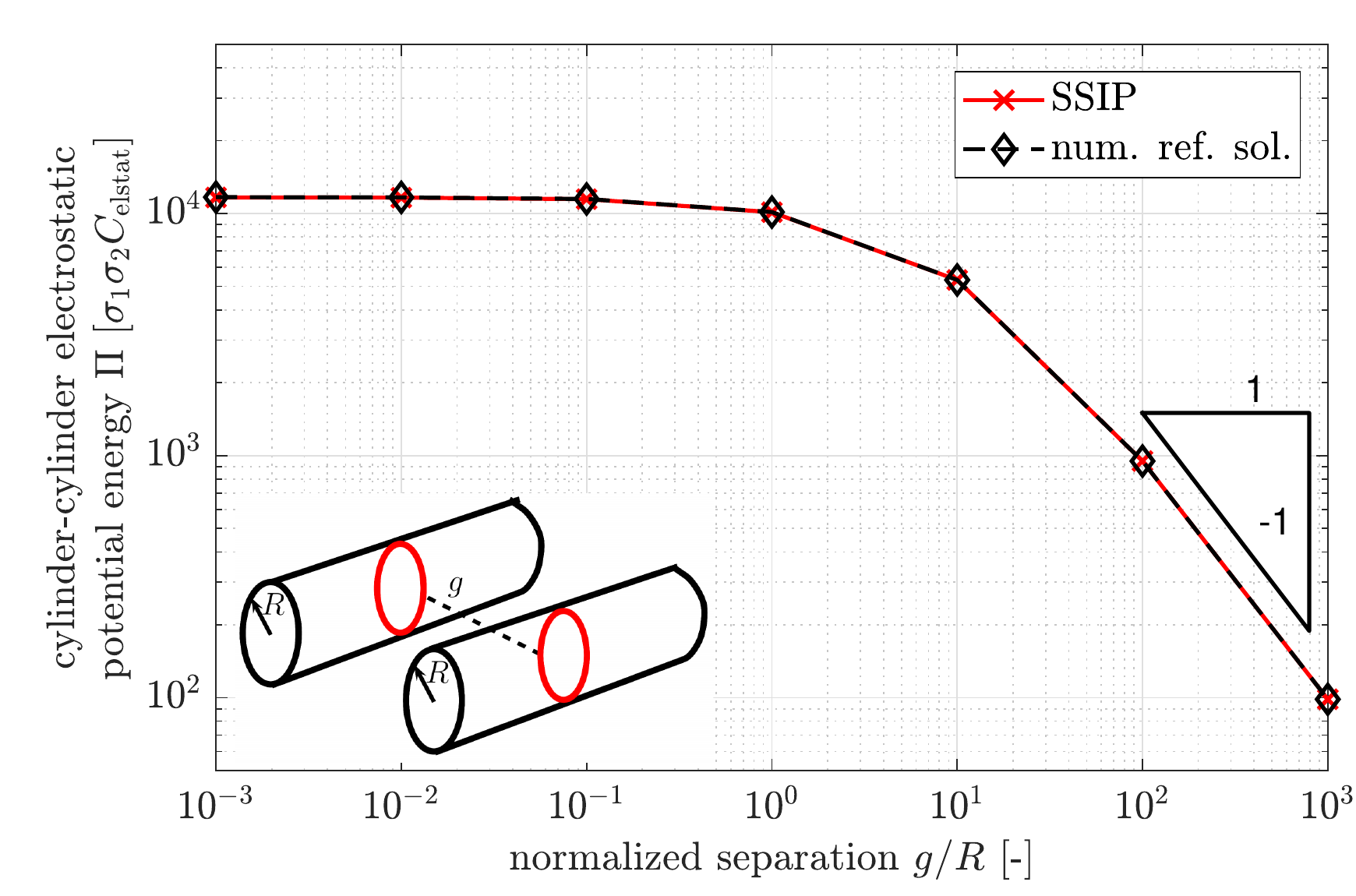}
    \label{fig::elstat_pot_over_gap_cylinders_parallel}
  }
  \hspace{-0.2cm}
  \subfigure[]{
    \includegraphics[width=0.49\textwidth]{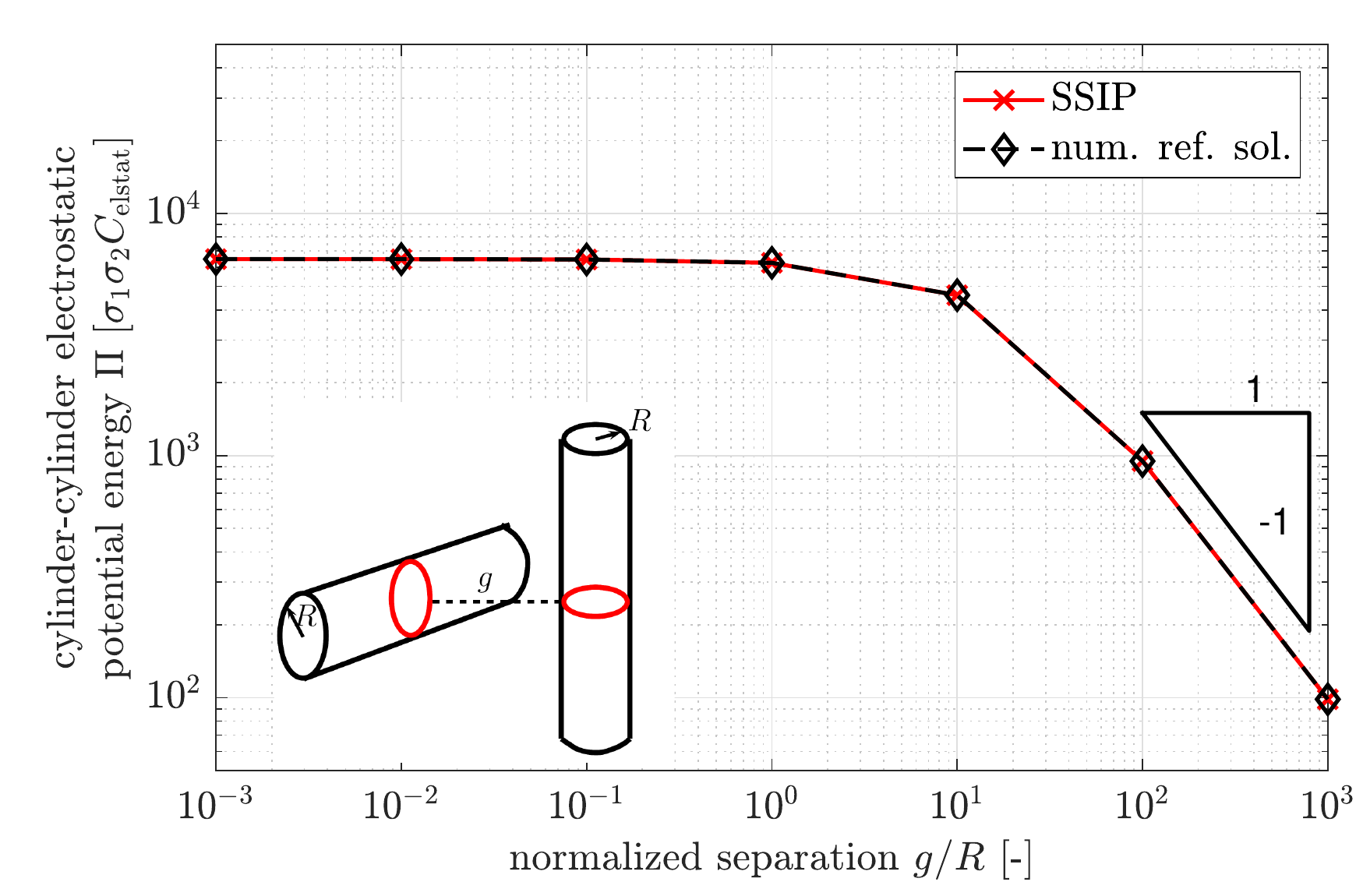}
    \label{fig::elstat_pot_over_gap_cylinders_perpendicular}
  }
  \caption{Electrostatic interaction potential~$\Pi$ of (a) two parallel cylinders and (b) two perpendicular cylinders, plotted over the normalized surface separation~$g/R$, respectively. The slenderness ratio of the cylinders is $\zeta=L/R=50$.}
  \label{fig::elstat_pot_over_gap_cylinders}
\end{figure}

\figref{fig::elstat_pot_over_gap_disks} shows the results for the simplified SSIP law obtained from the monopole-monopole interaction of two disk-shaped cross-sections in \secref{sec::ia_pot_double_length_specific_evaluation_elstat} (red line) and a numerical reference solution (black dashed line with diamonds).
As expected, the proposed SSIP law excellently matches the reference solution in the regime of large separations, both for the parallel as well as perpendicular configuration.
In both cases, the relative error is below $7\%$ already for $g/R=1$.
The most important and remarkable result of this section however is the following.
The inevitable error of the simplified SSIP law in the regime of small separations does not carry over to beam-to-beam interactions as shown in~\figref{fig::elstat_pot_over_gap_cylinders}.
For both parallel as well as perpendicular cylinders, the results from the SSIP approach using this simplified SSIP law from~\secref{sec::ia_pot_double_length_specific_evaluation_elstat} (red line with crosses) agree very well with the numerical reference solution (black dashed line with diamonds) over the entire range of separations.
This confirms the theoretical considerations from~\secref{sec::assumptions_simplifications} arguing that the beam-beam interaction will be dominated by the large number of section pairs with large separations, which outweigh the contributions of the few section pairs with smallest separations.

A closer look reveals that the relative error for the parallel cylinders is below $0.3\%$ even for the smallest separation~$g/R=10^{-3}$ considered here.
For the presumably worst case of perpendicular cylinders, this deviation is even smaller with a relative error of~$0.03\%$, which can be explained by the following two reasons.
First, a comparison of~\figref{fig::elstat_pot_over_gap_disks_parallel} and \ref{fig::elstat_pot_over_gap_disks_perpendicular} reveals that the accuracy of the simplified SSIP law in the regime of small separations is higher for perpendicular orientation, which can be regarded a happy coincidence.
And second, the large majority of all section pairs has a larger separation (which according to~\figref{fig::elstat_pot_over_gap_disks} is the regime of higher accuracy) as compared to the case of parallel cylinders.

Note that unlike in the case of short-range interactions, here the total interaction potential is considered also for the parallel cylinders.
Due to the long range of interactions, the interaction potential per length depends on the length of the cylinders and is thus no representative quantity.
For~\figref{fig::elstat_pot_over_gap_cylinders}, a slenderness ratio of~$\zeta=L/R=50$ is chosen exemplarily.
The two nested 1D integrals along the cylinder length dimensions are evaluated using $n_\text{GP,ele-length}=5$ Gauss points for each of the 64 elements used to discretize each cylinder.
Additionally, $n_\text{GP,circ}=8\times32=256$ Gauss points over the circumference of each disk are used to compute the numerical reference solution.
In all cases, it has been verified that the numerical integration error does not influence the results noticeably.

At this point, recall from~\secref{sec::ia_pot_double_length_specific_evaluation_elstat} that the accuracy of the applied SSIP law can still be increased whenever deemed necessary by including more terms of the multipole expansion of the cross-sections.
However, because the results of this section show a high level of accuracy and the resulting simplification is significant, the simplified SSIP law seems to be the best compromise for our purposes.
To conclude this section it can thus be stated that the novel SSIP approach as proposed in~\secref{sec::method_double_length_specific_integral} in combination with the simplified SSIP law from~\secref{sec::ia_pot_double_length_specific_evaluation_elstat} is a simple, efficient, and accurate computational model for long-range interactions of slender fibers.
In the following, it will be applied to first numerical examples of \textit{deformable} slender fibers in~\secref{sec::num_ex_elstat_attraction_twoparallelbeams} and~\ref{sec::num_ex_twocrossedbeams_elstat_snapintocontact}.
\subsection{Repulsive steric interaction between two contacting beams}\label{sec::example_contact_repusive_LJpot}
This numerical example aims to demonstrate the general ability of our proposed method to preclude penetration of two slender bodies that come into contact under arbitrary mutual orientation in 3D.
No adhesive forces are considered in this example.
The setup is inspired by Example 1 in~\cite{Meier2017a} where the macroscopic, so-called all-angle beam contact (ABC) formulation is used to account for the non-penetrability constraint.
Here, we model the contact interaction based on the repulsive part of the LJ interaction potential~\eqref{eq::pot_ia_LJ_pointpair}.
More specifically, we apply the novel SSIP approach as proposed in~\secref{sec::method_double_length_specific_integral} in combination with the SSIP law proposed in~\secref{sec::ia_pot_double_length_specific_evaluation_vdW}.
The parameter specifying the strength of repulsion is set to be~$k\rho_1\rho_2=10^{-16}$.
To be consistent throughout this article, we apply Hermitian Simo-Reissner beam elements instead of the torsion-free Kirchhoff elements used in~\cite{Meier2017a}.
As compared to the original example, this requires us to replace the hinged support of the upper beam by clamped end Dirichlet boundary conditions in order to eliminate all rigid body modes in this quasi-static example.
The same number of three finite elements for the upper, deformable beam and one element for the lower, rigid beam is used.
\begin{figure}[htpb]%
  \centering
  \subfigure[initial configuration]{
    \includegraphics[width=0.3\textwidth]{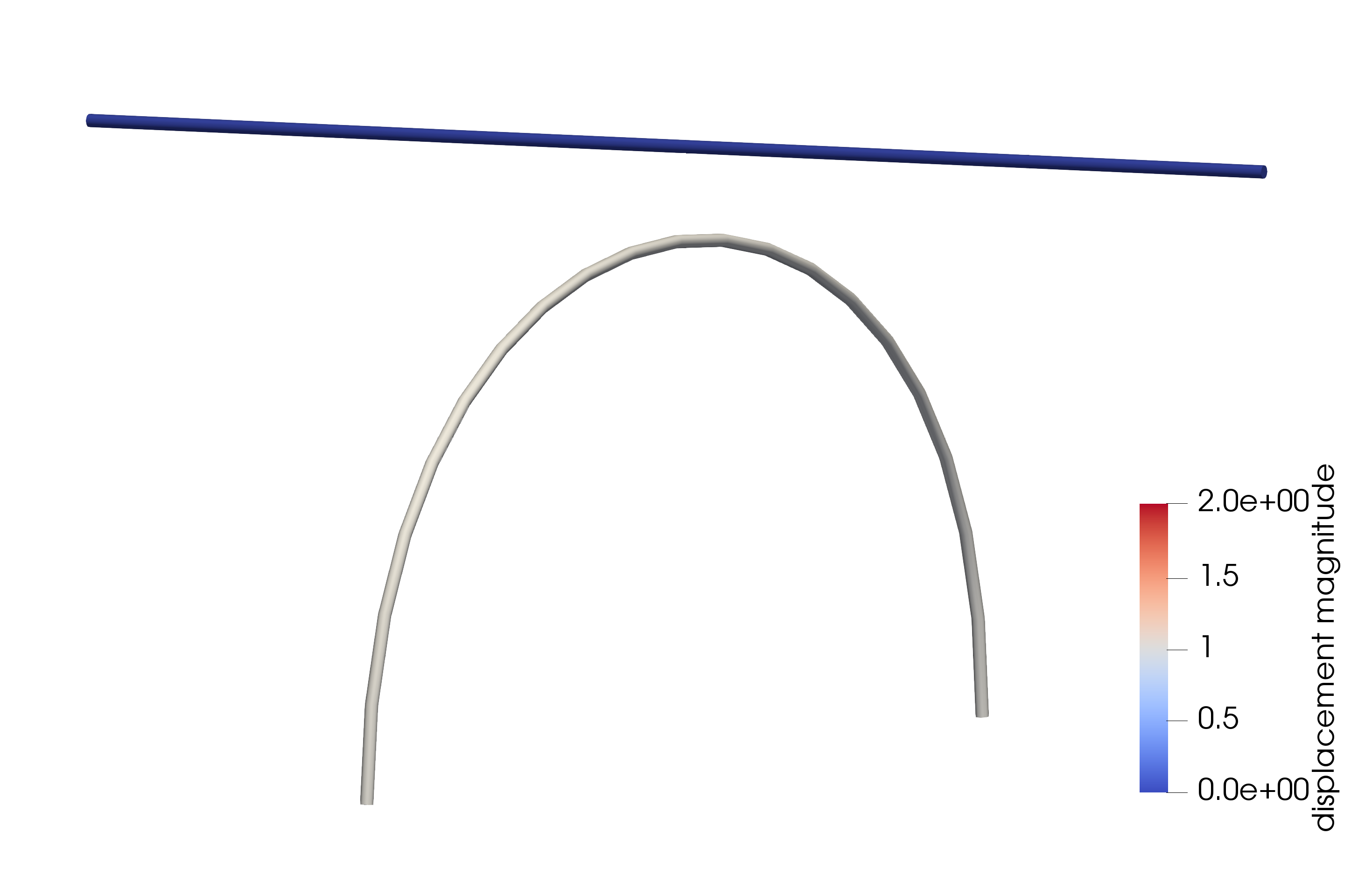}
    \label{fig::num_ex_beamrotatingonarc_snapshot_initial_config}
  }
  \subfigure[time~$t=1.0$]{
    \includegraphics[width=0.3\textwidth]{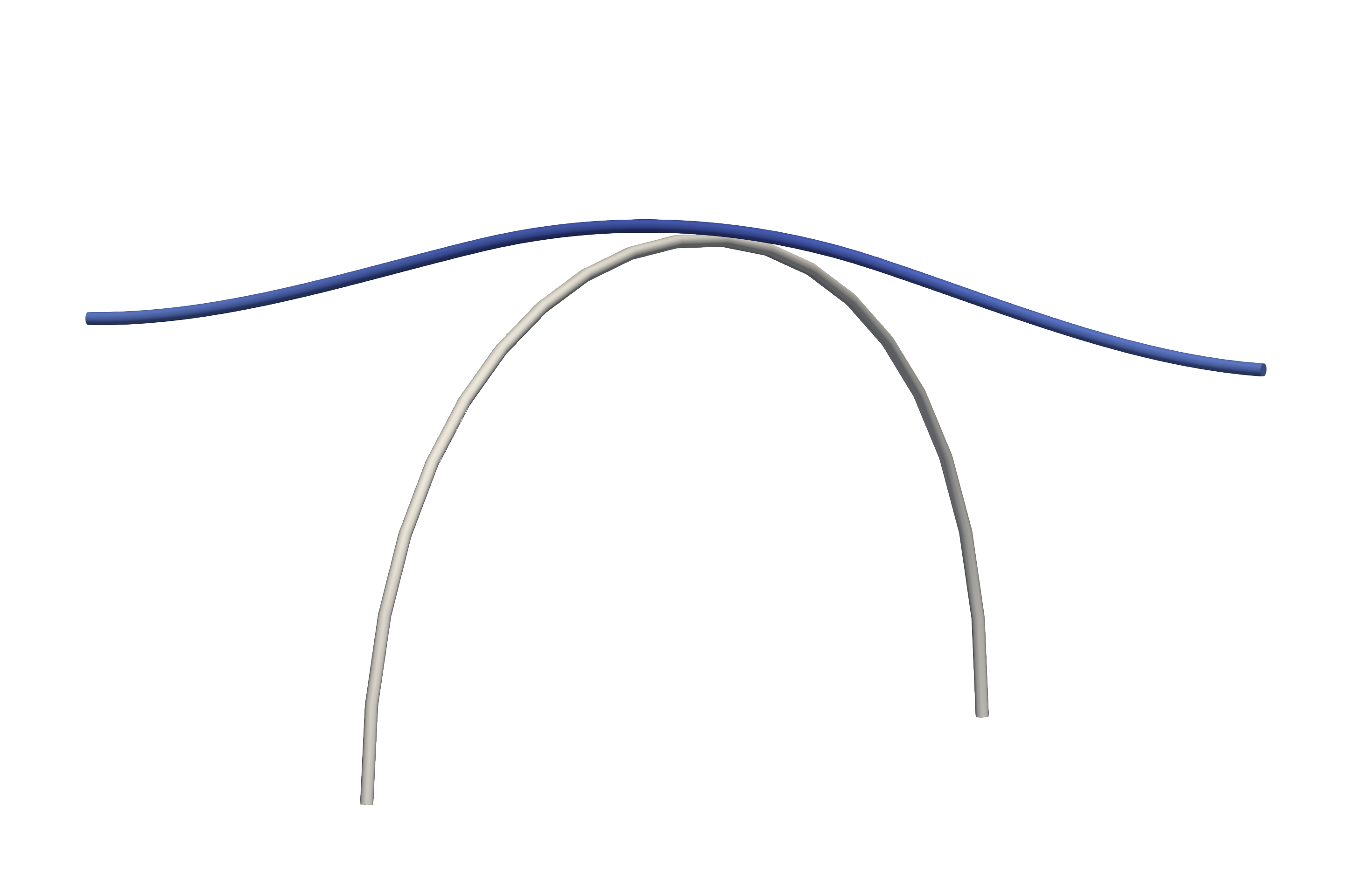}
    \label{fig::num_ex_beamrotatingonarc_snapshot_step1000}
  }
  \subfigure[time~$t=1.5$]{
    \includegraphics[width=0.3\textwidth]{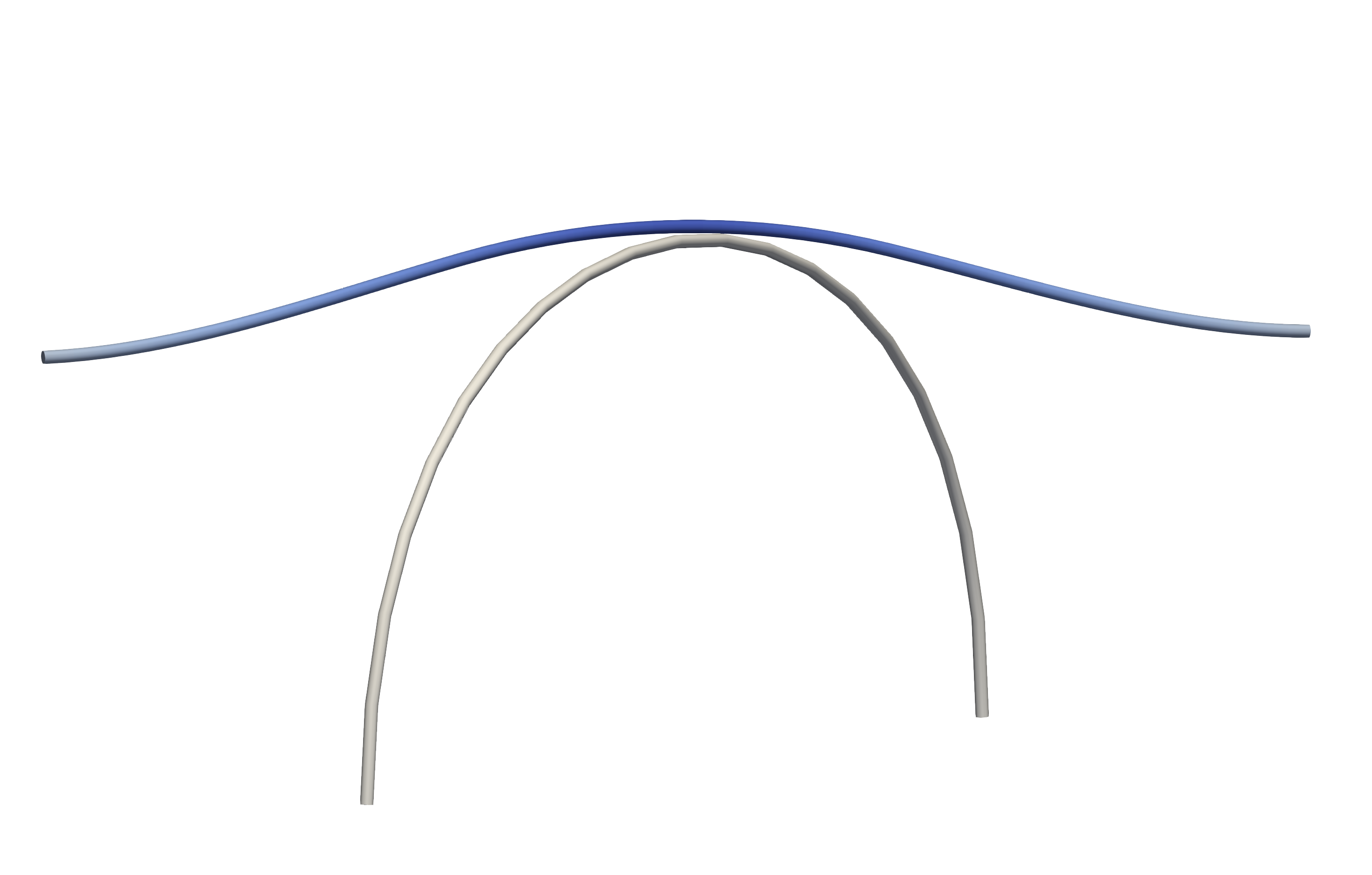}
    \label{fig::num_ex_beamrotatingonarc_snapshot_step1500}
  }
  \subfigure[time~$t=2.0$]{
    \includegraphics[width=0.3\textwidth]{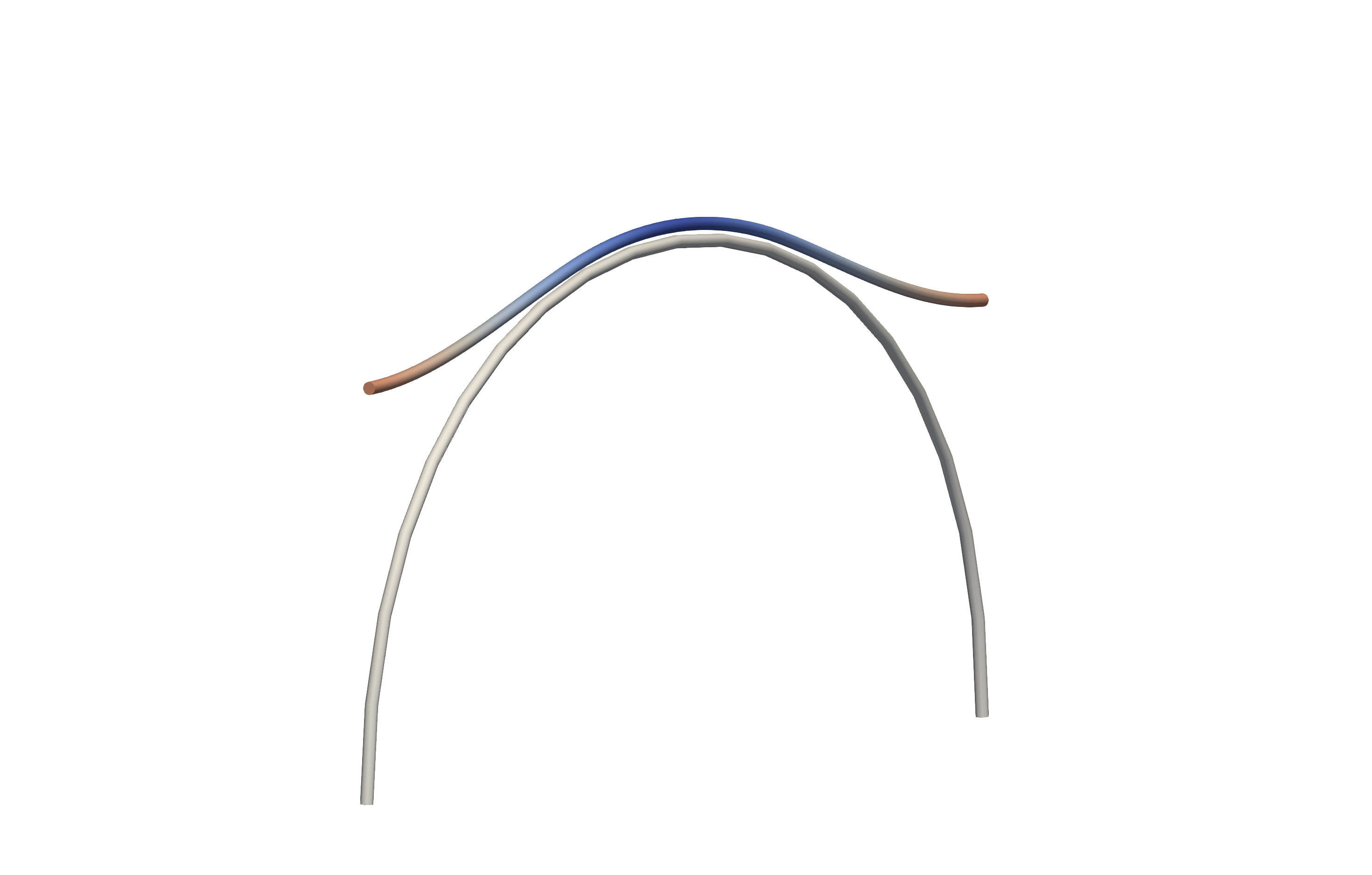}
    \label{fig::num_ex_beamrotatingonarc_snapshot_step2000}
  }
  \subfigure[time~$t=2.5$]{
    \includegraphics[width=0.3\textwidth]{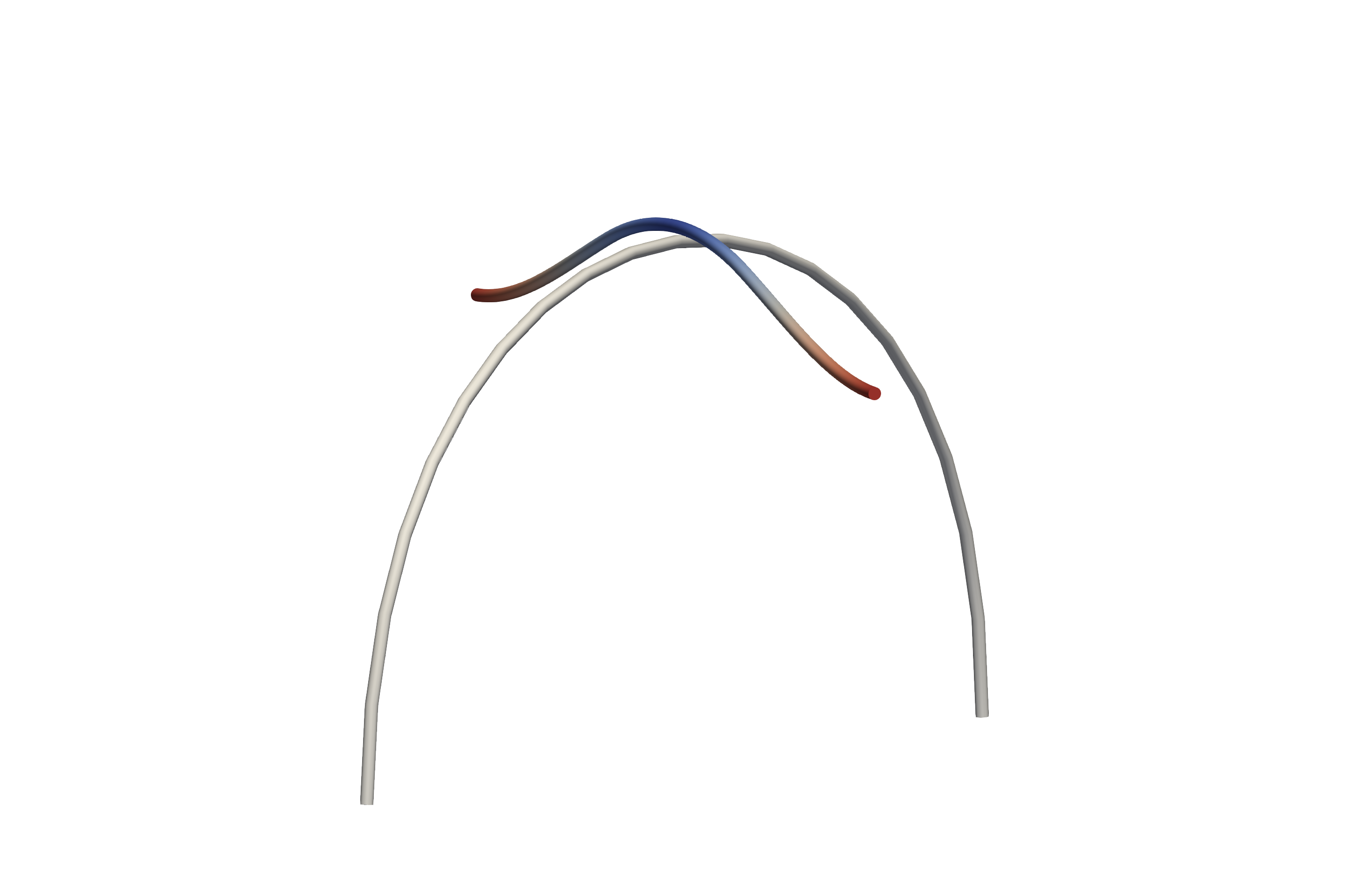}
    \label{fig::num_ex_beamrotatingonarc_snapshot_step2500}
  }
  \subfigure[time~$t=3.0$]{
    \includegraphics[width=0.3\textwidth]{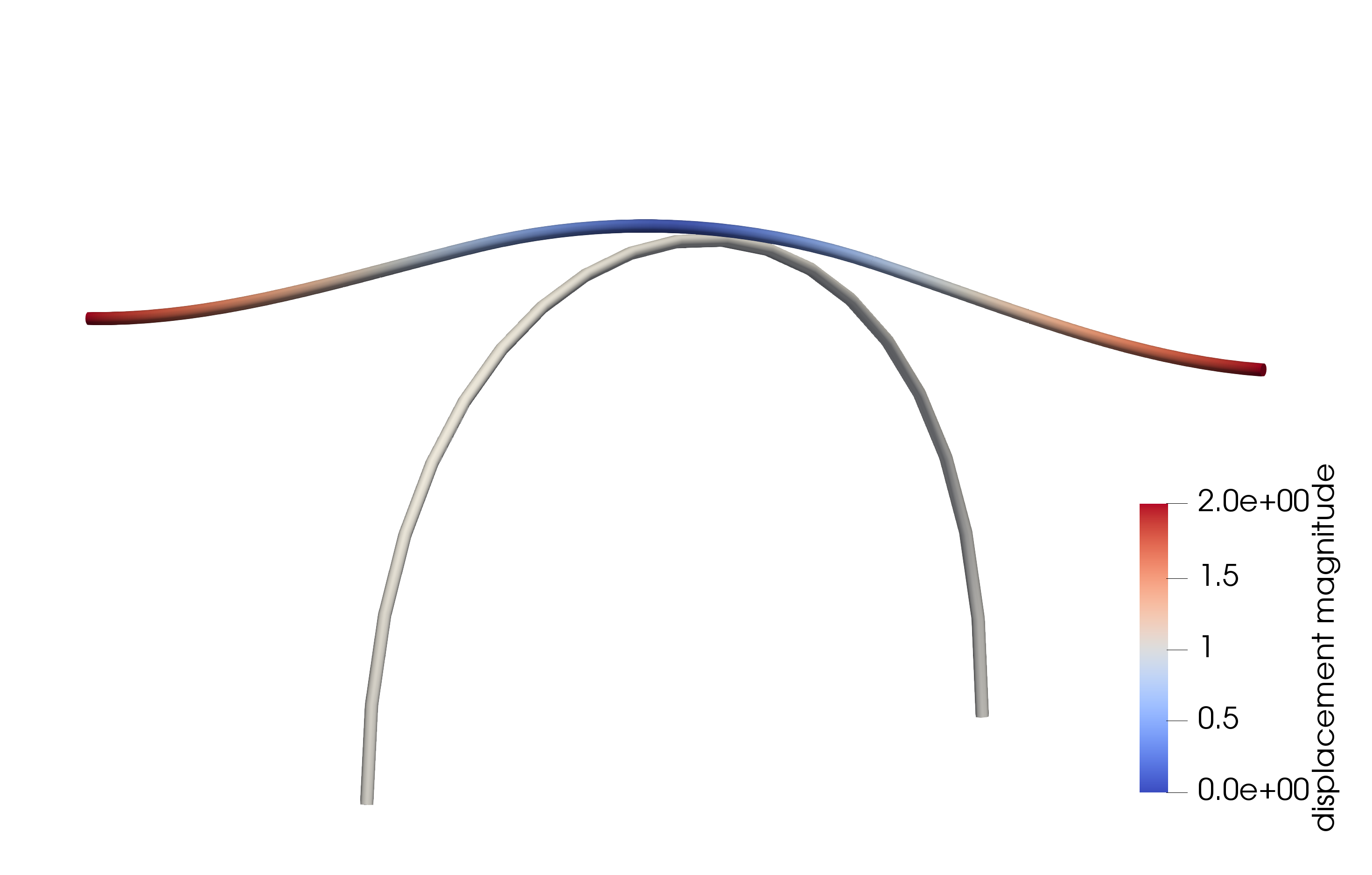}
    \label{fig::num_ex_beamrotatingonarc_snapshot_step3000}
  }
  \caption{Simulation snapshots: a straight deformable beam rotating on a rigid arc.}
  \label{fig::num_ex_beamrotatingonarc_snapshots}
\end{figure}

A sequence of the resulting simulation snapshots is shown in~\figref{fig::num_ex_beamrotatingonarc_snapshots}.
As expected, the two beams do not penetrate each other in any of the various mutual orientations throughout the simulation.
\begin{figure}[htpb]%
  \centering
  \subfigure[time~$t=1.95$]{
    \includegraphics[width=0.242\textwidth]{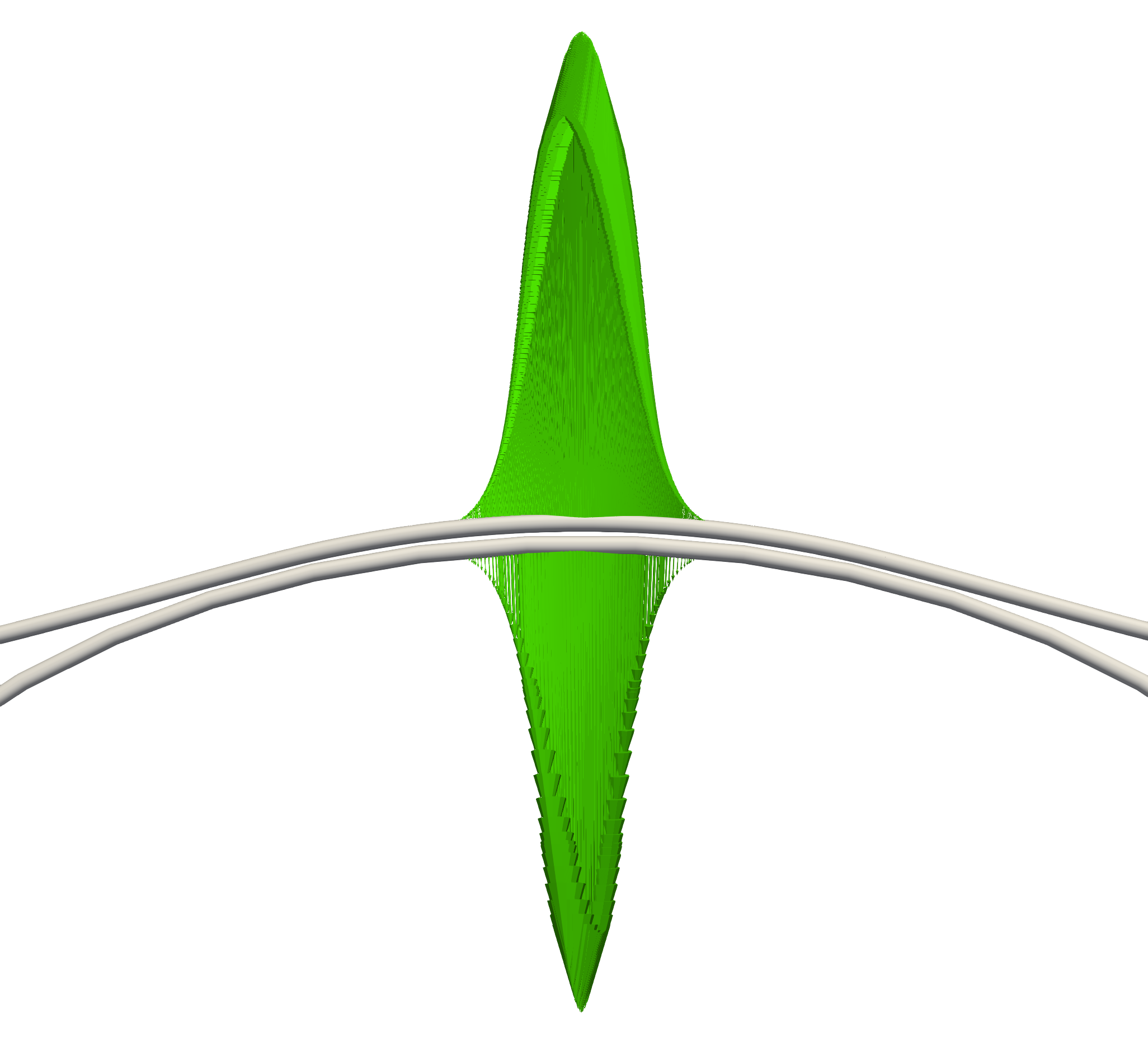}
    \label{fig::num_ex_beamrotatingonarc_LJpot_GPs_50_10_contactforce_distribution_step1950}
  }
  \hspace{-10pt}
  \subfigure[time~$t=1.97$]{
    \includegraphics[width=0.242\textwidth]{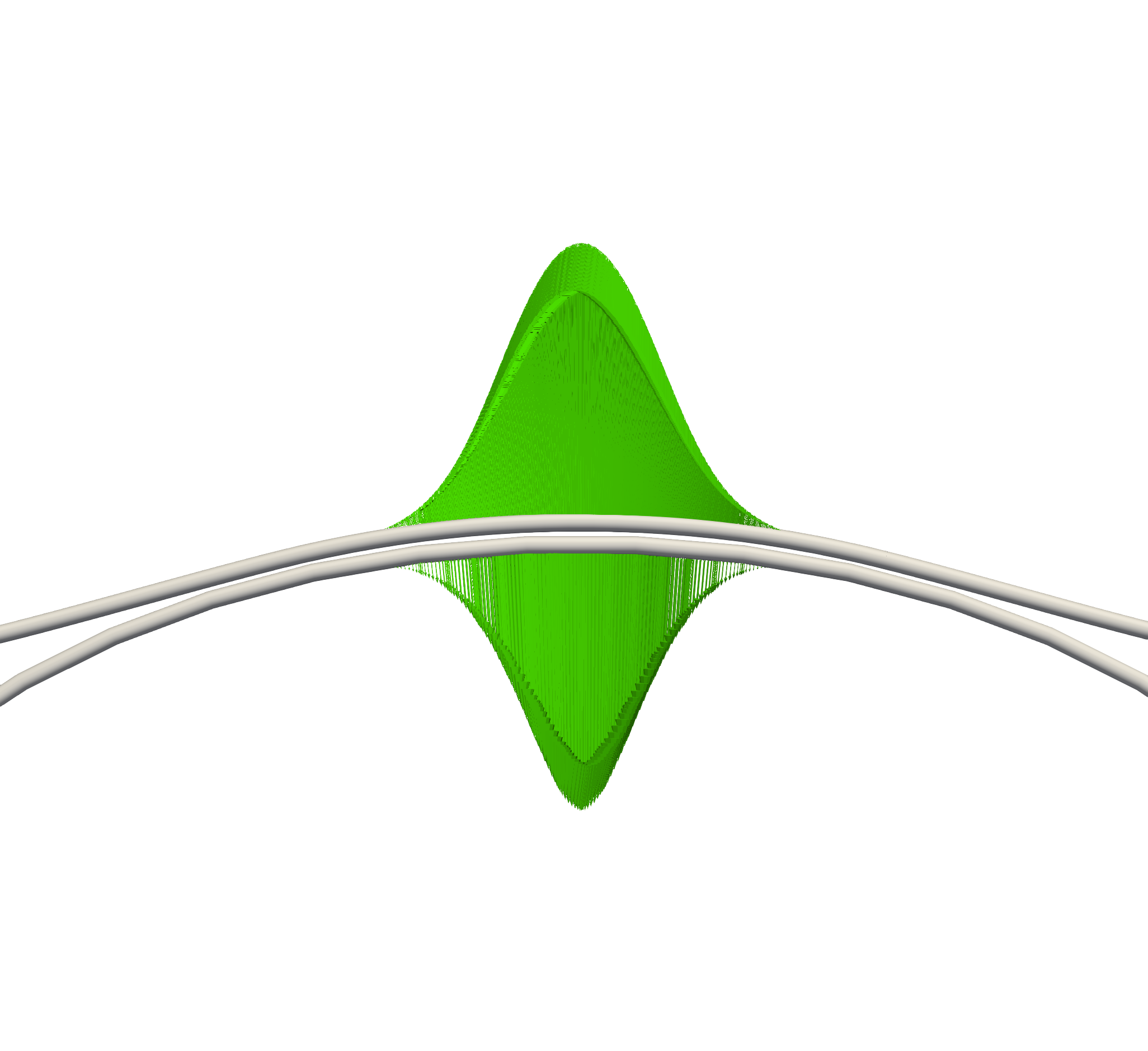}
    \label{fig::num_ex_beamrotatingonarc_LJpot_GPs_50_10_contactforce_distribution_step1970}
  }
  \hspace{-10pt}
  \subfigure[time~$t=1.99$]{
    \includegraphics[width=0.242\textwidth]{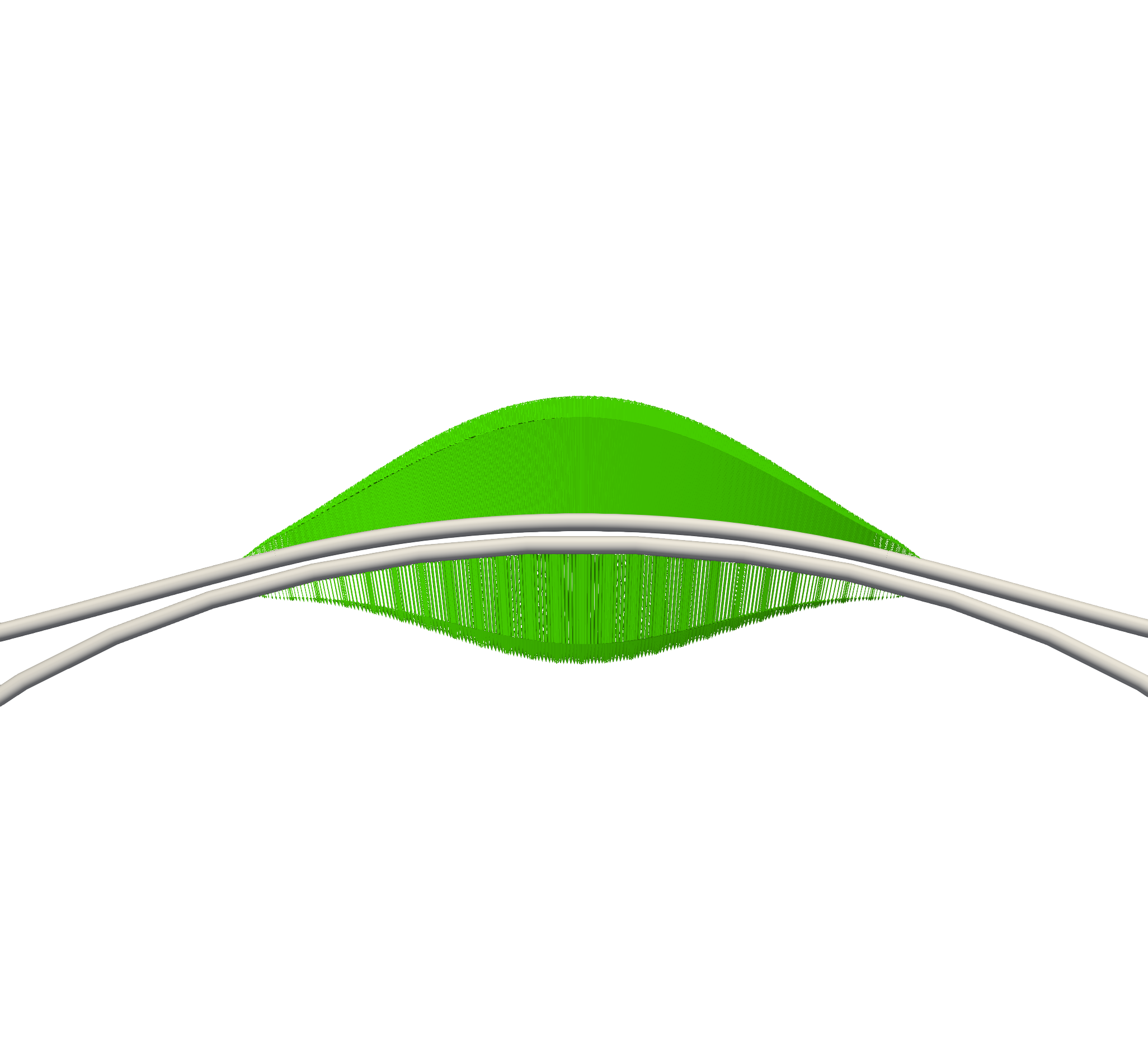}
    \label{fig::num_ex_beamrotatingonarc_LJpot_GPs_50_10_contactforce_distribution_step1990}
  }
  \hspace{-10pt}
  \subfigure[time~$t=2.0$]{
    \includegraphics[width=0.242\textwidth]{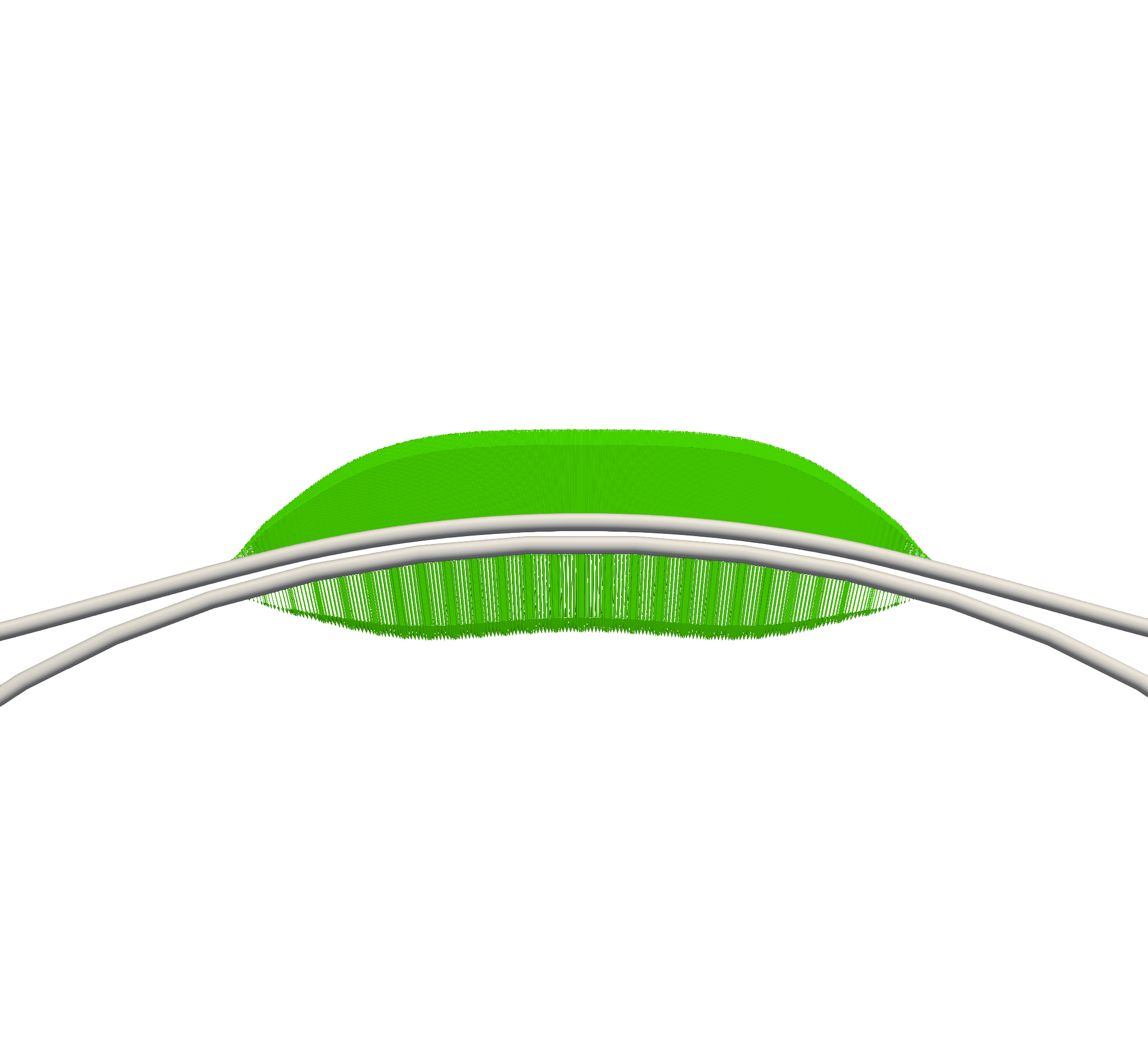}
    \label{fig::num_ex_beamrotatingonarc_LJpot_GPs_50_10_contactforce_distribution_step2000}
  }
  \caption{Evolution of contact force distribution in the regime of small contact angles between the beam axes. Each beam element is divided in~$100$ integration segments with~$10$ Gauss points each in the simulation shown here.}
  \label{fig::num_ex_beamrotatingonarc_contactforce_distribution}
\end{figure}
\figref{fig::num_ex_beamrotatingonarc_contactforce_distribution} visualizes the contact force distributions%
\footnote{More precisely, the vectorial line load with dimensions of force per unit length is visualized as an arrow at each integration point. The force resultant therefore equals the integral over the contour curve defined by the arrows' tips (i.\,e.~the area under this curve), and not the vector sum of all arrows shown. This is important to understand because the number of visible arrows per unit length depends on the discretization and is thus higher for the upper, deformable beam.}
in the most interesting time span before the beams reach the parallel orientation at time~$t=2.0$.
The force distribution quickly changes from a point-like force for large mutual angles to a broad distributed load for parallel beam axes.
Note also that the line load has a three dimensional shape where the out-of-plane component decreases with decreasing mutual angle until both beam axes and thus also the line loads lie in one plane at~$t=2.0$.
Another remarkable result is the symmetry between the line loads on both fibers.
It nicely confirms that the novel approach indeed fulfills the expected \textit{local} equilibrium of interaction forces in good approximation.
In contrast to existing, macroscopic formulations for beam contact, this is not postulated a priori in our approach and hence is a valuable verification at this point.
See~\cite{Sauer2013} for a comprehensive discussion of this important topic in the context of contact between 3D solids described by inter-surface potentials.
The \textit{global} equilibrium of contact forces on the other hand is fulfilled exactly, as can be concluded from the \text{global} conservation of linear momentum that can be shown analytically as outlined in ~\secref{sec::conservation_properties}.
In this numerical example, we found that the sum of all reaction forces in either of the spatial dimensions is indeed zero with a maximal residuum of~$10^{-10}$ throughout all simulations considered here, which confirms the statement numerically.
\begin{figure}[htpb]%
  \centering
  \subfigure[]{
    \includegraphics[width=0.4\textwidth]{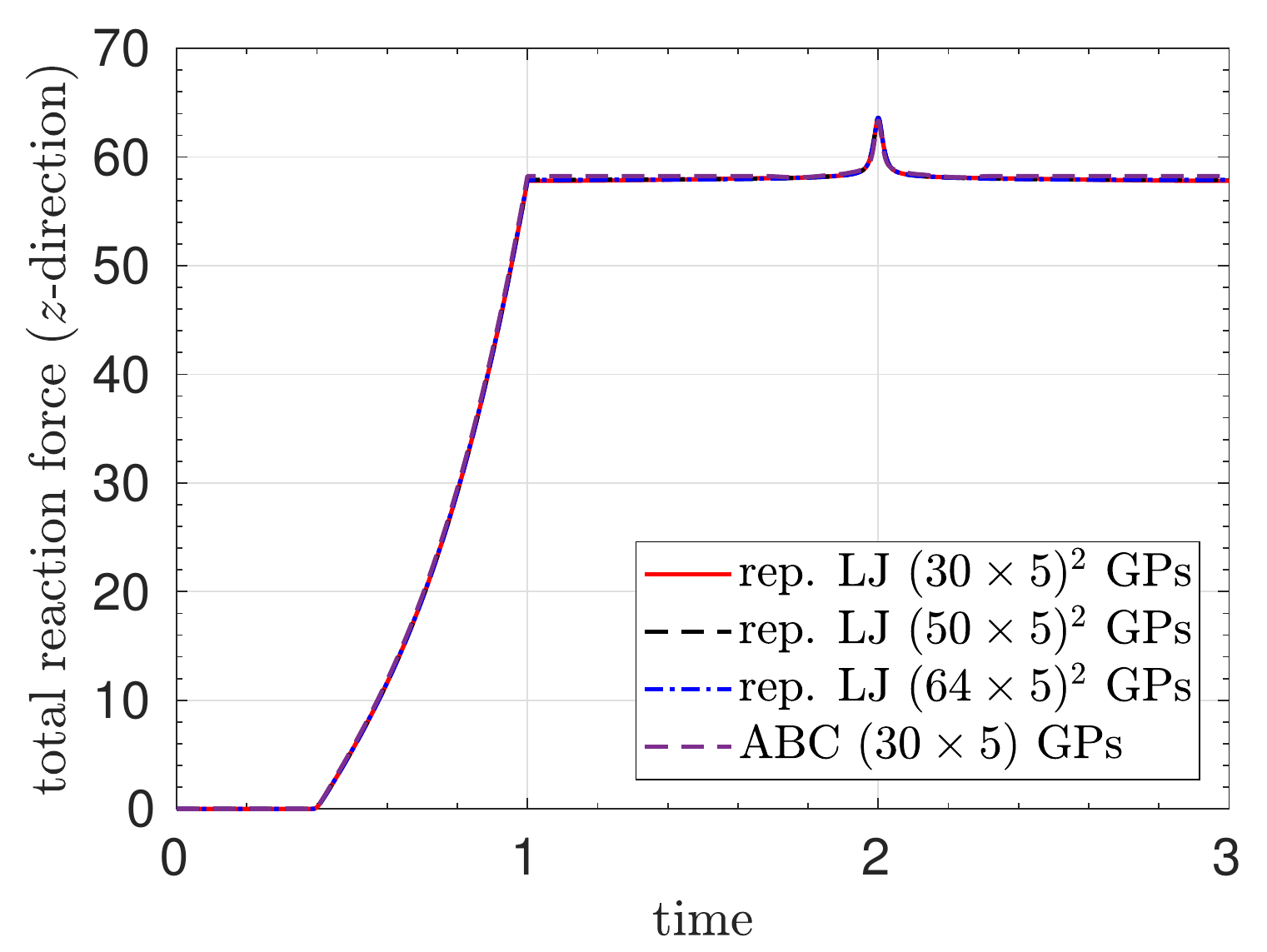}
    \label{fig::num_ex_beamrotatingonarc_reactionforce_z_over_time_LJ_numGP5_varying_num_integration_segments_vs_ABC}
  }
  \subfigure[]{
    \includegraphics[width=0.4\textwidth]{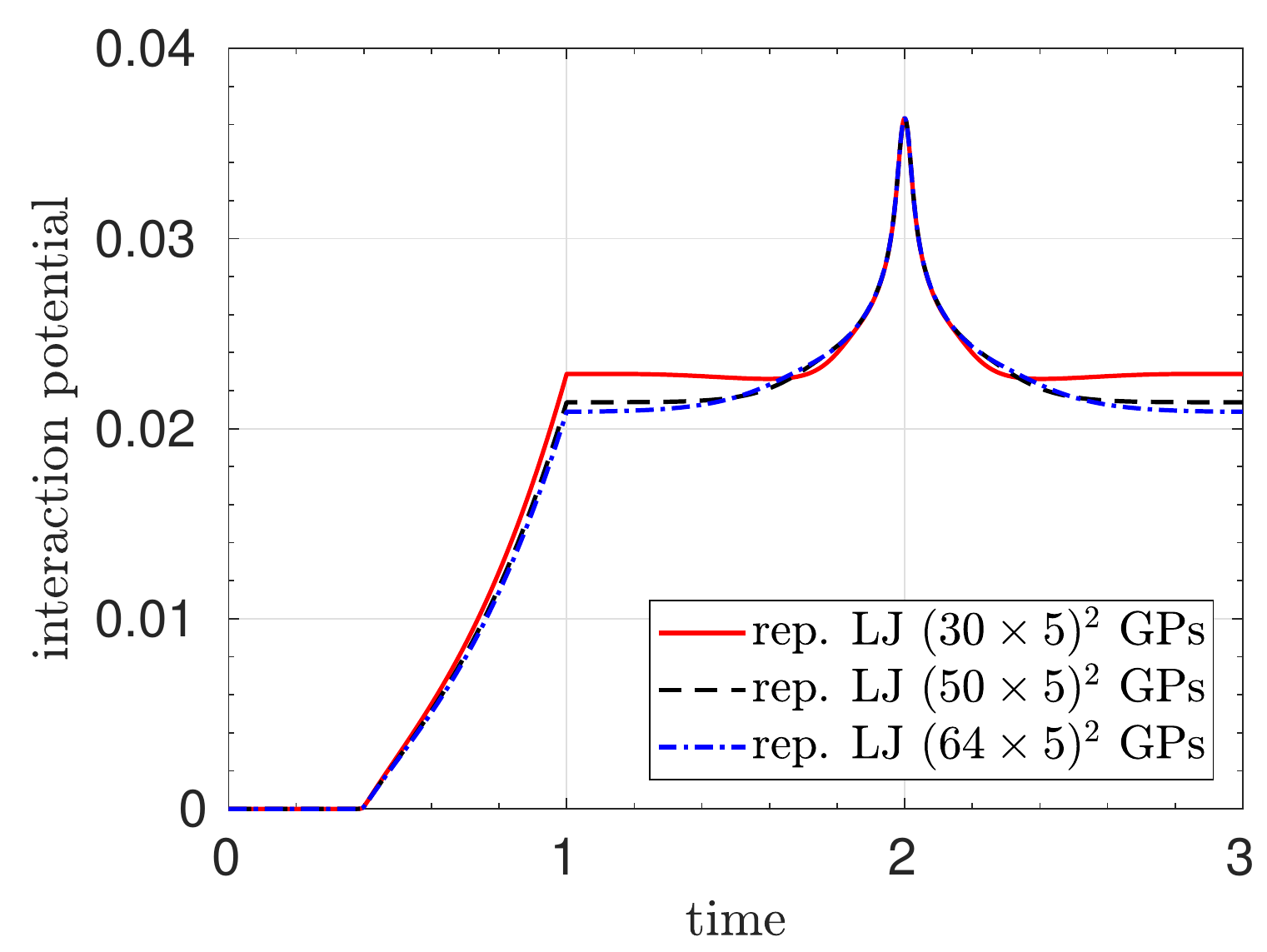}
    \label{fig::num_ex_beamrotatingonarc_interaction_potential_over_time_LJ_numGP5_varying_num_integration_segments}
  }
  \caption{(a) Reaction force and (b) interaction potential over time.}
  \label{fig::num_ex_beamrotatingonarc_reactionforce_ia_pot_quadrature}
\end{figure}

\figref{fig::num_ex_beamrotatingonarc_reactionforce_ia_pot_quadrature} shows the resulting vertical reaction force as well as the interaction potential over time.
Due to the inverse-twelve power law and the extremely small separations of the interacting bodies, the numerical integration of the disk-disk interaction forces is very challenging and we studied the influence of the number of Gauss points.
For this purpose, the number of integration segments per element with five Gauss points each is set to~$30$, $50$, or $64$, respectively.
Interestingly, the interaction potential shown in \figref{fig::num_ex_beamrotatingonarc_interaction_potential_over_time_LJ_numGP5_varying_num_integration_segments} seems to be more sensitive with respect to the integration error than the vertical reaction force shown in \figref{fig::num_ex_beamrotatingonarc_reactionforce_z_over_time_LJ_numGP5_varying_num_integration_segments_vs_ABC} despite the fact that the latter has a higher inverse power law exponent.
Presumably, this is due to the fact that the reaction force is dominated by the bending deformation of the beams.
For reference, the reaction force obtained by using the macroscopic ABC formulation is shown as well and is in excellent agreement with the one resulting from the repulsive part of the LJ interaction potential.

A more comprehensive comparison of this novel SSIP approach to model contact between beams based on (the repulsive part of) the molecular LJ interaction and existing, macroscopic formulations based on heuristic penalty force laws is a highly interesting subject that is worth to investigate in the future.

\subsection{Two initially straight, deformable fibers carrying opposite surface charge}\label{sec::num_ex_elstat_attraction_twoparallelbeams}
The following example consists of two initially straight and parallel, deformable fibers that attract each other due to their surface charge of opposite sign.
Its setup is kept as simple as possible to allow for an isolated and clear analysis of the physical effects as well as the main characteristics of the proposed SSIP approach.
In a first step presented here, the interplay of elasticity and electrostatic attraction in the regime of large separations is studied.
Additionally, the authors' recent contribution~\cite{GrillPeelingPulloff} considers the scenario of separating these adhesive fibers starting from initial contact and studies a variety of physical effects and influences in depth, which would go beyond the scope of this work.

\begin{figure}[htpb]%
  \centering
  \subfigure[Problem setup: undeformed configuration.]{
    \def\svgwidth{0.45\textwidth}
    \input{num_ex_elstat_attraction_twoparallelbeams_problem_setup.pdf_tex}
    \label{fig::num_ex_elstat_attraction_twoparallelbeams_problem_setup}
  }
  \hfill
  \subfigure[Static equilibrium configurations for varying attractive strength. Solution for beam centerlines and corresponding value of the potential law prefactor~$k$ shown in the same color.]{
    \includegraphics[width=0.5\textwidth]{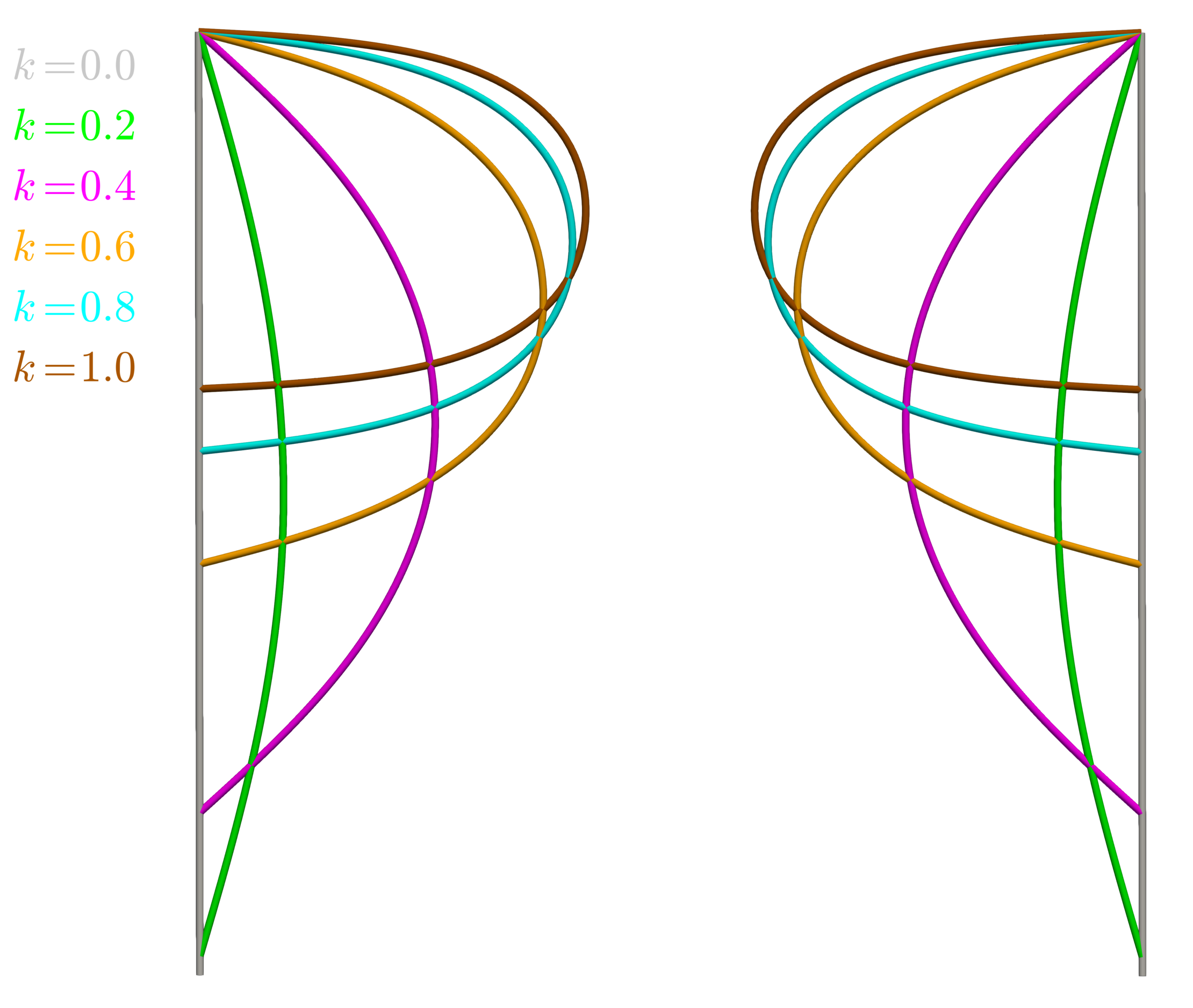}
    \label{fig::num_ex_elstat_attraction_twoparallelbeams_static_varying_prefactor}
  }
  \caption{Two parallel beams with constant surface charge density (left beam positive, right beam negative).}
  \label{fig::num_ex_elstat_attraction_twoparallelbeams_static}
\end{figure}

In this numerical example, we are interested in the static equilibrium configurations for varying attractive strength.
As shown in~\figref{fig::num_ex_elstat_attraction_twoparallelbeams_problem_setup}, two straight beams of length~$l=5$ are aligned with the global~$y$-axis at an inter-axis separation~$d=5$.
Both are simply supported and restricted to move only within the~$xy$-plane and rotate only around the global~$z$-axis.
The beams have a circular cross-section with radius~$R=0.02$ which results in a slenderness ratio of~$\zeta=250$.
Cross-section area, area moments of inertia and shear correction factor are computed using standard formula for a circle.
A hyperelastic material law with Young's modulus~$E=10^{5}$ and Poisson's ratio~$\nu=0.3$ is applied.
In terms of spatial discretization, we use five Hermitian Simo-Reissner beam elements per fiber (see \cite{Meier2017b} for details on this element formulation).
Electrostatic interaction is modeled via the SSIP approach as presented in~\secref{sec::method_double_length_specific_integral} and applied to long-range Coulomb interactions in~\secref{sec::ia_pot_double_length_specific_evaluation_elstat}.
Both beams are nonconducting with a constant surface charge density of~$\sigma_1=1.0$ and~$\sigma_2=-1.0$, respectively.
For simplicity, we vary the prefactor~$k$ of the underlying Coulomb law~$\Phi(r)=k\, r^{-1}$ to vary the strength of attraction.
However, as becomes clear from~\eqref{eq::var_pot_ia_powerlaw_large_sep_surface}, this is equivalent to a variation of surface charge densities because in our case the product of these quantities is a constant prefactor in all relevant equations.
In order to evaluate the electrostatic force and stiffness contributions, Gauss quadrature with two integration segments per element and ten Gauss points per integration segment is applied.
This turns out to be fine enough to not change the presented results perceptibly.
More precisely, the difference in the displacement of the beam midpoint for~$n_\text{GP}=(2 \times 10)^2$ as compared to~$(2 \times 32)^2$ is below~$10^{-8}$.
No cut-off radius is applied here, i.\,e.,~the contributions of all Gauss point pairs are evaluated and included.

\begin{figure}[htpb]%
  \centering
  \subfigure[Resulting electrostatic forces evaluated at the Gauss points.]{
    \includegraphics[width=0.6\textwidth]{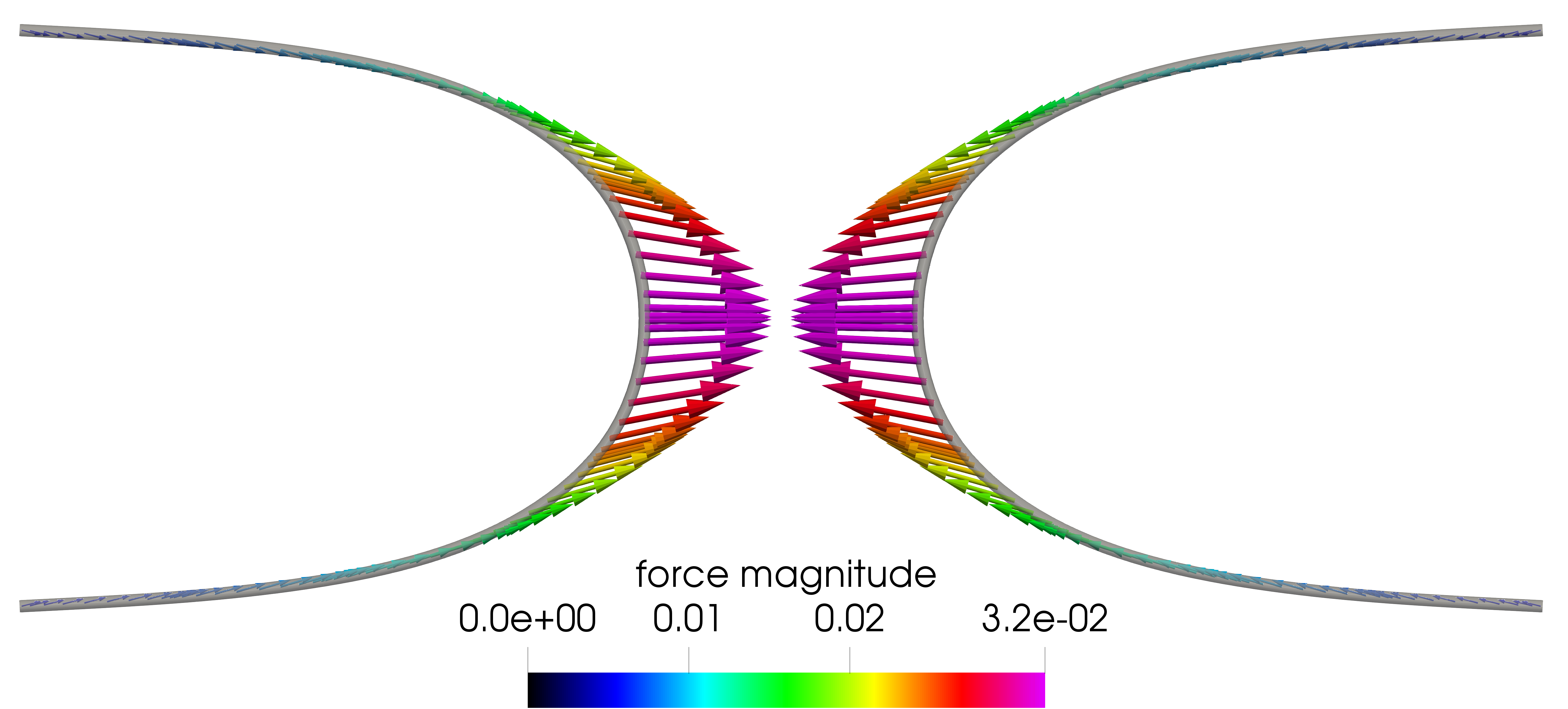}
    \label{fig::num_ex_elstat_attraction_twoparallelbeams_static_resulting_forces_all_elepairs}
  }
  \subfigure[Individual electrostatic force contributions of all element pairs.]{
    \includegraphics[width=0.6\textwidth]{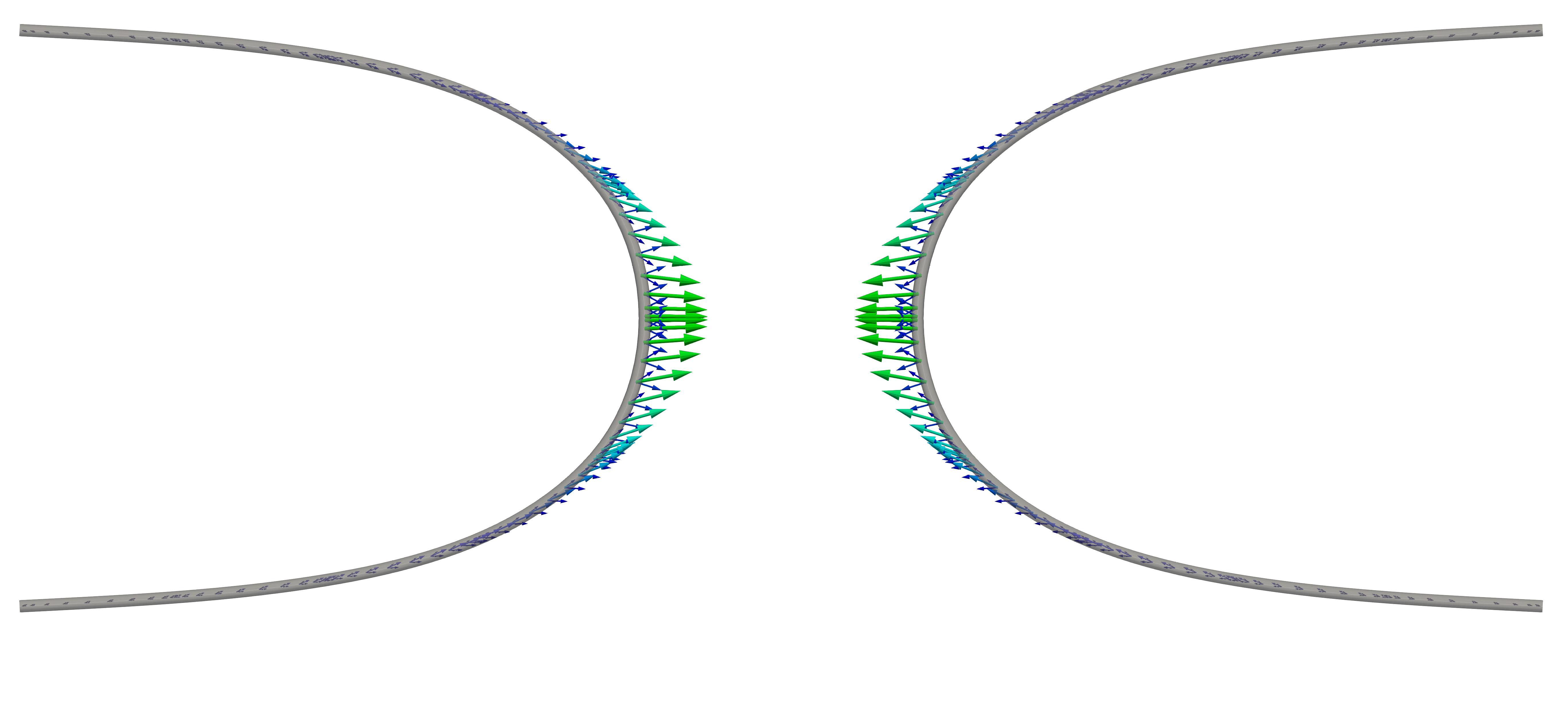}
    \label{fig::num_ex_elstat_attraction_twoparallelbeams_static_forces_all_elepairs}
  }
  \subfigure[Electrostatic force contributions of the closest element pair.]{
    \includegraphics[width=0.48\textwidth]{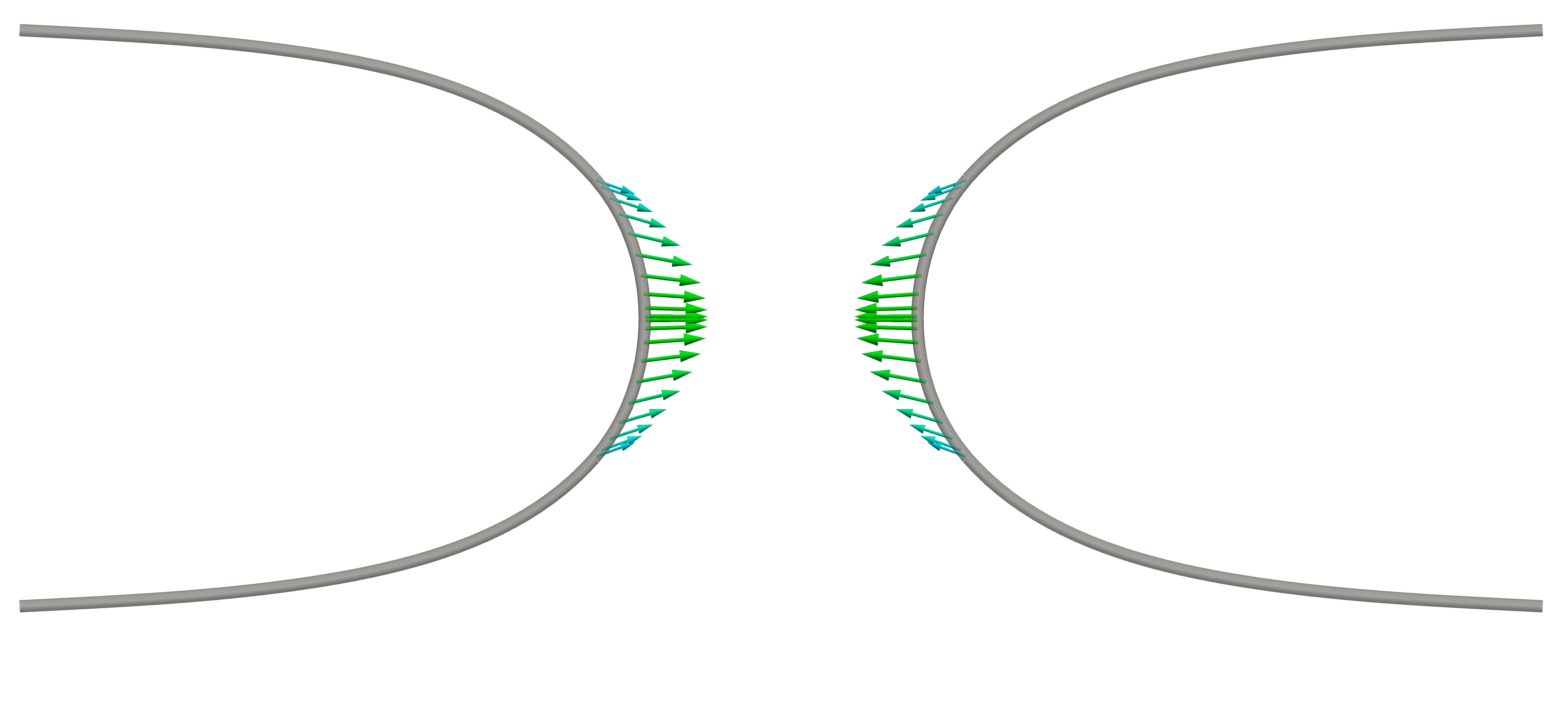}
    \label{fig::num_ex_elstat_attraction_twoparallelbeams_static_forces_elepair3_8}
  }
  \subfigure[Electrostatic force contributions of the next-nearest pair.]{
    \includegraphics[width=0.48\textwidth]{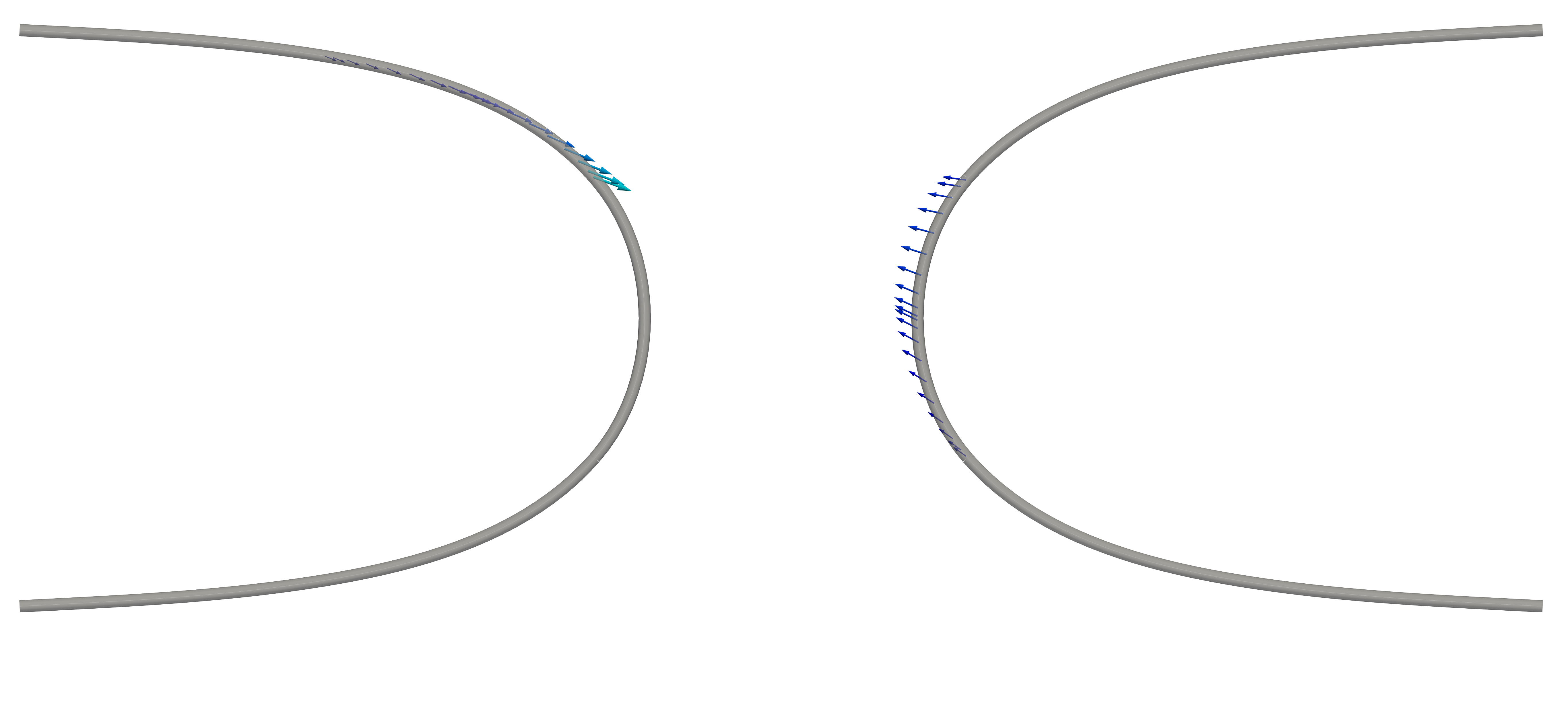}
    \label{fig::num_ex_elstat_attraction_twoparallelbeams_static_forces_elepair2_8}
  }
  \caption{Electrostatic forces acting on the beams for $k=1.0$. Color indicates force magnitude.}
  \label{fig::num_ex_elstat_attraction_twoparallelbeams_static_force_distribution}
\end{figure}

\figref{fig::num_ex_elstat_attraction_twoparallelbeams_static_varying_prefactor} finally shows the resulting static equilibrium configurations for different levels of attractive strength.
As expected, the beams are increasingly deflected and pulled towards each other if the prefactor of the applied Coulomb law~$k$ and thus the attractive strength is increased.
Like the problem definition, also all the solutions are perfectly symmetric with respect to the vertical axis of symmetry located at~$x=d/2$.
Moreover, the centerline curves of each individual solution show a horizontal axis of symmetry defined by the position of the two beam midpoints in the respective deformed state.
As a consequence, the vertical force components in the system cancel and the vertical reaction forces vanish.
This also becomes clear when looking at the visualization of the resulting electrostatic forces as shown exemplarily for~$k=1.0$ in \figref{fig::num_ex_elstat_attraction_twoparallelbeams_static_resulting_forces_all_elepairs}.
Additionally, the forces acting on the Gauss point of one beam caused by the interaction with one finite element of the other beam are visualized individually in~\figref{fig::num_ex_elstat_attraction_twoparallelbeams_static_forces_all_elepairs}.
This representation illustrates the nature of the SSIP approach, which is based on two nested 1D numerical integrals that are evaluated element pair-wise.
Accordingly, we can identify five force contributions at each Gauss point, one for each of the five beam elements on the opposing fiber.
As expected, the magnitude of these individual forces decays with the distance and the contributions of the closest element pair shown in an isolated manner in \figref{fig::num_ex_elstat_attraction_twoparallelbeams_static_forces_elepair3_8} constitute the largest part of the total electrostatic load on the beams and are clearly larger than the contributions of the next-nearest element pair shown in~\figref{fig::num_ex_elstat_attraction_twoparallelbeams_static_forces_elepair2_8}.
However, the comparatively long range of electrostatic forces yields a smooth force distribution along the centerlines and we can identify non-zero force contributions even at the most distant Gauss points right next to the supports in~\figref{fig::num_ex_elstat_attraction_twoparallelbeams_static_resulting_forces_all_elepairs}.
As mentioned above, a quantitative analysis of the resulting horizontal reaction forces is presented in~\cite{GrillPeelingPulloff}.

To conclude this example of two charged, attractive beams, we briefly look at the nonlinear solver.
Newton's method without any adaptations is used here to allow for a clear and meaningful analysis of nonlinear convergence behavior.
The solutions for~$k\leq0.4$ can be found within one load step which is a remarkable result given the resulting large deflection of the beams shown in~\figref{fig::num_ex_elstat_attraction_twoparallelbeams_static_varying_prefactor} and the strong nonlinear nature of the system.
For stronger attractive forces, the strength of electrostatic attraction was ramped up in up to ten equal steps~$\Delta k = 0.1$.
As convergence criteria, we enforced both for the Euclidean norm of the residual vector~$\norm{\vdR} < 10^{-10}$ and for the norm of the iterative displacement update vector~$\norm{\Delta \vdX} < 10^{-8}$.
In fact, this combination leads to~$\norm{\vdR} < 10^{-12}$ in almost all equilibrium configurations shown here.

\subsection{Two charged deformable fibers dynamically snap into contact}\label{sec::num_ex_twocrossedbeams_elstat_snapintocontact}
\begin{figure}[htpb]%
  \centering
  \subfigure[Problem setup.]{
    \def\svgwidth{0.3\textwidth}
    \small{
    \input{num_ex_elstat_attraction_twocrossedbeams_snapintocontact_problem_setup.pdf_tex}
    }
    \label{fig::twocrossedbeams_elstat_snapintocontact_problem_setup}
  }
  \subfigure[Energy over time.]{
    \includegraphics[width=0.6\textwidth]{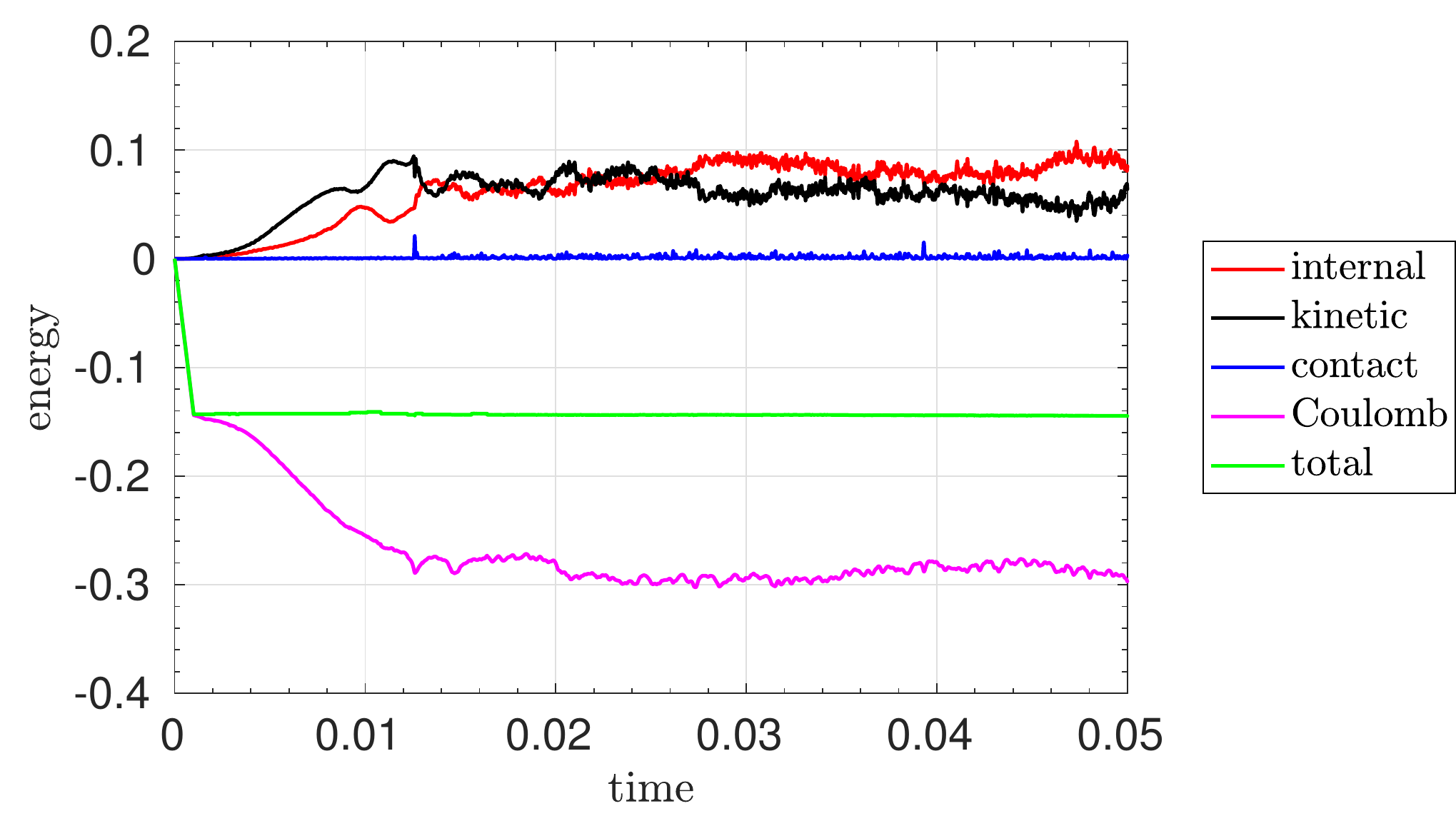}
    \label{fig::twocrossedbeams_elstat_snapintocontact_energy_over_time}
  }
  \caption{Two oppositely charged, crossed beams dynamically snap into contact.}
  \label{fig::twocrossedbeams_elstat_snapintocontact}
\end{figure}

Due to the high gradients in the inverse power laws, molecular interactions give rise to highly dynamic systems.
This is a first, simple example for a dynamic system consisting of two oppositely charged fibers with a hinged support at one end each, that will snap into contact.
In the initial configuration shown in~\figref{fig::twocrossedbeams_elstat_snapintocontact_problem_setup}, the straight fibers include an angle of~$45^\circ$ and their axes are separated by~$5R$ in the out-of-plane direction~$z$.
With a cross-section radius~$R=0.02$ and length~$l$ set to~$l=5$, they have a high slenderness ratio of~$\zeta=250$ and~$354$.
Each of the fibers is discretized by~$10$ Hermitian Simo-Reissner beam elements and the material parameters are chosen to be~$E=10^5$, $\nu=0.3$, and~$\rho=10^{-3}$.
The fibers carry a constant, opposite surface charge~$\sigma_{1/2}=\pm 1.0$ and interact via the Coulomb potential law stated in eq.~\eqref{eq::pot_ia_elstat_pointpair_Coulomb} with the prefactor set to~$C_\text{elstat}=0.4$.
In order to start from a stress-free initial configuration, the charge of one of the fibers is ramped up linearly within the first~$100$ time steps.
We apply the SSIP approach as proposed in~\secref{sec::method_double_length_specific_integral} and applied to Coulomb interactions in~\secref{sec::ia_pot_double_length_specific_evaluation_elstat}.
A total of~$5$ integration segments per element with~$10$ Gauss points each is used to evaluate these electrostatic contributions.
The contact interaction between the fibers is modeled by the line contact formulation proposed in~\cite{meier2016}, using a penalty parameter~$\varepsilon=10^3$ and~$20$ integration segments per element with~$5$ Gauss points each for numerical integration.
An undetected crossing of the fiber axes is prevented by applying the modified Newton method limiting the maximal displacement increment per nonlinear iteration to~$R/2$ (see \ref{sec::algorithm_implementation_aspects} for details).

\begin{figure}[htb]%
  \centering
  \subfigure[initial configuration]{
    \includegraphics[width=0.31\textwidth]{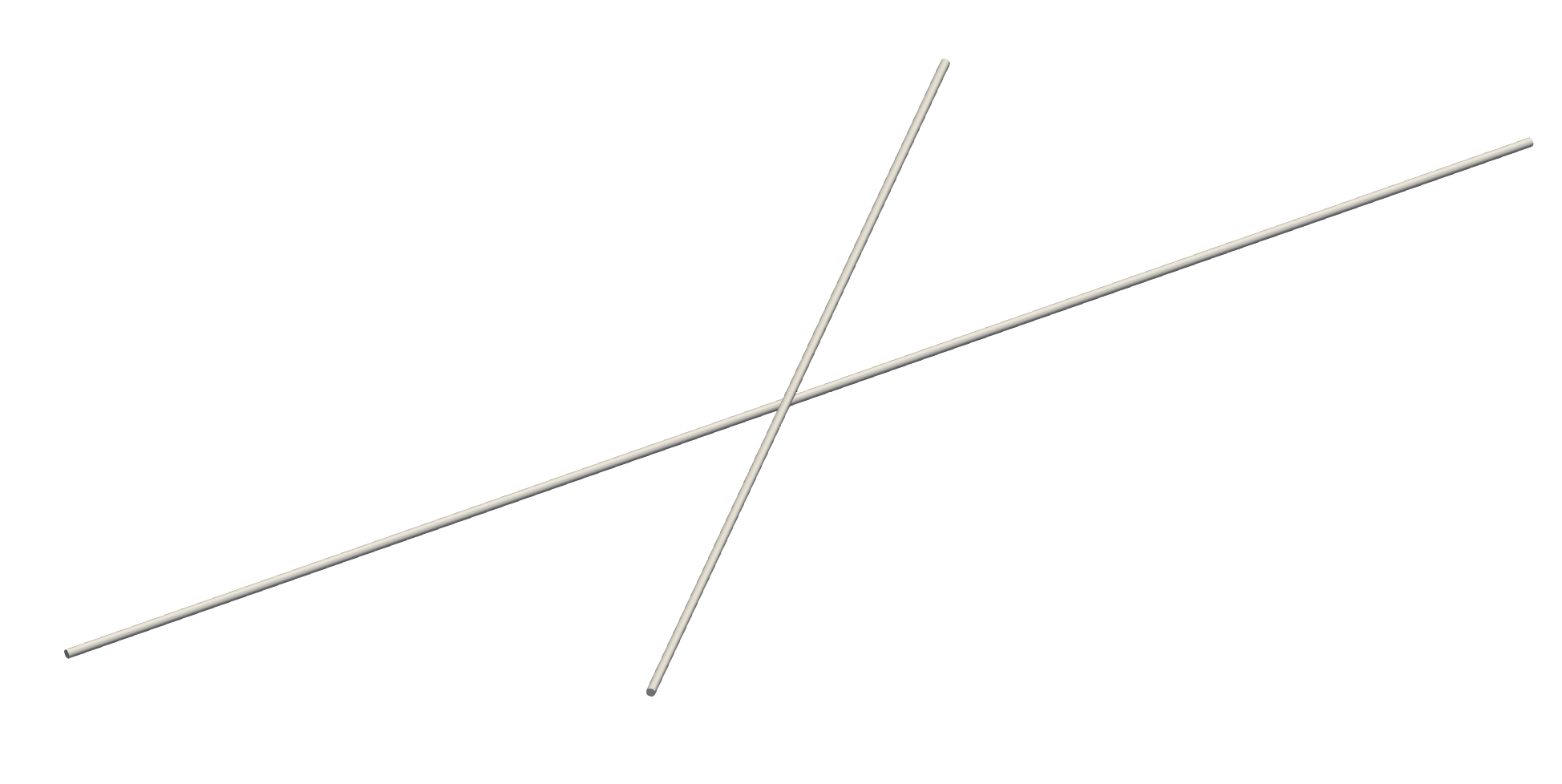}
    \label{fig::twocrossedbeams_elstat_snapintocontact_snapshots_initial}
  }
  \subfigure[time $t=1\times10^{-3}$, ramp-up of charge completed]{
    \includegraphics[width=0.31\textwidth]{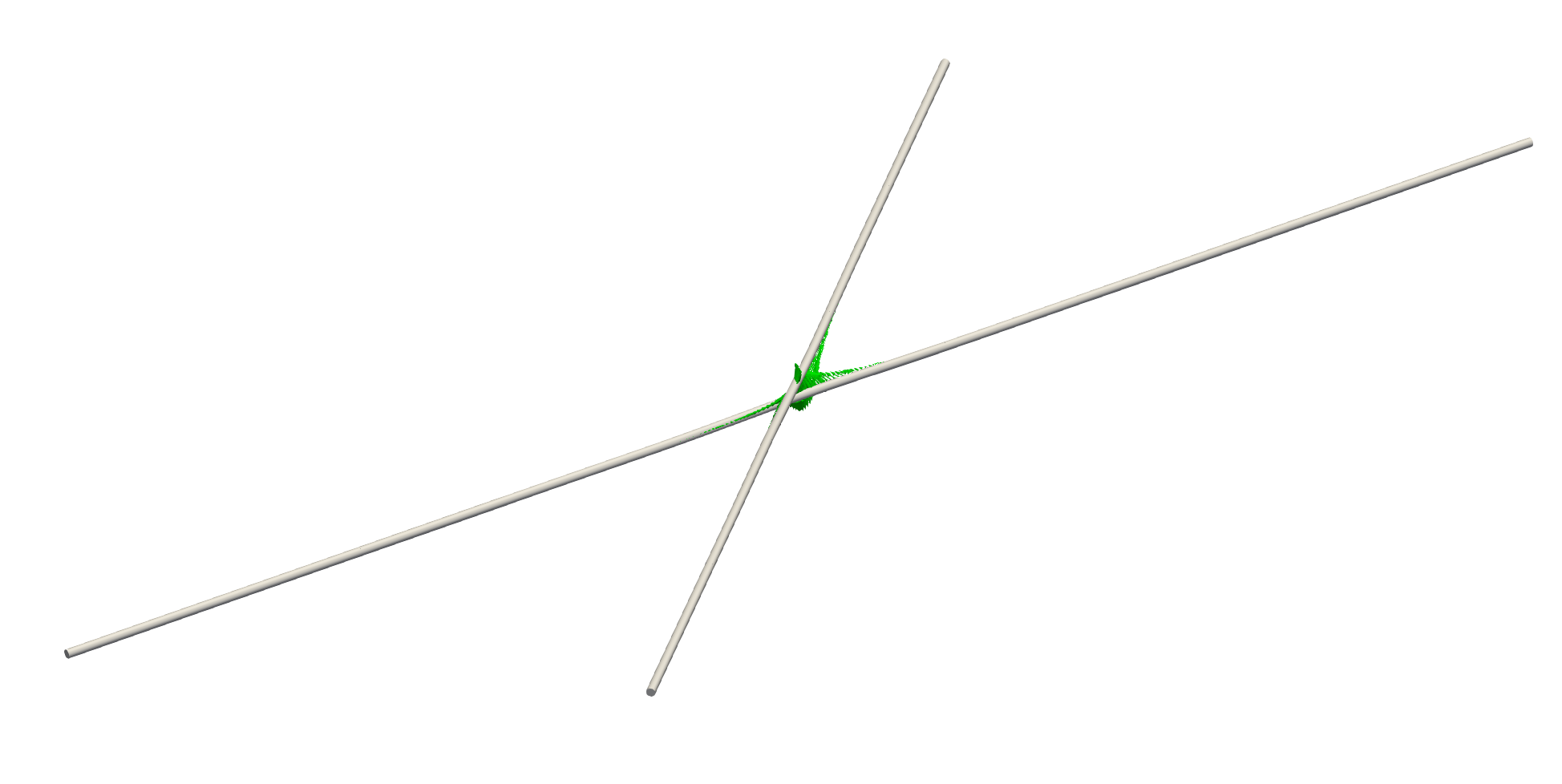}
    \label{fig::twocrossedbeams_elstat_snapintocontact_snapshots_0001}
  }
  \subfigure[time $t=5\times10^{-3}$]{
    \includegraphics[width=0.31\textwidth]{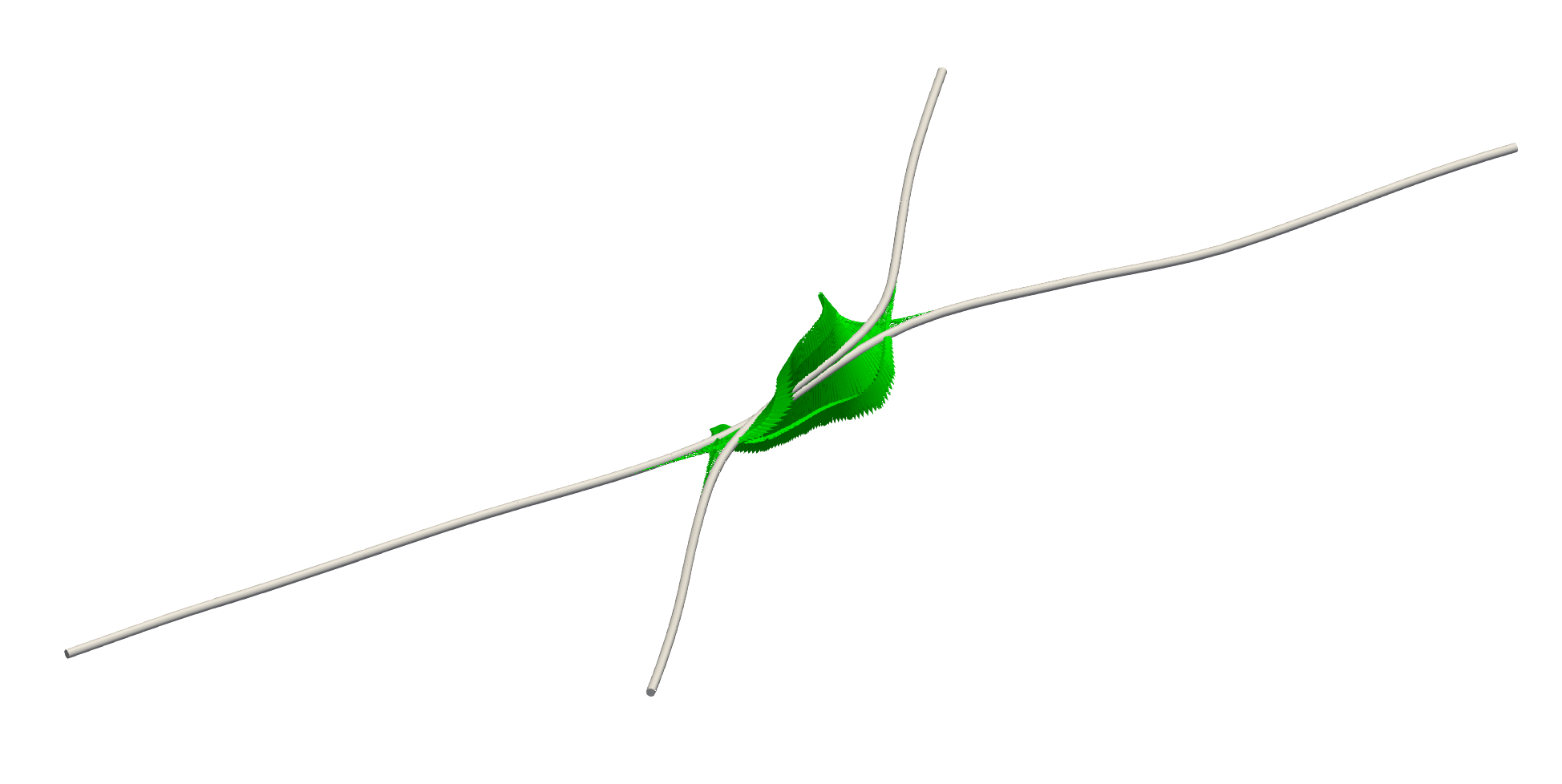}
    \label{fig::twocrossedbeams_elstat_snapintocontact_snapshots_0005}
  }
  \subfigure[time $t=1\times10^{-2}$]{
    \includegraphics[width=0.31\textwidth]{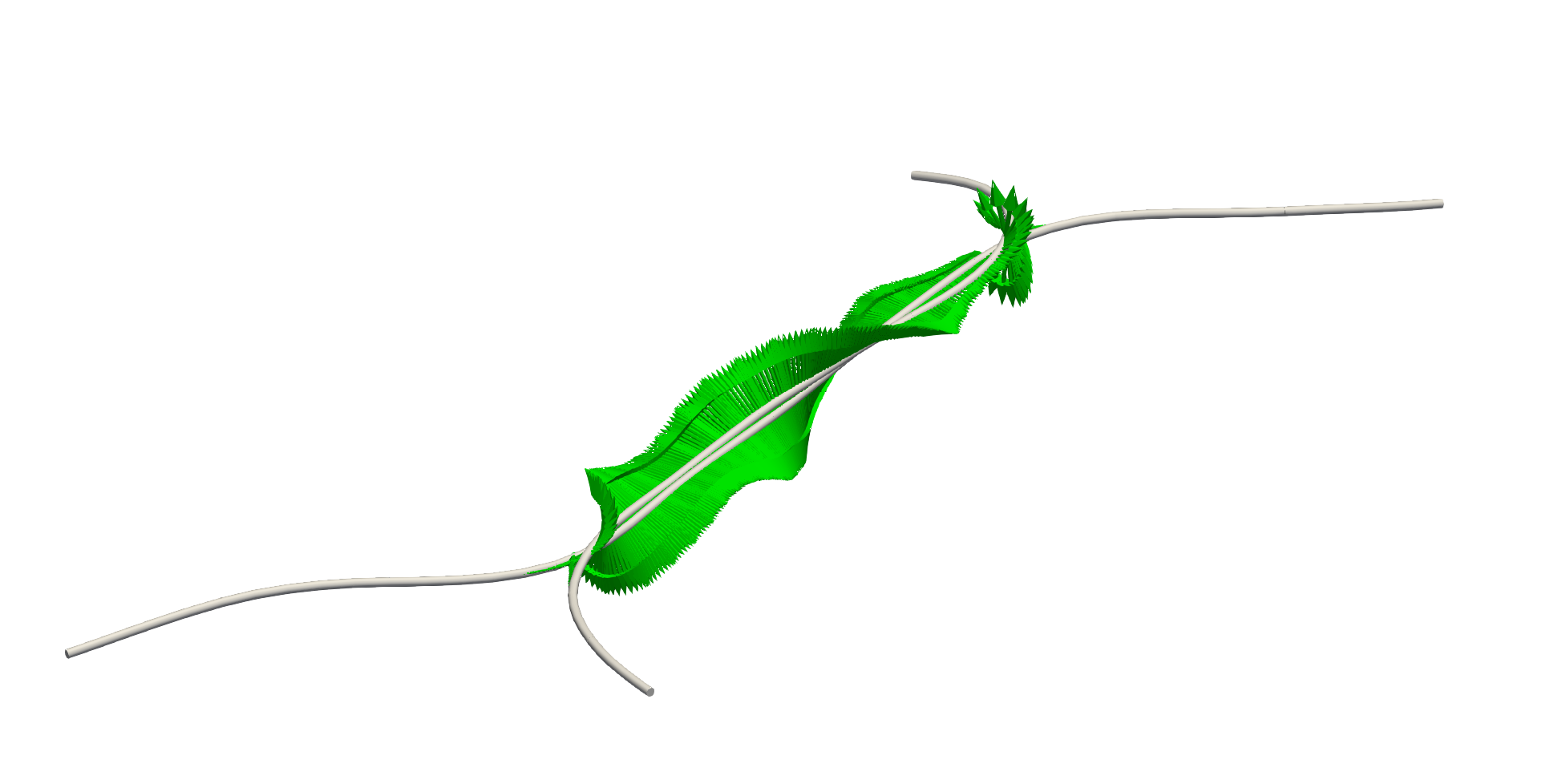}
    \label{fig::twocrossedbeams_elstat_snapintocontact_snapshots_0010}
  }
  \subfigure[time $t=2\times10^{-2}$]{
    \includegraphics[width=0.31\textwidth]{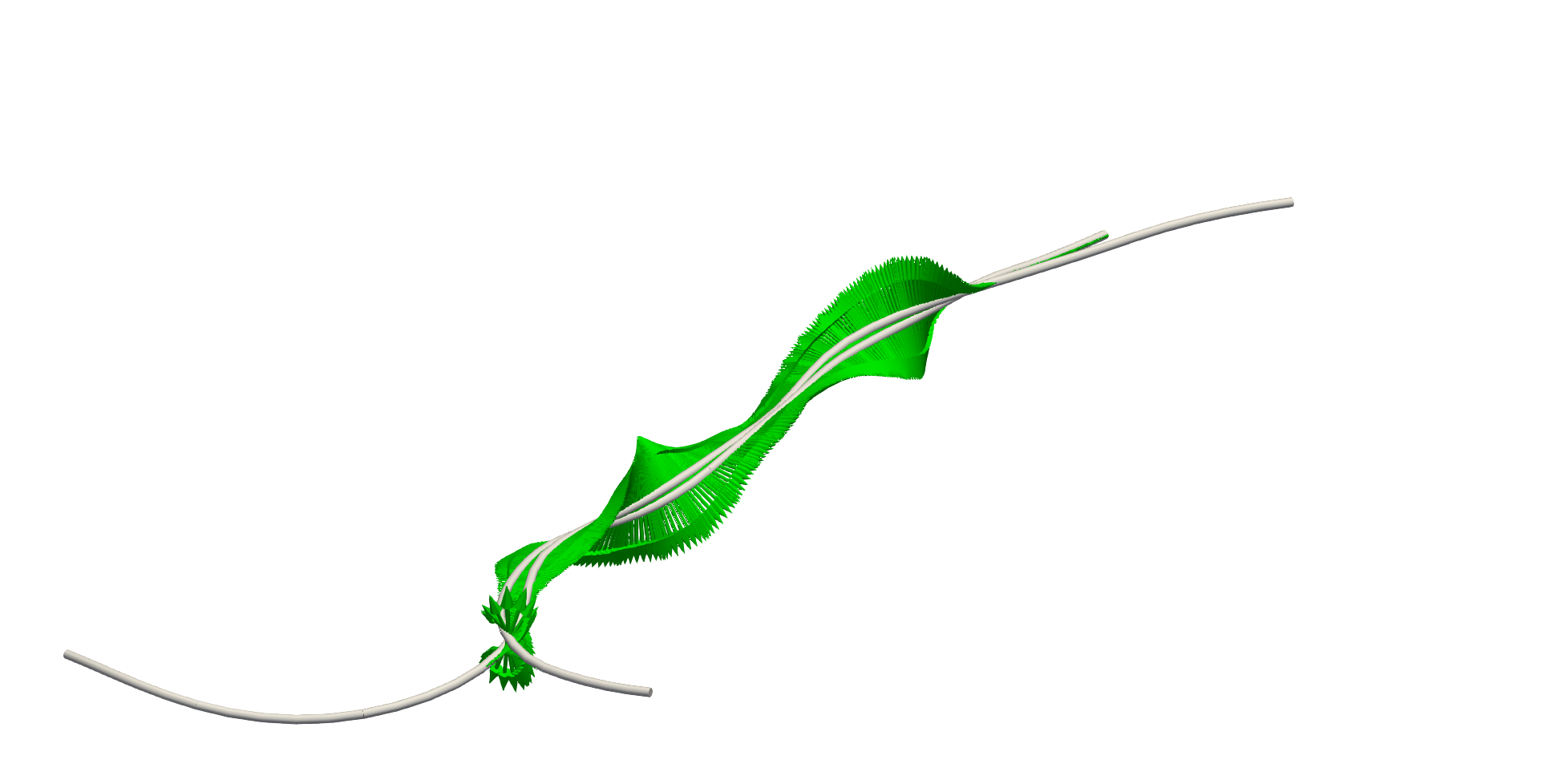}
    \label{fig::twocrossedbeams_elstat_snapintocontact_snapshots_0020}
  }
  \subfigure[time $t=4\times10^{-2}$]{
    \includegraphics[width=0.31\textwidth]{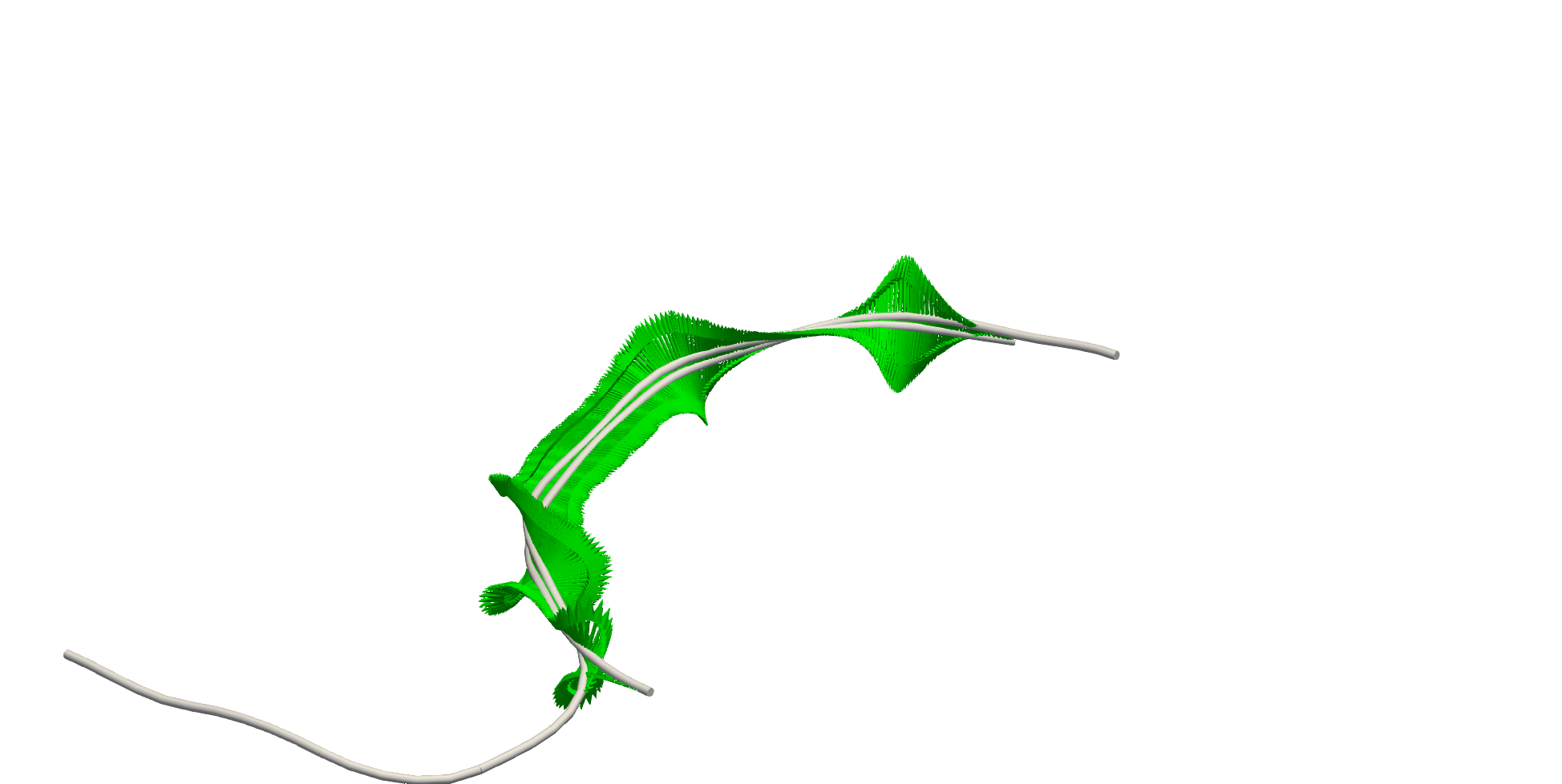}
    \label{fig::twocrossedbeams_elstat_snapintocontact_snapshots_0040}
  }
  \subfigure[time $t=6\times10^{-2}$]{
    \includegraphics[width=0.31\textwidth]{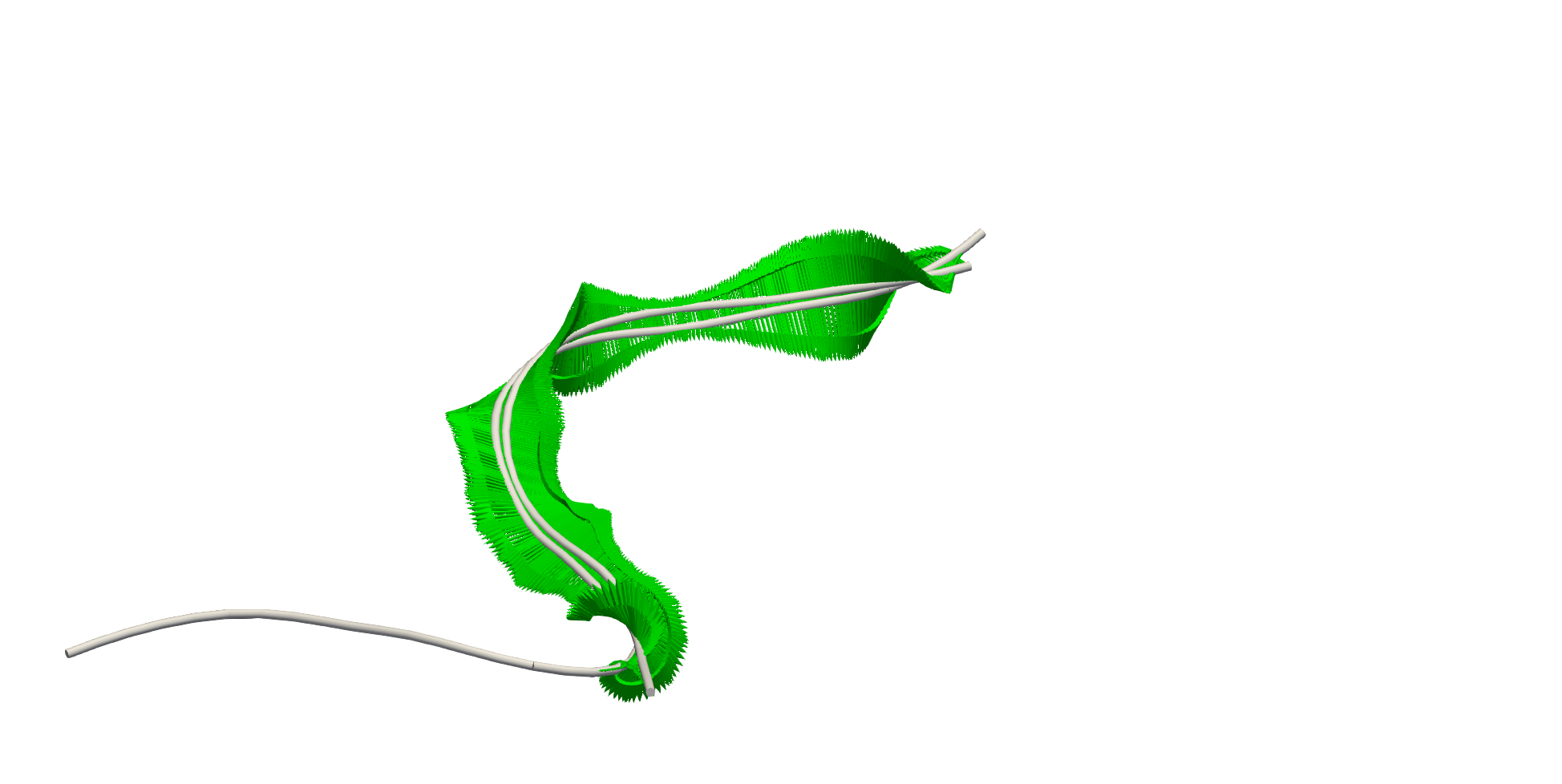}
    \label{fig::twocrossedbeams_elstat_snapintocontact_snapshots_0060}
  }
  \subfigure[time $t=8\times10^{-2}$]{
    \includegraphics[width=0.31\textwidth]{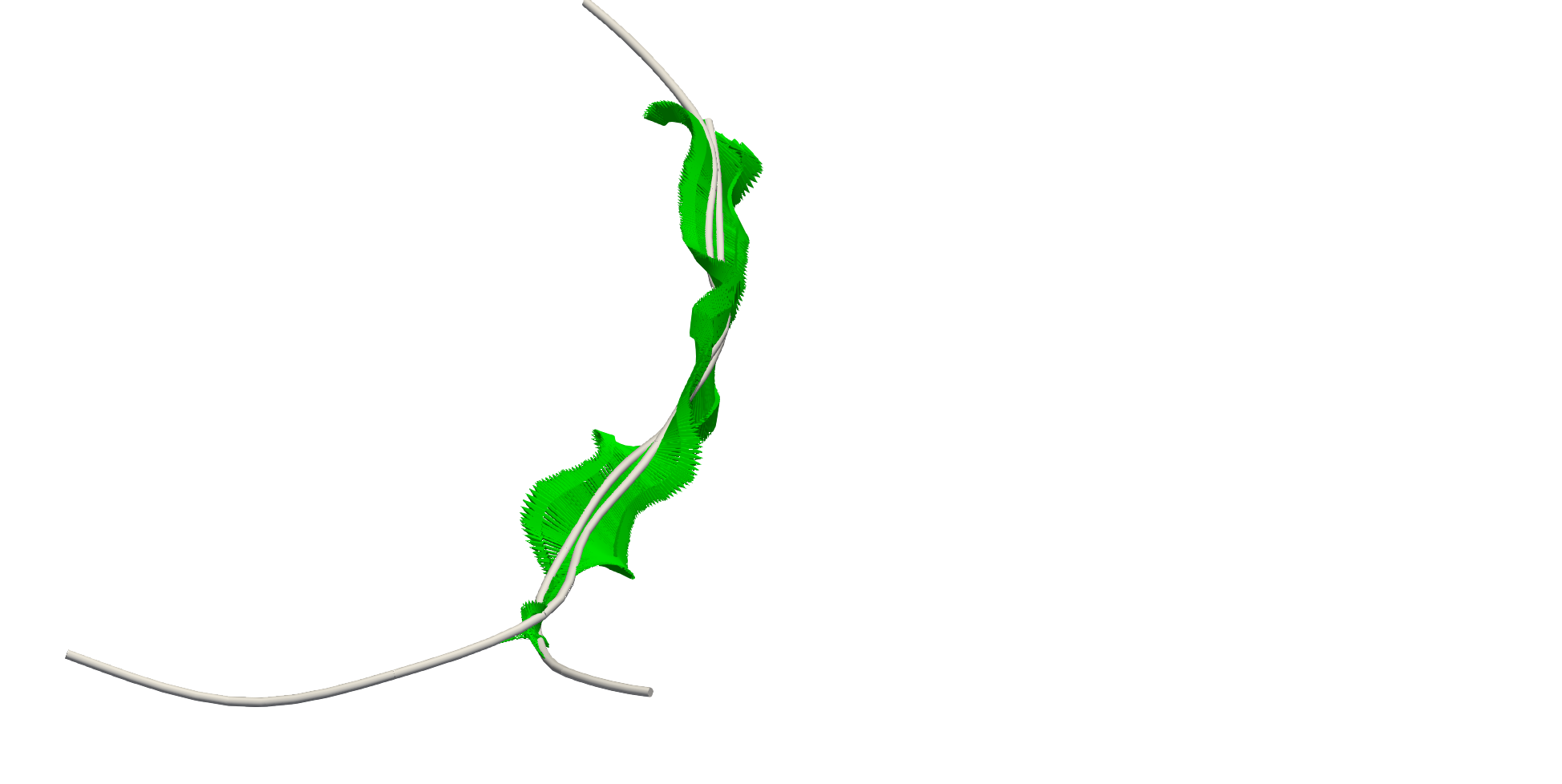}
    \label{fig::twocrossedbeams_elstat_snapintocontact_snapshots_0080}
  }
  \subfigure[time $t=1\times10^{-1}$]{
    \includegraphics[width=0.31\textwidth]{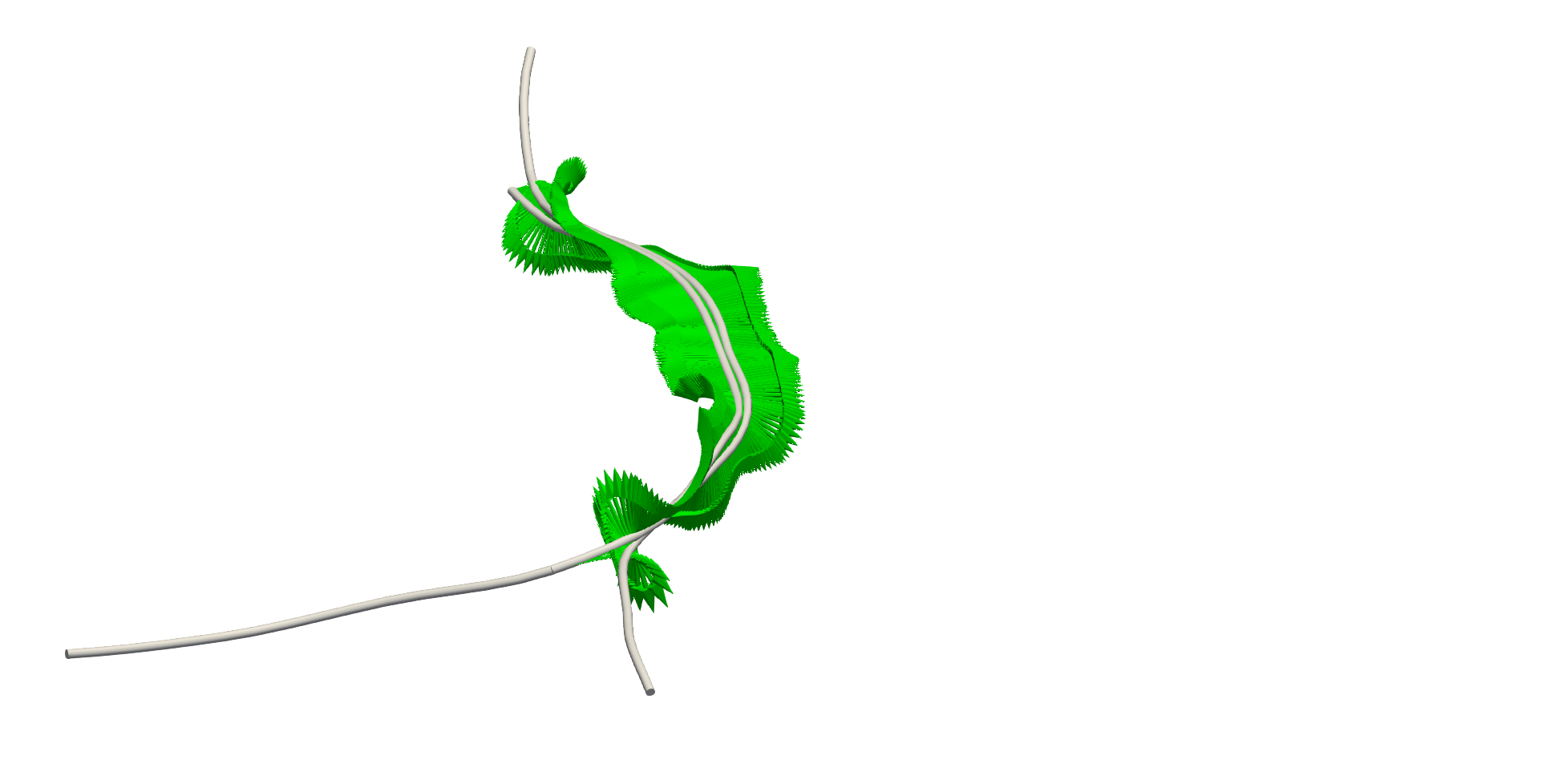}
    \label{fig::twocrossedbeams_elstat_snapintocontact_snapshots_0100}
  }
  \caption{Sequence of simulation snapshots. Electrostatic forces acting on both fibers shown in green.}
  \label{fig::twocrossedbeams_elstat_snapintocontact_snapshots}
\end{figure}

In terms of temporal discretization, we apply the Generalized-Alpha scheme for Lie groups as proposed in~\cite{bruels2010} and set the spectral radius at infinite frequencies to~$\rho_\infty=0.9$ for small numerical damping.
A small time step size of~$\Delta t=10^{-5}$ is applied to account for the highly dynamic behavior of this system.
\figref{fig::twocrossedbeams_elstat_snapintocontact_snapshots} shows a sequence of simulation snapshots where the electrostatic forces on both fibers are visualized as green arrows.
We observe a large variety of mutual orientations of the two fibers and a strong coupling of adhesive, repulsive and elastic forces that demonstrate the effectiveness and robustness of the proposed SSIP approach.
Most importantly, we see that the total system energy is preserved with very little deviation of~$\pm 2 \%$ as shown in~\figref{fig::twocrossedbeams_elstat_snapintocontact_energy_over_time}.
Note that the negative energy values result from defining the zero level of the interaction potential at infinite separation as described in~\secref{sec::molecular_interactions_classification_general_informations}.
Based on this numerical example, we can thus conclude that the novel SSIP approach proves to be effective as well as robust in a highly dynamic example with arbitrary mutual orientations of the fibers in three dimensions.

\section{Conclusions and Outlook}\label{sec::summary_outlook}
This contribution proposes the first 3D beam-to-beam interaction model for molecular interactions such as electrostatic, van der Waals (vdW) or repulsive steric forces between curved slender fibers undergoing large deformations.
While the general model is not restricted to a specific beam formulation, in the present work it is combined with the geometrically exact beam theory and discretized via the finite element method.
A direct evaluation of the total interaction potential for general 3D bodies requires the integration of contributions from molecule or charge distributions over the volumes of the interaction partners, leading to a 6D integral (two nested 3D integrals) that has to be solved numerically.
The central idea of our novel approach is to formulate reduced interaction laws for the resultant interaction potential between a pair of cross-sections of two slender fibers such that only the two 1D integrals along the fibers' length directions have to be solved numerically.
This section-to-section interaction potential (SSIP) approach therefore reduces the dimensionality of the required numerical integration from 6D to 2D and yields a significant gain in efficiency, which only enables the simulation of relevant time and length scales for many practical applications.
Being the key to this SSIP approach, the analytical derivation of the specific SSIP laws is based on careful consideration of the characteristics of the different types of molecular interactions, most importantly their point pair potential law and the range of the interaction.
In a first step, the most generic form of the SSIP law, which is valid for arbitrary shapes of cross-sections and inhomogeneous distribution of interacting points (e.\,g.~atoms or charges) within the cross-sections has been presented before the assumptions and resulting simplifications for the specific SSIP laws have been discussed in detail.
For the practically relevant case of homogeneous, disk-shaped cross-sections, specific, ready-to-use SSIP laws for short-range volume interactions such as vdW or steric interactions and for long-range surface interactions such as Coulomb interactions have been proposed.
We would like to stress that postulating the general structure of the SSIP law and fitting the free parameters to e.\,g.~experimental data is one of the promising alternatives to the strategy of analytical derivation of the SSIP law as applied in this article.
It is also important to emphasize that the general SSIP approach can be seamlessly integrated into an existing finite element framework for solid mechanics.
In particular, it does neither depend on any specific beam formulation nor the applied spatial discretization scheme and in the context of the present work, we have exemplarily used it with geometrically exact Kirchhoff-Love as well as Simo-Reissner type beam finite elements.
Likewise, it is independent of the temporal discretization and we have used it along with static and (Lie group) Generalized-Alpha time stepping schemes as well as inside a Brownian dynamics framework.

The accuracy of the proposed SSIP laws as well as the general SSIP approach has been studied in a thorough quantitative analysis using analytical as well as numerical reference solutions for the case of vdW as well as electrostatic interactions.
We find that a very high level of accuracy is achieved for long-range interactions such as electrostatics both for the entire range of separations as well as all mutual angles of the fibers from parallel to perpendicular.
In the case of short-range interactions, however, the derived SSIP law without cross-section orientation information slightly overestimates the asymptotic power-law exponent of the interaction potential over separation.
As a pragmatic solution, a calibration of the simple SSIP law has been proposed to fit a given reference solution in the small yet decisive range of separations around the equilibrium distance of the Lennard-Jones (LJ) potential.
In the authors' recent contribution~\cite{GrillPeelingPulloff}, this strategy led to very good agreement in the force response on the system level.
While this accuracy might already be sufficient for certain real world applications, our future research work will focus on the derivation of enhanced interaction laws including information about the cross-section orientation with the aim to achieve higher accuracy and the exact asymptotic scaling behavior.

The presented set of numerical examples finally demonstrates the effectiveness and robustness of the SSIP approach to model steric repulsion, electrostatic or vdW adhesion.
Several important aspects such as the influence of the Gauss integration error and the spatial discretization error as well as local and global equilibrium of forces and conservation of energy are studied in these simulations, including quasi-static and dynamic scenarios as well as arbitrary mutual orientations and separations of the interacting fibers.
In order to remedy the characteristic singularity of inverse power interaction laws in the limit of zero separation, we have proposed a numerical regularization of the LJ SSIP law, which leads to a significant increase in robustness and efficiency, saving a factor of five in the number of nonlinear iterations while yielding identical results.

\appendix
\section{Examples for the Derivation and Analysis of the Two-Body Interaction Potential and Force Laws for Parallel Disks and Cylinders}\label{sec::formulae_two_body_LJ_interaction}
The aim of this appendix is to present the mathematical background of analytical solutions for two-body interaction potential as well as force laws.
Generally, the strategy of pairwise summation, i.\,e., integration of a point pair potential, is applied.
See~\secref{sec::theory_molecular_interactions_twobody_vdW} for a discussion of the applicability of this approach.
Exemplarily, we consider the interaction between two parallel disks and two parallel cylinders since these scenarios proved to be most important throughout the derivation of SSIP laws as well as their verification in~\secref{sec::method_application_to_specific_types_of_interactions} and~\ref{sec::verification_methods}, respectively.
In addition, we are interested in the total LJ interaction potential and force law in the limit of small separations, because the regularization proposed in~\secref{sec::regularization} is based on these theoretical considerations.
Finally, also the equilibrium spacing~$g_\text{LJ,eq,cyl$\parallel$cyl}$ of two infinitely long cylinders interacting via the LJ potential will be derived and has proven helpful in order to choose an almost stress-free initial configuration of two deformable, straight fibers e.\,g.~in the authors' recent contribution~\cite{GrillPeelingPulloff} studying the peeling and pull-off behavior.

\subsection{A generic interaction potential described by an inverse power law}\label{sec::derivation_pot_ia_two_body_powerlaw}
Instead of $\Phi_\text{vdW}(r)$ from \eqref{eq::pot_ia_vdW_pointpair} or any other particular interaction type, here, we rather use the more general power law~$\Phi_m(r)=k_m \, r^{-m}$ for the point pair potential.
As noted already in \cite{langbein1972}, this does not introduce any additional complexity in the derivations and the solutions can directly be used for other exponents~$m$.
We will make use of this fact when considering LJ interaction between two disks and two cylinders analytically in~\ref{sec::derivation_pot_force_ia_LJ_disks} and \ref{sec::derivation_pot_force_ia_LJ_cylinders}, respectively.
These findings are to be used in the context of deriving a proper regularization of the potential laws in \secref{sec::regularization}.

\subsubsection{Disk-disk interaction}\label{sec::derivation_pot_ia_powerlaw_disks}
The following refers to the analytical solutions for the disk-disk vdW interaction potential from literature that is summarized in~\secref{sec::theory_molecular_interactions_twobody_vdW}.
Let us first state the underlying mathematical problem.
We would like to find an analytical solution for the required 4D integral~$C_m$ over the circular area of each disk
\begin{equation}\label{eq::cross_sec_integral_langbein}
  C_m \defvariable \iint_{A_1,A_2} \Phi_m(r) \dd A_2 \dd A_1 \qquad \text{with} \quad \Phi_m(r)=k_m \, r^{-m}
\end{equation}
in order to arrive at the disk-disk interaction potential
\begin{equation}
  \tilde{\tilde{\pi}}_m = \rho_1 \rho_2 \, C_m.
\end{equation}

\paragraph{Details on 2(a) in~\secref{sec::theory_molecular_interactions_twobody_vdW}: The regime of large separations}$\,$\\
For the limit of large separations $g \gg R_1,R_2$, the solution is quite straightforward and can be explained in simple words as follows.
The distance of any point in a disk to its center is of order $\bigO(R)$ and thus much smaller than the disks' surface-to-surface separation~$g$:
\begin{equation}
 \tilde{r}_{1/2} = \bigO(R_{1/2}) \ll \bigO(g)
\end{equation}
\begin{figure}[htp]%
  \centering
    \includegraphics[width=0.35\textwidth]{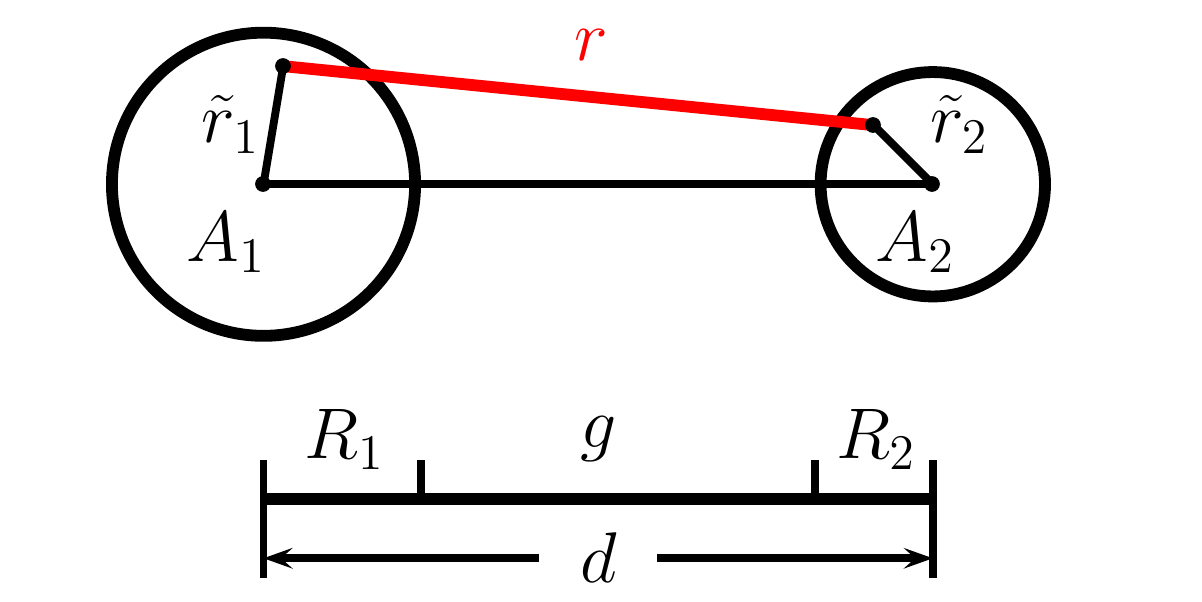}
  \caption{Two circular cross-sections, i.\,e.~disks in parallel alignment}
  \label{fig::cross_sec_parallel_integral}
\end{figure}
Figure \ref{fig::cross_sec_parallel_integral} illustrates the introduced geometrical quantities.
The distance $r$ between any two points $\vx_1$ in disk~$1$ and $\vx_2$ in disk~$2$ may therefore be approximated by the inter-axis distance $d=g+R_1+R_2$:
\begin{equation}
 r = \norm{\vx_1 - \vx_2} = \norm{ \vr_1 + \tilde{\vr}_1 - \vr_2 - \tilde{\vr}_2 } \approx \norm{\vr_1 - \vr_2} = d
\end{equation}
Double integration over both disks is hence equivalent to a multiplication with the disks' areas~$A_1,A_2$
\begin{equation}
  C_{m\text{,ls}} \approx A_1 A_2 \, \Phi_m(r=d)
\end{equation}
and finally we end up with the sought-after expression for the general disk-disk interaction potential in the limit of large separations
\begin{equation}\label{eq::approx_large_sep}
  \tilde{\tilde{\pi}}_{m,\text{disk$\parallel$disk,ls}} \approx \rho_1 \rho_2 \, A_1 A_2 \, \Phi_m(r=d).
\end{equation}
Note that this approximation is valid for arbitrary pair interaction functions $\Phi(r)$.
Moreover, this solution does not even depend on the parallel orientation of the disks.
It is valid for all mutual angles of the disks which is important because we will apply it to arbitrary configurations of deflected beams.
For the special case of parallel disks, this result can alternatively be obtained by the sound mathematical derivation of \cite[eq. (10)]{langbein1972}.
The leading term of his hypergeometric series is identical to the right hand side of equation \eqref{eq::approx_large_sep}.

\paragraph{Details on 1(a) in~\secref{sec::theory_molecular_interactions_twobody_vdW}: The regime of small separations}$\,$\\
Now, we consider the limit of small separations $g \ll R_1,R_2$.
The problem has been studied by Langbein \cite{langbein1972} in the context of vdW attraction of rigid cylinders, rods or fibers.
In the following, we will briefly present the central mathematical concept of his derivations.
\begin{figure}[htpb]%
  \centering
    \includegraphics[width=0.5\textwidth]{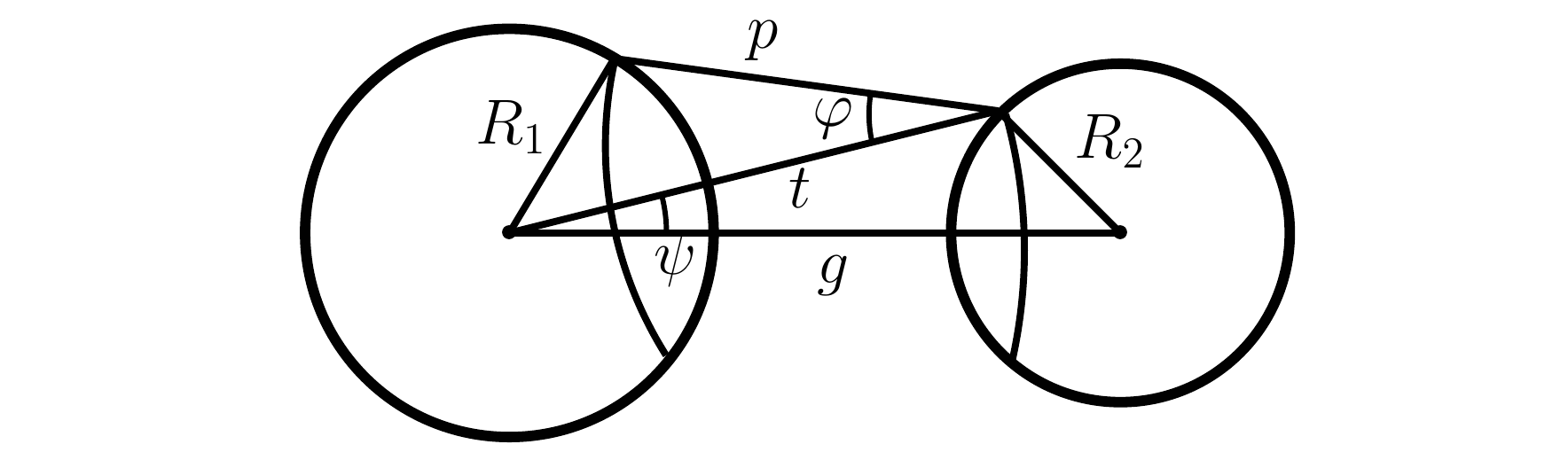}
  \caption{Integration over the cross-sections at small separations, figure taken from \cite{langbein1972} with adapted notation.}
  \label{fig::cross_sec_intgral_langbein}
\end{figure}%
The basic idea is to choose a favorable set of integration variables $p,t,\varphi,\psi$ as shown in \figref{fig::cross_sec_intgral_langbein}.
In this way, the four dimensional integral \eqref{eq::cross_sec_integral_langbein} can be reduced to a double integral because the integrand $\Phi_m(p)$ does not depend on the angles $\varphi$ and $\psi$:
\begin{align}
 C_m &= \int_{A_1} \int_{A_2} \Phi_m(p) \dd A_2 \dd A_1 = \int_p \int_t \int_{\varphi} \int_{\psi} \Phi_m(p) \dd \psi \dd \varphi \dd t \dd p \nonumber \\ & \qquad \qquad \qquad \qquad \quad=\int_p \int_t \Phi_m(p) \, 2 p \varphi(p,t) \,2 t \psi(p,t) \dd t \dd p \label{eq::dimred_langbein} \\
 & \text{where} \quad  \cos (\varphi) = \frac{p^2 + t^2 - R_1^2}{2pt}, \quad \cos(\psi) = \frac{t^2 + d^2 - R_2^2}{2td}, \quad d=g+R_1+R_2 \nonumber
\end{align}
For a general potential law $\Phi_m(r=p) = k_m p^{-m}$, this reads
\begin{equation}
 C_m = 4 k_m \int_p \int_t p^{-m+1} \varphi t \psi \dd t \dd p
\end{equation}
Making use of $g \ll R_1,R_2$ and introducing reduced variables $\bar{p}= p/g$ and $\bar{t}= t/g$ leads to
\begin{align}
 C_{m\text{,ss}} &= 4 k_m \sqrt{ \frac{2 R_1 R_2}{R_1+R_2} } \, \int_{g}^{g+2R_1+2R_2}  p^{-m+1} \int_{g}^{p}  \sqrt{t-g} \, \arccos \left( \frac{t}{p} \right) \dd t \dd p \\
 &= 4 k_m \sqrt{ \frac{2 R_1 R_2}{R_1+R_2} } \, g^{-m+7/2} \, \int_{1}^{\infty}  \bar{p}^{-m+1} \int_{1}^{\bar{p}}  \sqrt{\bar{t}-1} \, \arccos \left( \frac{\bar{t}}{\bar{p}} \right) \dd \bar{t} \dd \bar{p}
\end{align}
Another substitution of variables $x=\bar{t}/\bar{p}$ and interchanging the order of integration finally yields the solution%
\footnote{Note that in the original article \cite{langbein1972}, the final form of $C_m$ (eq.~(15) on p.~65) seems to be incorrect.
A comparison with \cite[p. 172]{parsegian2005} for the case of vdW potential with~$m=6$ confirms the solution presented here.
Additionally, this solution is verified by means of numerical quadrature in \secref{sec::verif_approx} (cf.~\figref{fig::vdW_pot_over_gap_disks_parallel}).
}\\
\begin{equation}\label{eq::langbein_Cm_smallsep}
 C_{m\text{,ss}} = g^{-m+\tfrac{7}{2}} \quad \frac{2 k_m \pi}{(m-2)^2} \quad \sqrt{ \frac{2 R_1 R_2}{R_1+R_2} } \quad \frac{\Gamma(m-\tfrac{7}{2}) \,\Gamma(\tfrac{m-1}{2})}{\Gamma(m-2) \, \Gamma(\tfrac{m}{2}-1)} \qquad \text{for} \quad m>\tfrac{7}{2}
\end{equation}
Here, $\Gamma$ denotes the gamma function which is defined by $\Gamma(z) = \int_0^{\infty} w^{z-1} e^{-w} \dd w$.

Multiplication with the particle densities finally results in the sought-after general disk-disk interaction potential for the regime of small separations
\begin{equation}\label{eq::approx_small_sep}
  \tilde{ \tilde{ \pi}}_{m,\text{disk$\parallel$disk,ss}} = \rho_1 \rho_2 \, C_{m\text{,ss}} \qquad \text{for} \quad m>\tfrac{7}{2}
\end{equation}
that can be further specified by means of~$m=6$ and~$k_6 = -C_\text{vdW}$ to end up with~$\tilde{ \tilde{ \pi}}_\text{vdW,disk$\parallel$disk,ss}$ as in \eqref{eq::pot_ia_vdW_disk_disk_parallel_smallseparation}.

\paragraph{Remarks}
\begin{enumerate}
  \item Note that this solution is valid for exponents $m>7/2$ only. This is in contrast to the approximation for large separations~\eqref{eq::approx_large_sep} which is valid for arbitrary forms of the pair interaction potential~$\Phi(r)$.
  \item Note however the conceptual similarity of this expression to the one valid for the limit of large separations \eqref{eq::approx_large_sep}.
Here, we also find a power law, however in the surface-to-surface distance $g$ instead of the inter-axis distance $d$ and with a different exponent.
\end{enumerate}

\subsubsection{Cylinder-cylinder interaction}\label{sec::derivation_pot_ia_powerlaw_cylinders}
Considering the case of two parallel cylinders, we are interested in the length-specific interaction potential
\begin{equation}\label{eq::def_length_specific_pot_powerlaw}
  \tilde{\pi}_{m,\text{cyl$\parallel$cyl}} = \lim_{l_1 \to \infty} \frac{1}{l_1} \, \int_{-l_1/2}^{l_1/2} \int_{-\infty}^\infty \iint_{A_1,A_2} \rho_1 \rho_2 \, \Phi_m(r) \dd A_2 \dd A_1 \dd s_2 \dd s_1  \qquad \text{with} \quad \Phi_m(r)=k_m \, r^{-m}.
\end{equation}
The integral over~$s_1=-l_1/2 \ldots l_1/2$ yields a factor of~$l_1$ since the integrand is constant along~$s_1$ and thus immediately cancels with the normalization factor~$1/l_1$.

Exemplarily, we want to discuss the more interesting and challenging regime of small separations here.
Following \cite[p.63]{langbein1972}, one can interchange the order of integration, solve the integral over the infinitely long cylinder length analytically in a first step, and then make use of the generic solution for~$C_{m,\text{ss}}$ from \eqref{eq::langbein_Cm_smallsep}, but this time with reduced exponent~$m-1$, to end up with
\begin{align}
  \tilde{\pi}_{m,\text{cyl$\parallel$cyl},ss} &= \iint_{A_1,A_2} \int_{-\infty}^\infty \rho_1 \rho_2 \, \Phi_m(r) \dd s_2 \dd s_1 \dd A_2 \dd A_1 \\
  &= \frac{3 \pi}{8} \, \rho_1 \rho_2 \, k_m \, \frac{C_{m-1,\text{ss}}}{k_{m-1}}. \label{eq::pot_ia_LJ_cylinders_parallel_smallsep}
\end{align}
Plugging in~$m=6$ for vdW interaction directly yields the two-body interaction potential per unit length for two parallel cylinders in the regime of small separations~$\tilde \pi_\text{vdW,cyl$\parallel$cyl,ss}$ as stated in eq.~\eqref{eq::pot_ia_vdW_cyl_cyl_parallel_smallseparation}.
This generic expression~\eqref{eq::pot_ia_LJ_cylinders_parallel_smallsep} will be exploited when deriving the total LJ interaction law in \ref{sec::derivation_pot_force_ia_LJ_cylinders}.

\subsection{Lennard-Jones force laws in the regime of small separations}\label{sec::derivation_pot_force_ia_LJ}
As compared to the preceding sections, we now want to turn to the LJ interaction consisting of two power law contributions, one adhesive and one repulsive, respectively.
Our motivation is to study the characteristics of the resulting, superposed force laws for disk-disk as well as cylinder-cylinder interactions by means of theoretical analysis of the analytical expressions.
These findings shall prove valuable when deriving an effective yet accurate regularization of the LJ potential law for the limit of zero separation in \secref{sec::regularization}.
We therefore focus on the regime of small separations throughout this section.

Coming from the expressions for the two-body interaction potential~$\tilde{\tilde{\pi}}_{m,\text{disk$\parallel$disk,ss}}$ and~$\tilde{\pi}_{m,\text{cyl$\parallel$cyl,ss}}$ derived for a generic point pair potential~$\Phi_m$ in \ref{sec::derivation_pot_ia_two_body_powerlaw}, we will now sum the adhesive contribution~$m=6$ and the repulsive contribution~$m=12$ and differentiate once to arrive at the desired LJ force laws.

\subsubsection{Disk-disk interaction}\label{sec::derivation_pot_force_ia_LJ_disks}
As outlined above, we make use of~\eqref{eq::approx_small_sep} for both parts of the LJ interaction and immediately obtain
\begin{equation}
  \tilde{\tilde{\pi}}_\text{LJ,disk$\parallel$disk,ss} = \tilde{k}_6 \, g^{-\frac{5}{2}} + \tilde{k}_{12} \, g^{-\frac{17}{2}}
\end{equation}
where the following abbreviations for the constant prefactors have been introduced:
\begin{align}
  \tilde{k}_6 \defvariable \frac{\pi}{8} k_6 \rho_1 \rho_2 \sqrt{\frac{2 R_1 R_2}{R_1+R_2}} \, \Gamma^2\left(\frac{5}{2}\right)
  \qquad \text{and} \qquad
  \tilde{k}_{12} \defvariable k_{12} \rho_1 \rho_2 \sqrt{\frac{2 R_1 R_2}{R_1+R_2}} \, \num{5.30e-3}
\end{align}
For later use in the analysis of the force law, let us restate the conversion from one set of parameters~$k_6, k_{12}$ specifying the point pair LJ potential to the other commonly used set~$\Phi_\text{LJ,eq}, r_\text{LJ,eq}$ according to \eqref{eq::pot_ia_LJ_pointpair}:
\begin{equation}
  k_6 = 2 \Phi_\text{LJ,eq} r_\text{LJ,eq}^{6} \qquad \text{and} \qquad k_{12} = - \Phi_\text{LJ,eq} r_\text{LJ,eq}^{12}
\end{equation}
Differentiation with respect to the separation yields the disk-disk LJ force law
\begin{equation}\label{eq::force_LJ_disks_parallel_smallsep}
  \tilde{\tilde{f}}_\text{LJ,disk$\parallel$disk,ss} = -\diff{ \tilde{\tilde{\pi}}_\text{ LJ,disk$\parallel$disk,ss } }{ g } =
  \frac{5}{2} \, \tilde{k}_6 \, g^{-\frac{7}{2}} + \frac{17}{2} \, \tilde{k}_{12} \, g^{-\frac{19}{2}}.
\end{equation}
See~\secref{sec::regularization} for a plot of the function.
This expression allows us to determine some very interesting, characteristic quantities like the equilibrium spacing~$g_\text{LJ,eq,disk$\parallel$disk}$, i.\,e., the distance where the force vanishes:
\begin{equation}\label{eq::equilibrium_spacing_LJ_disks_parallel_smallsep}
  g_\text{LJ,eq,disk$\parallel$disk} = \left( - \frac{17}{5} \, \frac{\tilde k_{12}}{\tilde k_6} \right)^\frac{1}{6} \approx \num{0.653513} \, r_\text{LJ,eq}.
\end{equation}
Due to the fact, that repulsive contributions from proximate point pairs decay faster than the adhesive contributions, we obtain a smaller equilibrium spacing as compared to the scenario of a point pair.
Another differentiation allows us to determine the value of the force minimum, i.\,e., the maximal adhesive force, and the corresponding separation
\begin{align}
  \tilde{\tilde{f}}_\text{LJ,disk$\parallel$disk,min} &\approx \num{0.904115} \, \rho_1 \rho_2 \sqrt{\frac{2 R_1 R_2}{R_1+R_2}} \, r_\text{LJ,eq}^{\frac{5}{2}} \, \Phi_\text{LJ,eq}\\
  g_{\tilde{\tilde{f}}_\text{LJ,disk$\parallel$disk,min}} &= \left( - \frac{323}{35} \frac{\tilde k_{12}}{\tilde k_6} \right)^\frac{1}{6}
  \approx \num{0.7718448} \, r_\text{LJ,eq}
  \approx \num{1.18107} \, g_\text{LJ,eq,disk$\parallel$disk}. \label{eq::gap_force_min_LJ_disks_parallel}
\end{align}
These quantities turn out to be decisive for the choice of a regularized, i.\,e., altered force law that is to be used instead of the original one in order to cure the numerical problems that come with the singularity at zero separation~$g=0$.

In summary, we have found an analytical, closed-form expression for the disk-disk LJ force law~\eqref{eq::force_LJ_disks_parallel_smallsep}, valid in the regime of small separations and for parallel disks.
By means of elementary algebra, we were thus able to determine analytical expressions for the characteristic equilibrium spacing as well as value and spacing of the force minimum.

\subsubsection{Cylinder-cylinder interaction}\label{sec::derivation_pot_force_ia_LJ_cylinders}
As in the previous section (\ref{sec::derivation_pot_ia_powerlaw_cylinders}), we want to restrict ourselves to parallel, infinite cylinders and consider the length-specific interaction potential as well as force law.
Again, starting from the expression for a generic interaction potential~\eqref{eq::pot_ia_LJ_cylinders_parallel_smallsep}, superposition yields
\begin{equation}
  \tilde{\pi}_\text{LJ,cyl$\parallel$cyl,ss} = \tilde{k}_{\text{cyl},6} \, g^{-\frac{3}{2}} + \tilde{k}_{\text{cyl},12} \, g^{-\frac{15}{2}}
\end{equation}
where the following abbreviations have been introduced:
\begin{align}
  \tilde{k}_{\text{cyl},6} \defvariable \frac{\pi^2}{24} k_6 \rho_1 \rho_2 \sqrt{\frac{2 R_1 R_2}{R_1+R_2}}
  \qquad \text{and} \qquad
  \tilde{k}_{\text{cyl},12} \defvariable \num{5.81868e-4} \, k_{12} \pi^2 \rho_1 \rho_2  \sqrt{\frac{2 R_1 R_2}{R_1+R_2}}
\end{align}
Differentiation with respect to the separation yields the cylinder-cylinder LJ force law
\begin{equation}\label{eq::force_LJ_cylinders_parallel_smallsep}
  \tilde{f}_\text{LJ,cyl$\parallel$cyl,ss} = -\diff{ \tilde{\pi}_\text{LJ,cyl$\parallel$cyl,ss} }{ g } =
  \frac{3}{2} \, \tilde{k}_{\text{cyl},6} \, g^{-\frac{5}{2}} + \frac{15}{2} \, \tilde{k}_{\text{cyl},12} \, g^{-\frac{17}{2}}.
\end{equation}
that shall be further analyzed in the following.
To begin with, the equilibrium spacing for two parallel cylinders interacting via a LJ potential can be derived as
\begin{equation}\label{eq::equilibrium_spacing_LJ_cylinders_parallel_smallsep}
  g_\text{LJ,eq,cyl$\parallel$cyl} = \left( - 5 \, \frac{\tilde k_{\text{cyl},12}}{\tilde k_{\text{cyl},6}} \right)^\frac{1}{6} \approx \num{0.57169} \, r_\text{LJ,eq}.
\end{equation}
This is an extremely interesting and important result, since it leads the way to the non-trivial stress-free configuration of two flexible, initially straight fibers.
We make use of this knowledge e.\,g.~in~\cite{GrillPeelingPulloff}.
Again, since the repulsive contribution of proximate point pairs decays faster than the adhesive contribution, this equilibrium spacing is smaller than~$g_\text{LJ,eq,disk$\parallel$disk}$ for the disks, that in turn is smaller than~$r_\text{LJ,eq}$ in the fundamental case of a point pair.
The very same value of~$57\%$ of the point pair equilibrium spacing has already been mentioned as a side note by Langbein \cite[p.~62]{langbein1972}, however, without presenting the detailed, comprehensive derivation.
In addition to the equilibrium spacing, we can again determine and look at the value and location of the force minimum
\begin{align}\label{eq::minimal_force_LJ_cylinders_parallel}
  \tilde{f}_\text{LJ,cyl$\parallel$cyl,min} &\approx \num{2.11634} \, \rho_1 \rho_2 \sqrt{\frac{2 R_1 R_2}{R_1+R_2}} \, r_\text{LJ,eq}^{\frac{7}{2}} \, \Phi_\text{LJ,eq}\\
  g_{\tilde{f}_\text{LJ,cyl$\parallel$cyl,min}} &= \left( - \frac{255}{15} \, \frac{\tilde k_{\text{cyl},12}}{\tilde k_{\text{cyl},6}} \right)^\frac{1}{6}
  \approx \num{0.70104} \, r_\text{LJ,eq}
  \approx \num{1.22625} \, g_\text{LJ,eq,cyl$\parallel$cyl}.
\end{align}
Here, we find that the force minimum, i.\,e., the maximal adhesive force is slightly shifted towards a smaller separation as compared to the disk-disk interaction.
However, expressed in terms of the respective equilibrium spacing~$g_\text{LJ,eq,cyl$\parallel$cyl}$, the value is slightly larger as compared to~$\num{1.18} \, g_\text{LJ,eq,disk$\parallel$disk}$ from~\eqref{eq::gap_force_min_LJ_disks_parallel}.
With these results we conclude the derivation and analysis of LJ force laws in the regime of small separations and summarize the most important results in the following table to serve as a quick access reference.

\subsubsection{Summary}\label{sec::derivation_pot_force_ia_LJ_summary}
The following table gives an overview of some important quantities characterizing the LJ force laws for point-point, parallel disk-disk, and parallel cylinder-cylinder interaction.
\begin{table}[htpb]
  \begin{center}
    \begin{tabular}{|c|c|c|c|}\hline
      &&&\vspace{-1em}\\
      & equilibrium spacing & location of force min. & min.~force value\\
      & $r_\text{LJ,eq}$ / $g_\text{LJ,eq}$ & $r_{f_\text{LJ,min}}$ / $g_{f_\text{LJ,min}}$ & $f_\text{LJ,min}$  / $\tilde{\tilde{f}}_\text{LJ,min}$ / $\tilde f_\text{LJ,min}$\\
      &&&\vspace{-1em}\\\hline
      &&&\vspace{-1em}\\
      point-point & $1 \, [r_\text{LJ,eq}]$ & $\num{1.11} \, [r_\text{LJ,eq}]$ & $\num{2.69} \, \left[\frac{\Phi_\text{LJ,eq}}{r_\text{LJ,eq}}\right]$\\
      &&&\vspace{-1em}\\\hline
      &&&\vspace{-1em}\\
      disk$\parallel$disk & $\num{0.65} \, [r_\text{LJ,eq}]$ & $\num{0.77} \, [r_\text{LJ,eq}]$ & $\num{0.90} \, \left[\rho_1 \rho_2 \sqrt{\frac{2 R_1 R_2}{R_1+R_2}} \, r_\text{LJ,eq}^{\frac{5}{2}} \, \Phi_\text{LJ,eq} \right]$\\
      &&&\vspace{-1em}\\\hline
      &&&\vspace{-1em}\\
      cylinder$\parallel$cylinder & $0.57 \, [r_\text{LJ,eq}]$ & $0.70 \, [r_\text{LJ,eq}]$ & $\num{2.12} \, \left[\rho_1 \rho_2 \sqrt{\frac{2 R_1 R_2}{R_1+R_2}} \, r_\text{LJ,eq}^{\frac{7}{2}} \, \Phi_\text{LJ,eq} \right]$\\\hline
    \end{tabular}
  \end{center}
  \caption{Comparison of characteristic quantities of LJ force laws for a pair of points, parallel disks and parallel cylinders.}
  \label{tab::LJ_force_laws_analysis_comparison}
\end{table}

\section{Linearization of the Virtual Work Contributions from Molecular Interactions}\label{sec::linearization}
Generally, the discrete residual vectors~$\vdr_{\text{ia},j}$ from molecular interactions between two beam elements~$j=1,2$ depend on the primary variables~$\hat \vdx_k$ of both beam elements~$k=1,2$.
Consistent linearization thus yields the following four sub-matrices~$\vdk_{jk}$ to be considered and assembled into the global stiffness matrix, i.\,e., system Jacobian~$\vdK$:
\begin{equation}
  \vdk_{11} \defvariable \diff{ \vdr_{\text{ia},1} }{ {\hat \vdx_1} }, \quad
  \vdk_{12} \defvariable \diff{ \vdr_{\text{ia},1} }{ {\hat \vdx_2} }, \quad
  \vdk_{21} \defvariable \diff{ \vdr_{\text{ia},2} }{ {\hat \vdx_1} }, \quad
  \vdk_{22} \defvariable \diff{ \vdr_{\text{ia},2} }{ {\hat \vdx_2} }
\end{equation}
Note that the linearization with respect to the primary variables~$\hat \vdx_k$ of both interacting beam elements simplifies due to the fact that the residuals~$\vdr_{\text{ia},j}$ do not depend on the cross-section rotations as discussed along with the derivation of the specific SSIP laws in~\secref{sec::method_application_to_specific_types_of_interactions}.
Thus, only the linearization with respect to the centerline degrees of freedom~$\hat \vdd_k$ yields non-zero entries and are therefore presented in the remainder of this section.

\subsection{Short-range volume interactions such as van der Waals and steric repulsion}\label{sec::method_vdW_linearization}
The linearization of the residual contributions with respect to the primary variables~$\hat \vdx$ of both interacting beam elements is directly obtained from differentiation of eq.~\eqref{eq::res_ia_pot_smallsep_ele1} and~\eqref{eq::res_ia_pot_smallsep_ele2}:
\begin{align}
  \vdk_{\text{m,ss},11} &= (m-\tfrac{7}{2}) \int_0^{l_1} \int_0^{l_2} c_\text{m,ss} \left( - d^{-1} \, g^{-m+\tfrac{5}{2}} \, \vdH_1^\text{T} \vdH_1  + \right. & \nonumber \\  & \left. \left(  d^{-3} \, g^{-m+\tfrac{5}{2}} + (m-\tfrac{5}{2}) \, d^{-2} \, g^{-m+\tfrac{3}{2}} \right) \vdH_1^\text{T} \left( \vr_1 - \vr_2 \right) \otimes \left( \vr_1 - \vr_2 \right)^T \vdH_1 \right)  \dd s_2 \dd s_1 & \\
  \vdk_{\text{m,ss},12} &= (m-\tfrac{7}{2}) \int_0^{l_1} \int_0^{l_2} c_\text{m,ss} \left( d^{-1} \, g^{-m+\tfrac{5}{2}} \, \vdH_1^\text{T} \vdH_2  - \right. & \nonumber \\  & \left. \left(  d^{-3} \, g^{-m+\tfrac{5}{2}} + (m-\tfrac{5}{2}) \, d^{-2} \, g^{-m+\tfrac{3}{2}} \right) \vdH_1^\text{T} \left( \vr_1 - \vr_2 \right) \otimes \left( \vr_1 - \vr_2 \right)^T \vdH_2 \right)  \dd s_2 \dd s_1 & \\
  \vdk_{\text{m,ss},21} &= (m-\tfrac{7}{2}) \int_0^{l_1} \int_0^{l_2} c_\text{m,ss} \left( d^{-1} \, g^{-m+\tfrac{5}{2}} \, \vdH_2^\text{T} \vdH_1  - \right. & \nonumber \\  & \left. \left(  d^{-3} \, g^{-m+\tfrac{5}{2}} + (m-\tfrac{5}{2}) \, d^{-2} \, g^{-m+\tfrac{3}{2}} \right) \vdH_2^\text{T} \left( \vr_1 - \vr_2 \right) \otimes \left( \vr_1 - \vr_2 \right)^T \vdH_1 \right)  \dd s_2 \dd s_1 & \\
  \vdk_{\text{m,ss},22} &= (m-\tfrac{7}{2}) \int_0^{l_1} \int_0^{l_2} c_\text{m,ss} \left( - d^{-1} \, g^{-m+\tfrac{5}{2}} \, \vdH_2^\text{T} \vdH_2  + \right. & \nonumber \\  & \left. \left(  d^{-3} \, g^{-m+\tfrac{5}{2}} + (m-\tfrac{5}{2}) \, d^{-2} \, g^{-m+\tfrac{3}{2}} \right) \vdH_2^\text{T} \left( \vr_1 - \vr_2 \right) \otimes \left( \vr_1 - \vr_2 \right)^T \vdH_2 \right)  \dd s_2 \dd s_1. &
\end{align}
See eq.~\eqref{eq::vdW_small_sep_def_constant} for the definition of the constant~$c_\text{m,ss}$ and eq.~\eqref{eq::centerline_discretization} for the definition of the shape function matrices~$\vdH_j$.
Note that the `mixed' matrix products $\vdH_1^\text{T} (\ldots) \vdH_2$ and $\vdH_2^\text{T} (\ldots) \vdH_1$ lead to off-diagonal entries in the tangent stiffness matrix of the system which couple the corresponding degrees of freedom.
This is reasonable and necessary because these couplings represent the interaction between the respective bodies.

\subsection{Long-range surface interactions such as electrostatics}\label{sec::method_elstat_linearization}
In analogy to the previous section, differentiation of eq.~\eqref{eq::res_ia_pot_largesep} yields
\begin{align}
  \vdk_{\text{m,ls},11} &=  - \int_0^{l_1} \int_0^{l_2} c_\text{m,ls} \, \vdH_1^\text{T} \frac{1}{d^{2m+4}} \left(  \vdH_1 d^{m+2} - (m+2) d^{m} \left( \vr_{1} - \vr_{2} \right) \otimes \left( \vr_{1} - \vr_{2} \right)^T \vdH_1 \right) \dd s_2 \dd s_1 & \nonumber \\
  & = \int_0^{l_1} \int_0^{l_2} c_\text{m,ls} \left( - \frac{1}{d^{m+2}} \vdH_1^\text{T} \vdH_1  +  \frac{\left(m+2\right)}{d^{m+4}} \vdH_1^\text{T} \left( \vr_{1} - \vr_{2} \right) \otimes \left( \vr_{1} - \vr_{2} \right)^T \vdH_1 \right)  \dd s_2 \dd s_1 &\\
  \vdk_{\text{m,ls},12} &= \int_0^{l_1} \int_0^{l_2} c_\text{m,ls} \left( \frac{1}{d^{m+2}} \vdH_1^\text{T} \vdH_2 - \frac{\left(m+2\right)}{d^{m+4}} \vdH_1^\text{T} \left( \vr_{1} - \vr_{2} \right) \otimes \left( \vr_{1} - \vr_{2} \right)^T  \vdH_2 \right) \dd s_2 \dd s_1 &\\
  \vdk_{\text{m,ls},21} &= \int_0^{l_1} \int_0^{l_2} c_\text{m,ls} \left( \frac{1}{d^{m+2}} \vdH_2^\text{T} \vdH_1  - \frac{\left(m+2\right)}{d^{m+4}} \vdH_2^\text{T} \left( \vr_{1} - \vr_{2} \right) \otimes \left( \vr_{1} - \vr_{2} \right)^T  \vdH_1 \right) \dd s_2 \dd s_1 &\\
  \vdk_{\text{m,ls},22} &= \int_0^{l_1} \int_0^{l_2} c_\text{m,ls} \left( - \frac{1}{d^{m+2}} \vdH_2^\text{T} \vdH_2  + \frac{\left(m+2\right)}{d^{m+4}} \vdH_2^\text{T} \left( \vr_{1} - \vr_{2} \right) \otimes \left( \vr_{1} - \vr_{2} \right)^T  \vdH_2 \right) \dd s_2 \dd s_1 &
\end{align}
See again eq.~\eqref{eq::centerline_discretization} for the definition of the shape function matrices~$\vdH_j$.
As mentioned before, the discrete element residual vectors in the specific case of Coulombic interactions directly follow for~$m=1$ and~$c_\text{m,ls} = C_\text{elstat} \lambda_1 \lambda_2$.
See \secref{sec::theory_electrostatics_pointcharges} for the definition of~$C_\text{elstat}$ and~\secref{sec::ia_pot_double_length_specific_evaluation_elstat} for the definition of the linear charge densities~$\lambda_i$.
Again, as mentioned already in~\secref{sec::ia_pot_double_length_specific_evaluation_elstat}, the case of long-range \textit{volume} interactions only requires to adapt the constant prefactor via~$c_\text{m,ls}=k m A_1 A_2 \rho_1 \rho_2$.

\section{Supplementary information on algorithms and code framework used for the simulations}\label{sec::algorithm_implementation_aspects}

\begin{description}
  \item[\normalfont{\textit{Implementation}}]
  All novel methods have been implemented in~C++ within the framework of the multi-purpose and multi-physics in-house research code BACI~\cite{BACI2018}.
  \item[\normalfont{\textit{Integration into existing code framework}}]
  The novel SSIP approach can be integrated very well in an existing nonlinear finite element solver for solid mechanics.
  In particular, it does not depend on a specific beam (finite element) formulation and has been used with geometrically exact Kirchhoff-Love as well as Simo-Reissner beam elements.
  Also, it is independent of the temporal discretization and has been used along with statics, Lie group Generalized-Alpha as well as Brownian dynamics.
  \item[\normalfont{\textit{Load/time stepping}}]
  We either applied a fixed step size or an automatic step size adaption that is outlined in the following.
  Starting from a given initial step size, a step is repeated with half of the previous step size if and only if the nonlinear solver did not converge within a prescribed number of iterations.
  This procedure may be repeated until convergence is achieved (or until a given finest step size is reached which aborts the algorithm).
  After four subsequent converging steps with a new, small step size, the step size is doubled.
  Again, this is repeated until the initial step size is reached.
  \item[\normalfont{\textit{Nonlinear solver}}]
  The Newton-Raphson algorithm used throughout this work is based on the package NOX which is part of the Trilinos project~\cite{Trilinos2012}.
  Unless otherwise stated, the Euclidean norms of the displacement increment vector and of the residual vector are used as convergence criteria.
  Typically, the corresponding tolerances were chosen as~$10^{-10}$ and~$10^{-7}$, respectively.
  In some of the numerical examples, an additional Newton step size control is applied.
  It restricts the step size such that a specified upper bound of the displacement increment per nonlinear iteration is not exceeded.
  In simple terms, it is meant to prevent any two points on two beams from moving too far and eventually crossing each other without being detected from one iteration to the other.
  For this reason, the value for this upper bound is typically chosen as half of the beam radius.
  \item[\normalfont{\textit{Linear solver}}]
  We use the algorithm UMFPACK~\cite{UMFPACK2004} which is a direct solver for sparse linear systems of equations based on LU-factorization and included in the package Amesos which is part of the Trilinos project~\cite{Trilinos2012}.
  \item[\normalfont{\textit{Parallel computing}}]
  The implementation of the novel methods supports parallel computing and is based on the package Epetra which is part of the Trilinos project~\cite{Trilinos2012}.
  See~\secref{sec::search_parallel_computing} for details on the partitioning of the problem in the context of the search algorithm applied to identify spatially proximate interaction partners.
  \item[\normalfont{\textit{Post-processing and visualization}}]
  The computer program MATLAB~\cite{MATLAB2017b} was used to post-process and plot simulation data.
  All visualizations of the simulation results were generated using Paraview~\cite{Paraview}.
\end{description}

\bibliography{library_MolecularInteractionBeamFEMethods}

\end{document}